\documentclass{article}
\usepackage[latin1]{inputenc}
\usepackage{graphics}
\usepackage{rotating}

\begin{document}

%%%%%%%%%%%%%%%%%%%%%%%%%

%% Macros for bibliography

%

\def\ea{{\textsl{et al.}}}

\def\aip{AIP Conf. Proc. }

\def\anp{Adv. Nucl. Phys. }

\def\arnps{Annu. Rev. Nucl. Part. Sci. }

\def\arns{Annu. Rev. Nucl.  Sci. }

\def\cjp{Can. J. Phys. }

\def\epja{Eur. Phys. J. A }

\def\fbs{Few-Body Systems }

\def\jpg{J. Phys. G: Nucl. Part. Phys. }

\def\nim{Nucl. Instr. Meth. Phys. Res. {\bf A}}

\def\np{Nucl. Phys. }

\def\nc{Nuovo Cimento }

\def\pl{Phys. Lett. }

\def\pr{Phys. Rev. }

\def\pra{Phys. Rev. A }

\def\prc{Phys. Rev. C }

\def\prd{Phys. Rev. D }

\def\prl{Phys. Rev. Lett. }

\def\prp{Phys. Rep. }

\def\prs{Proc. Roy. Soc. }

\def\rmp{Rev. Mod. Phys. }

\def\sjnp{Sov. J. Nucl. Phys. }

\def\zpa{Z. Phys. A }

\def\zpc{Z. Phys. C }

\def\up{ unpublished}

%

%Macros for variables and units

%

\def\t20{$t_{20}$}

\def\t20t{$\tilde{t}_{20}$}

\def\gev2{(GeV/c)$^2$}

\def\fm{fm$^{-1}$}

%%%%%%%%%%%%%%%%%%%%%%

\hfill \raisebox{1cm}{\parbox{1.5in}{DAPNIA/SPHN-01-02
                      JLAB-THY-01-6}}

\begin{center}
\huge The deuteron: structure and form factors
\end{center}

\vspace*{12pt}
\begin{center}
M. Gar\c{c}on\( ^{a} \) and J.W. Van Orden\( ^{b} \)
\end{center}
%\maketitle

\vspace*{12pt}
\begin{center}
{ \( ^{a} \)\textsl{\small DAPNIA/SPhN, CEA-Saclay, 91191
Gif-sur-Yvette, France }}\\

{ \( ^{b} \)\textsl{\small Old Dominion University, Norfolk, VA
23529  and Thomas Jefferson National Accelerator Facility, Newport
News, VA 23606, USA} }
\end{center}
\tableofcontents{}

\section{A historical introduction\label{sec:history}}

Diplon, deuton, deuteron: under different names, the nucleus of
deuterium, or diplogen, has been the subject of intense studies
since its discovery in 1932. As the only two-nucleon bound state,
its properties have continuously been viewed as important in
nuclear theory as the hydrogen atom is in atomic theory. Yet,
ambiguities remain in the relativistic description of this system
and the two-nucleon picture is incomplete: meson exchange and
nucleon excitation into resonances should be considered in the
deuteron description. The question of rare configurations where
the two nucleons overlap and loose their identity is still under
debate. We are still looking for the elusive effects of quarks in
the nuclear structure.

In this year of the millenium, the present article will first attempt to recall
the early discoveries, measurements and theories. It will then boldly jump over
decades of continuous efforts, building upon these, to present not an exhaustive
review but an up-to-date status of our understanding of the deuteron, with a
special emphasis on its electromagnetic form factors. To do justice to some
seventy years of activity in this field is an immense task which is more easily
approached by quoting here several reviews along this path~\cite{Bet36,Hul57,Gou66,Bro76,Rho79,Eri84,Fru84,Arv87,Eri88,Mac89,Won94,Car97,CarS98}.
The important subject of electro- and photo-disintegration of the deuteron will
be only partly covered, referring the reader to~\cite{Gil01}.

\subsection{Discovery of the deuteron\label{ssec:discovery}}

The existence of the first isotope of hydrogen was suggested in
1931 by Birge and Menzel~\cite{Bir31} in order to remove
discrepancies between two different measurements of the atomic
mass of hydrogen. A first estimate of an abundance ratio \( ^{1}
\)H/\( ^{2} \)H = 4500 was inferred from this hypothesis, close
indeed to the actual value of 6700. The stable isotope was
discovered by Urey and collaborators~\cite{Ure32} a few months
later, investigating distilled samples of natural hydrogen for the
optical atomic spectrum of \( ^{2} \)H in a discharge tube.
Isotopic separation to study the properties of deuterium quickly
became an intense activity. Its mass was measured by
Bainbridge~\cite{Bai32}. While Chadwick was discovering the
neutron, several tens of papers were written, in one years time,
devoted to the study of deuterium. An illuminating summary of this
early research was made by Bleakney and Gould~\cite{Ble33}. In
1933, {}``deutons{}'' were used as accelerated projectiles first
at Berkeley~\cite{Lew33}, then at Caltech and at Cavendish.
Chadwick and Goldhaber~\cite{Cha34} measured the first
photodisintegrations \( \gamma d\rightarrow pn \) in 1934.

\subsection{Early theories}

In 1932, there was no satisfactory theory of the nucleus. The
nucleus was thought to be composed of protons and electrons since
these were the only known charged particles and nuclei were seen
to emit electrons (\( \beta  \) decay). The electrons were needed
to cancel the positive charge of some of the protons in order to
account for nuclei with identical charges, but with different
masses, and to allow for the possibility of binding of the nucleus
by means of electric forces. This was clearly unsatisfactory
because the Coulomb force could not account for the binding
energies of nuclei and the attempt to construct the nuclei from
the incorrect number of spin-1/2 particles could not produce the
correct nuclear spins.

The discovery of the neutron, shortly after that of the deuteron,
did not immediately eliminate the confusion since the previous
model persisted by simply describing the neutron as a bound system
of a proton and an electron. Based on this faulty assumption,
Heisenberg produced the first model of proton-neutron
force~\cite{Hei32}. Since it was not possible to actually
construct a description of the neutron with the \( ep \) model,
Heisenberg simply assumed that the \( pn \) force could be
described by a phenomenological potential and that the neutron was
a spin-1/2 object like the proton. Based on an analogy with the
binding of the H\( ^{+}_{2} \) ion by electron sharing, Heisenberg
proposed that the force must involve the exchange of both spin and
charge in the form of \( \sigma ^{(1)}\cdot \sigma ^{(2)}\tau
^{(1)}\cdot \tau ^{(2)} \). Forces containing the remaining forms
of spin and isospin operators were soon introduced by
Wigner~\cite{Wig33}, Majorana~\cite{Maj33} and
Bartlett~\cite{Bar36}. In all cases the spatial form of the
potentials was to be determined phenomenologically to reproduce
the deuteron properties and the available nucleon-nucleon (\( NN
\)) scattering data. In 1935Bethe and Peierls~\cite{Bet35} wrote
the Hamiltonian of the {}``diplon{}'' with an explicit
introduction of a short range interaction. This approach became
the mainstay of nuclear physics which has produced considerable
success in describing nuclear systems and reactions. The \( ep \)
model of the neutron was not completely abandoned until after the
Fermi theory~\cite{Fer34} of \( \beta  \) decay became widely
accepted.

The progress in discoveries and understanding was then so great that, in spite
of an otherwise bleak social or political situation in many countries involved,
this period is recalled as {}``The Happy Thirties{}'' from a physicist's point
of view~\cite{Bet77}.

One of the other great theoretical preoccupations of the late
1920's and the 1930's was the development of quantum field theory
starting with the first works of Dirac on quantum electrodynamics
(QED)~\cite{Dir27}, the Dirac equation for the
electron~\cite{Dir28} and the Dirac hole theory~\cite{Dir29} with
field theory reaching its final modern form with
Heisenberg~\cite{Hei34}. QED at this time was very successful at
tree-level but the calculation of finite results from loops was
not really tractable until the introduction of systematic
renormalization schemes in the late 1940's. The first attempt to
apply quantum field theory to the strong nuclear force was
Yukawa's suggestion~\cite{Yuk35} that the force was mediated by a
new strongly coupling massive particle which became known as the
pion. This started another strong thread in the theoretical
approach of the nucleus by using meson-nucleon theory to obtain
nuclear forces consistent with the phenomenological potential
approach. The primary attraction of this approach is that a more
microscopic description of the degrees of freedom of the problem
is provided and that additional constraints are imposed on the
theory by the necessity of simutaneously describing
nucleon-nucleon and meson-nucleon scattering. Ultimately, as it
became clear that the mesons and nucleons were themselves
composite particles, meson-nucleon theories were replaced as
fundamental field theories of the strong interactions by quantum
chromodynamics (QCD). However, the meson-nucleon approach is still
a strong element in nuclear physics as a basis for phenomenology
and is making a potentially more rigorous comeback in the form of
the effective field theories associated with chiral perturbation
theory. This situation is unlikely to change until it becomes
possible to at least describe the \( NN \) force and the deuteron
directly from QCD.

\subsection{Spin\label{ssec:spin}}

Breit and Rabi \cite{Bre31} first suggested the use of magnetic
deflection of an atomic beam in an inhomogeneous field to measure
nuclear spins. The coupling of electronic (\( J_{e} \)) and
nuclear (\( J \)) spins is not totally negligible compared to the
coupling of the electronic spin to the external magnetic field,
provided the latter is weak enough. One then observes \(
(2J_{e}+1)\times (2J+1) \) lines with a predicted intensity
pattern. The atomic and molecular beam studies were to be
implemented with great success (see Sec.~\ref{sec:static}), but
the first determination of the deuteron spin used other methods.

Farkas and collaborators \cite{Far34} demonstrated the ortho-para
conversion in the diplogen (as they called the deuterium molecule)
and determined the spin and statistics of the nucleus from the
equilibrium ratio between these two states at different
temperatures. They concluded that the diplogen nucleus must obey
Bose-Einstein statistics, that the most probable value of its spin
was 1, and that its magnetic moment was about one fifth of that of
the proton.

Using photographic photometry, the alternating intensities in the molecular
spectrum of deuterium were investigated by Murphy and Johnston~\cite{Mur34},
who concluded that indeed \( J=1 \) for the {}``deuton{}''.

\subsection{Connection with OPE}

The deuteron thus quickly appeared as a loosely bound pair of nucleons with
spins aligned (spin triplet state). The existence of a small quadrupole moment
(see Sec.~\ref{sssec:Qd}) implies that these two nucleons are not in a pure
\( S \) state of relative orbital angular momentum, and that the force between
them is not central. Taking into account total spin and parity, an additional
\( D \) wave component is allowed. Such a \( D \) wave can be generated by
the tensor part of the one-pion exchange (OPE) potential~\cite{Bro76,Gle62}.

\section{The nonrelativistic two-nucleon bound state\label{sec:dwf}}

\subsection{The potential model of the deuteron}

The potential model of the deuteron is described by the Hamiltonian \begin{equation}
\hat{H}=\hat{T}_{1}+\hat{T}_{2}+\hat{V}
\end{equation}
 where \( \hat{T}_{i} \) is the kinetic energy operator for particle \( i \)
and \( \hat{V} \) is the two-body potential. Successful \( NN \)
potentials must have several basic characteristics in order to
satisfactorily describe the deuteron static properties and the \(
NN \) scattering data. The long distance part of the potentials is
described by one-pion exchange while the intermediate and short
range parts may be either parameterized in terms of simple
functional forms, or obtained from models involving meson
exchanges. The very strong anticorrelation of nucleons requires
that these potentials be repulsive at short distances. The
potentials must have terms involving scalar, spin-spin, tensor and
spin-orbit forces. The tensor force is of particular importance in
producing the single spin-1, iso-singlet deuteron bound state. The
long range tensor force is provided automatically by the exchange
of the pseudoscalar pion. Modern phenomenological potentials also
include additional nonlocalities by means of terms quadratic in
the relative momentum and/or quadratic spin-orbit terms. Improved
fits to scattering data also require that isospin symmetry
breaking be imposed via the inclusion of electromagnetic
interactions between nucleons and by additional explicit isospin
symmetry breaking terms in the potential. By fitting the
potentials directly to the scattering data, several
phenomenological potentials have been contructed that fit the
scattering database with \( \chi ^{2} \) very close to 1.

An discussion of the most commonly used \( NN \) potentials may be found in
the review~\cite{CarS98}. These include the so-called Reid-SC~\cite{RSC},
Paris~\cite{Paris}, Bonn~\cite{Bonn}, CD-Bonn~\cite{CD-Bonn}, Nijmegen~\cite{Nijm},
Reid93~\cite{Nijm} and Argonne \( v_{18} \)~\cite{Wir95} potentials.

Given a potential, the resolution of the Schrödinger equation in the \( T=0,J=1 \)
\( np \) channel leads to the bound state wave function discussed hereafter.

\subsection{The deuteron wave function\label{ssec:dwf}}

\begin{figure}
{\par\centering \resizebox*{6cm}{!}{\includegraphics{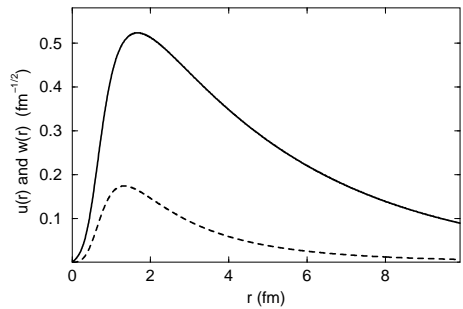}} \par}

\caption{The deuteron reduced radial wave functions \protect\( u\protect \) (solid
line) and \protect\( w\protect \) (dashed) for the Argonne \protect\( v_{18}\protect \)
potential, as a function of the relative coordinate.\label{fig:uw}}
\end{figure}

The tensor force requires that the deuteron wave function be a mixture of \( ^{3}S_{1} \)
and \( ^{3}D_{1} \) components, so the deuteron wave function is of the form
\begin{equation}
\psi _{^{^{M}}}({\textbf {x}})=\frac{u(r)}{r}{\mathcal{Y}}_{101}^{M}(\theta ,\phi )+\frac{w(r)}{r}{\mathcal{Y}}_{121}^{M}(\theta ,\phi )\, ,
\end{equation}
 where \begin{equation}
{\mathcal{Y}}_{JLS}^{M}(\theta ,\phi )=\sum _{m_{L},m_{S}}\left< J,M\left| L,m_{L};S,m_{S}\right. \right> Y_{LM}(\theta ,\phi )\left| S,m_{s}\right>
\end{equation}
 are the spin-spherical harmonics. The reduced radial wave functions \( u(r) \)
and \( w(r) \) correspond to the \( S \) and \( D \) waves respectively.
The \( S \) and \( D \) state probability densities are defined as \begin{equation}
\rho _{S}(r)=u^{2}(r)\quad \hbox {and}\quad \rho _{D}(r)=w^{2}(r)\, .
\end{equation}
 The corresponding \( S \) and \( D \) state probabilities are then \begin{equation}
P_{S}=\int _{0}^{\infty }\rho _{S}(r)dr\quad \hbox {and}\quad P_{D}=\int _{0}^{\infty }\rho _{D}(r)dr
\end{equation}
 and the normalization of the wave function requires that \begin{equation}
P_{S}+P_{D}=1\, .
\end{equation}
 The reduced radial wave functions for the Argonne \( v_{18} \) potential are
shown in Fig.~\ref{fig:uw}. The wave functions for other modern potentials
are very similar.

From the wave function, a characteristic size of the deuteron \(
r_{m} \) is defined as the rms-half distance between the two
nucleons : \begin{equation} \label{eq:rm}
r^{2}_{m}=\frac{1}{4}\int _{0}^{\infty }\left[
u^{2}(r)+w^{2}(r)\right] r^{2}dr
\end{equation}

\begin{figure}
{\par\centering \resizebox*{10cm}{!}{\includegraphics{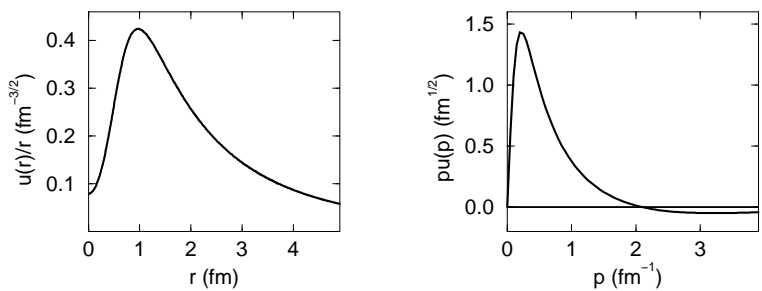}} \par}

\caption{The deuteron \protect\( S\protect \) wave function in configuration space
and in momentum space: \protect\( u(r)/r\protect \) and \protect\( pu(p)\protect \)
(calculated from the Argonne \protect\( v_{18}\protect \) potential).\label{fig:pu}}
\end{figure}
\begin{figure}
{\par\centering
\resizebox*{11cm}{!}{\includegraphics{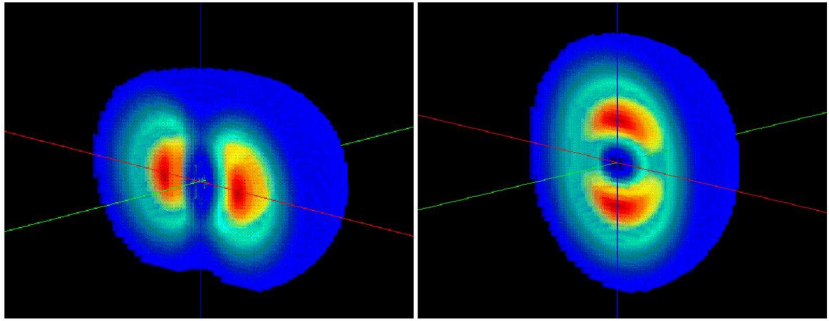}} \par}

\caption{Deuteron densities in \protect\( M=0\protect \) (left) and \protect\( M=1\protect \)
(right) magnetic substates. The red spots correspond to the maximal nucleonic
densities, while the dark volumes correspond to lower densities (outer surface
is for 10\% of maximal density). See~\cite{Eri88,For96} for equivalent representations.\label{fig:pretty_picture}}

% The following two lines should replace the sentence {}``The red spots...{}'' if this figure appears in black and white)

% The dark spots inside the clear blobs correspond to the maximal nucleonic densities.

% Other dark volumes correspond to lower densities (outer surface is for 10\% of maximal density).
\end{figure}

A conspicuous feature of the nucleon-nucleon interaction is the short range
repulsion, which leads the radial \( S \) wave function \( u(r)/r \) to be
significantly reduced at distances smaller than approximately 1 fm (see Fig.~\ref{fig:pu}).
This introduces a distance scale in the wave function in addition to the overall
deuteron size. This small distance behaviour is the subject of most of the experimental
and theoretical studies which will be presented in Secs.~\ref{sec:e-d} and~\ref{sec:theory}.
As a result of this dip at small \( r, \) the Fourier transform \( u(p) \)
contains a node at approximately 2 fm\( ^{-1} \), as seen also in Fig.~\ref{fig:pu}.
The \( u \) and \( w \) wave functions are given in momentum space by :\begin{equation}
\label{eq:uwp}
u(p)=\int _{0}^{\infty }u(r)j_{0}(pr)rdr\quad \hbox {and}\quad w(p)=-\int _{0}^{\infty }w(r)j_{2}(pr)rdr\, .
\end{equation}
To conclude this presentation of the size and shape of the deuteron, the densities
\( |\psi ^{0}(\mathbf{x})|^{2} \) and \( |\psi ^{1}(\mathbf{x})|^{2} \) are
illustrated in Fig.~\ref{fig:pretty_picture}.

In this simple potential model of the deuteron, the magnetic moment of the deuteron
is determined entirely by the \( D \) state probability \( P_{D} \)~: \begin{equation}
\mu _{d}=\mu _{s}-\frac{3}{2}\left( \mu _{s}-\frac{1}{2}\right) P_{D}\, ,
\end{equation}
 where \( \mu _{s}=\mu _{n}+\mu _{p} \) is the isoscalar nucleon magnetic moment.
The deuteron electric quadrupole moment is also determined from the wave functions:
\begin{equation}
Q_{d}=\frac{1}{\sqrt{50}}\int _{0}^{\infty }w(r)\left[ u(r)-\frac{1}{\sqrt{8}}w(r)\right] r^{2}dr\, .
\end{equation}
 In both cases, these quantities are modified by extensions to the basic potential
model (see Sec.~\ref{sec:theory}). In particular the direct relationship between
the magnetic moment and the \( D \) state probability will be broken by such
extensions and, therefore, this probability is not an observable~\cite{Fri79}.

Since the nuclear force is of finite range, it is easy to determine the asymptotic
form of the wave functions\begin{equation}
\label{eq:asymp_uw}
u(r)\sim A_{S}e^{-\gamma r}\quad \hbox {and}\quad w(r)\sim A_{D}e^{-\gamma r}\left[ 1+\frac{3}{\gamma r}+\frac{3}{(\gamma r)^{2}}\right] \quad {\textrm{as}}\; r\rightarrow \infty
\end{equation}
 where \( \gamma \simeq \sqrt{\varepsilon m} \), with \( m \) being the reduced
\( np \) mass and \( \varepsilon  \) the deuteron binding energy (see Ref.~\cite{Sch98}
for a relativistic definition of \( \gamma  \)). \( A_{S} \) and \( A_{D} \)
are the asymptotic normalization factors, determined by matching the asymptotic
form (\ref{eq:asymp_uw}) to the calculated wave functions in the interior region
where the potential is nonvanishing. \( A_{S} \) and the ratio \begin{equation}
\label{eq:etad}
\eta _{d}=\frac{A_{D}}{A_{S}}
\end{equation}
 are directly related to observables as discussed in the next section.

\section{Static and low energy properties\label{sec:static}}

A review of the measured static properties of the deuteron, together with the
low energy neutron proton (\( np \)) scattering parameters, was given by Ericson
and Rosa-Clot~\cite{Eri84,Eri83}, who studied in detail their connection with
\( NN \) potential models. An updated status of the experimental information
on the deuteron is given in Table~\ref{tab:static} and discussed hereafter.

\subsection{Deuteron static properties (experiment) \label{ssec:static_exp}}

\vspace{0.3cm}
{
\begin{table}

\caption{Experimental determinations of the deuteron static properties \label{tab:static}}
{\centering \begin{tabular}{|c|c|c|}
\hline
Quantity&
Most recent &
Value\\
&
determination&
\\
\hline
\hline
Mass \( M_{d} \) &
\cite{Dif94,Moh00}&
1875.612762 (75) MeV\\
\hline
Binding energy \( \varepsilon  \)&
\cite{Kes99}&
2.22456612 (48) MeV\\
\hline
Magnetic dipole moment \( \mu _{d} \)&
\cite{Moh00}&
0.8574382284 (94) \( \mu _{N} \)\\
\hline
Electric quadrupole moment \( Q_{d} \)&
\cite{Eri83,Cod71,Bis79}&
0.2859 (3) fm\( ^{2} \)\\
\hline
Asymptotic ratio \( \eta _{d}=A_{D}/A_{S} \)&
\cite{Rod91}&
0.0256 (4)\\
\hline
Charge radius \( r_{ch} \)&
 \cite{Sic98}&
2.130 (10) fm\\
Matter radius \( r_{m} \)&
\cite{Hub98,Mar95} &
1.975 (3) fm\\
\hline
Electric polarizability \( \alpha _{E} \)&
\cite{Rod82,Fri83}&
0.645 (54) fm\( ^{3} \)\\
\hline
\end{tabular}\par}\end{table}
\par}
\vspace{0.3cm}

\subsubsection{Mass and binding energy}

In the past ten years, significant progress in the precision of the measurement
of some atomic masses has been made by comparing cyclotron frequencies of different
pairs of ions in a Penning trap. In this way, the deuterium atomic mass is known
with a relative precision of \( 10^{-8} \)~\cite{Dif94}.

The deuteron binding energy is best determined by measuring the energy of the
gamma-rays coming from radiative \( np \) capture with thermal neutrons. This
energy is now measured to a relative accuracy of \( 2\times 10^{-7} \), using
a crystal diffraction spectrometer~\cite{Kes99}. At this precision, even for
such a low energy process, some of the earlier work may have to be corrected
for relativistic kinematics and the Doppler effect~\cite{Yon99}.

The binding energy and the mass measurements can be combined for the most precise
determination of the neutron mass~\cite{Kes99}. The errors on the values reported
in Table~\ref{tab:static} include the uncertainty in the atomic mass unit
\( u=931.494013(37) \) MeV~\cite{Moh00}. They are well beyond the accuracy
of nuclear models.

\subsubsection{Magnetic dipole moment}

The first measurement of the deuteron magnetic moment was performed by Rabi
in 1934~\cite{Rab34}, based on a principle~\cite{Bre31} already alluded
to. From the deflection of an atomic beam in an inhomogeneous magnetic field
to the use of molecular beam resonance and other methods, these techniques were
continuously improved~\cite{Ram53}. Precise measurements of nuclear magnetic
resonance frequencies of the deuteron and proton in the HD molecule give the
ratio of deuteron to proton magnetic moments. However, the adopted value in
Table~\ref{tab:static} results from a simultaneous determination of the electronic
and nuclear Zeeman energy levels splittings in the deuterium atom, yielding
the ratio of deuteron to electron magnetic moments~\cite{Moh00}.

\subsubsection{Electric quadrupole moment\label{sssec:Qd}}

The deuteron was found to possess an electric quadrupole moment in
1939~\cite{Kel39}. This discovery had far reaching consequences:
it meant that nuclear forces were not central and were more
complex that previously thought. It was to become the best
qualitative and quantitative evidence for the role of pions in
nuclear physics~\cite{Eri84}.

In contrast to the case of the magnetic moment which is determined
through its coupling to an external applied magnetic field, the
quadrupole moment does not couple to an external electric field.
One measures instead, in HD or D\( _{2} \) molecules, the
interaction of the deuteron quadrupole moment with the electric
field gradient created along the molecular axis by the
neighbouring atom. The experiment provides an electric quadrupole
interaction constant~\cite{Cod71} which must be divided by the
theoretically calculated field gradient~\cite{Bis79} to obtain the
quadrupole moment.

\subsubsection{Asymptotic ratio \protect\( D/S\protect \)\label{sssec:etad}}

The ratio~(\ref{eq:etad}) is deduced from measurements of tensor analyzing
powers in sub-Coulomb \( (d,p) \) reactions on heavy nuclei by comparison with
calculations in the distorted wave Born approximation (DWBA)~\cite{Rod91}.
The value of \( \eta _{d} \) is then directly proportional to the analyzing
powers. Other determinations based on \( dp \) elastic scattering rely on pole-extrapolation
and are somewhat less precise.

\subsubsection{Radius and size\label{sssec:radius}}

The deuteron size may be characterized by a charge radius \( r_{ch} \) and
by a matter radius \( r_{m} \). The latter is defined from the deuteron wave
function~(\ref{eq:rm}).

Elastic electron scattering has been used since the early
fifties~\cite{Hof56} to measure the shape of nuclei. This topic
will be discussed at length in Sec.~\ref{sec:e-d}. At low momentum
transfer, the cross section data yield the charge rms-radius of
the target nucleus through the relation \( r_{ch}^{2}=-6\left.
dG_{C}/dQ^{2}\right| _{Q^{2}=0} \) (see Sec.~\ref{ssec:deff} for
the definition of \( G_{C} \)). It was demonstrated recently that
a precision extraction of the deuteron rms charge radius from
electron scattering data requires taking into account the Coulomb
distortion of the incoming and outgoing electrons~\cite{Sic98}. A
new analysis of the world data was then performed, yielding the
value in Table~\ref{tab:static}. The quoted uncertainty combines
quadratically the fit statistical error and the dominant
systematic error, the latter coming mostly from experimental
normalization uncertainties. The usual radiative corrections to
electron scattering do not include the contribution of hadronic
vacuum polarization, but that effect should be smaller than the
present uncertainties when extracting charge radii~\cite{Fri99}.

The rms charge radius \( r_{ch} \) is related to the matter (or
rather nucleonic) rms-radius \( r_{m} \)~\cite{Kla86}
by\begin{equation} \label{eq:r2} r_{ch}^{2}=r_{m}^{2}+\Delta
r_{m}^{2}+r_{p}^{2}+r_{n}^{2}+\frac{3}{4}\left( \frac{\hbar
}{m_{p}}\right) ^{2},
\end{equation}
where \( r_{p}=0.862(12) \) fm is the proton charge rms-radius~\cite{Sim80}
and \( r_{n}^{2}=-0.113(5) \) fm\( ^{2} \) is the neutron charge ms-radius~\cite{Kop95}.
\( \Delta r_{m}^{2} \) is a contribution from non-nucleonic degrees of freedom,
close to 0 but with an uncertainty estimated to \( \pm 0.01 \) fm\( ^{2} \).
The quantity \( r_{d} \) given by \( r_{d}^{2}=r_{m}^{2}+\Delta r^{2}_{m} \)
is usually defined as the deuteron radius. The last term in (\ref{eq:r2}) is
of relativistic origin~\cite{Bea94}. Note that the above quoted value of \( r_{p} \),
as extracted from \( ep \) elastic scattering, is in slight disagreement (\( 2\sigma  \)
difference) with recent Lamb shift measurements~\cite{Kar99}. Finally, the
theoretical uncertainty in the deuteron radius associated with the correction
due to the nucleon finite size has been estimated to about 0.002 fm~\cite{Eri85}.

The nuclear-dependent correction to the Lamb shift in hydrogen and
deuterium atoms is directly proportional to the nuclear
mean-square radius. From the isotope shifts in the pure optical
frequency of \( 1S-2S \) two-photon transitions in atomic hydrogen
and deuterium, the difference \( r_{ch}^{2}-r_{p}^{2} \) of
mean-square charge radii for the deuteron and proton is accurately
determined~\cite{Hub98}. Small corrections due to the deuteron
polarizability seem to be under control~\cite{Mar95}. Then from
(\ref{eq:r2}) and the value of \( r_{n}^{2} \), \( r_{m} \) is
extracted with a better precision than \( r_{ch} \) from \( ed \)
scattering. Our quoted uncertainty is larger than in
Ref.~\cite{Hub98} because of the use of a larger uncertainty in \(
r_{n}^{2} \) and the addition of the uncertainty due to \( \Delta
r_{m}^{2} \).

Taking into account additional small corrections summarized in Ref.~\cite{Sic98},
the two results given in Table~\ref{tab:static} are quite compatible, in the
sense that they satisfy Eq.(\ref{eq:r2}). Furthermore, the value of \( r_{m} \)
follows the expectations from modern \( NN \) potentials.

\subsubsection{Electric polarizability}

The electric polarizability \( \alpha _{E} \) characterizes how the deuteron
charge distribution can be stretched and acquire an electric dipole moment under
the influence of an external electric field. It was determined through elastic
scattering of deuterons from \( ^{208}Pb \) well below the Coulomb barrier~\cite{Rod82}
and extracted from low energy photoabsorption~\cite{Fri83}. The two results
are slightly incompatible (\( 2\sigma  \) difference). The value in Table~\ref{tab:static}
is our average.

\subsection{Low energy \protect\( np\protect \) scattering parameters}

The deuteron may also be viewed as a pole in the \( S \)-matrix describing
the \( np \) scattering in the coupled \( ^{3}S_{1} \) and \( ^{3}D_{1} \)
channels. This \( S \)-matrix can be experimentally determined from a phase-shift
analysis of the scattering data. An extrapolation to negative energies down
to the measured deuteron binding energy, either by an effective range expansion~\cite{Eri84}
or a \( P \)-matrix approach~\cite{Sto88}, allows to extract \( \eta _{d} \)
and \( A_{S} \) . The asymptotic ratio \( \eta _{d}=0.0254(2) \) is given
by the extrapolated mixing parameter \( \varepsilon _{1} \) while the asymptotic
\( S \) state normalization \( A_{S}=0.8847(8) \) fm\( ^{-1/2} \) is essentially
related to the effective range and thus to the triplet scattering length \( a_{t} \)
(numerical values from~\cite{Sto94}).

\subsection{Static properties and the \protect\( NN\protect \) potential}

All deuteron static properties discussed above are well reproduced
by \( NN \) potential calculations such as Argonne \( v_{18} \),
Nijmegen II, Reid93 or CD-Bonn, with the notorious exception of
the quadrupole moment, which is always a few percent too low (see
also Sec.~\ref{ssec:potmod}). Meson exchange contributions, to be
discussed later, must be taken into account for a better agreement
with data. \textit{}The binding energy \( \varepsilon  \) is taken
as a constraint in the determination of all potentials.

Compilations of deuteron static properties caculated with recent \( NN \) interaction
models appear in~\cite{CarS98,Lev99}. Note that most potentials result in
a \( D \) wave probability \( P_{D} \) between 5.6 and 5.8\%, except for the
CD-Bonn potential where \( P_{D}=4.83\% \).

Various correlations were established between the calculated static properties,
independently of the \( NN \) potential used. For example, linear relationships
between \( A_{S} \) and \( r_{m} \)~\cite{Eri84}, or \( A_{S}^{2}(1+\eta ^{2}) \)
and \( r_{m}^{2} \)~\cite{Mar95}, and \( Q_{d}/A_{S}^{2} \) and \( \eta  \)~\cite{Eri84}
were established, the latter depending on the value of the \( \pi NN \) coupling
constant used in the potential calculation. For more recent potentials, linear
relationships between \( \mu _{d} \) and \( \eta  \) on one hand, \( Q_{d} \)
and \( \eta  \) on the other hand, are illustrated in~\cite{Lev99}. Finally
the electric polarizability \( \alpha _{E} \) is directly proportional to \( r^{2}_{m} \)~\cite{Mar95}.

\section{Elastic electron-deuteron scattering\label{sec:e-d}}

\subsection{Deuteron electromagnetic form factors\label{ssec:deff}}

Invoking Lorentz invariance, current conservation, parity and
time-reversal invariance, the general form of the electromagnetic
current matrix element for elastic electron scattering from the
spin-1 deuteron can be shown to have the general form
\cite{Arn81}:
\begin{eqnarray}
G^{\mu }_{\lambda _{d}'\lambda _{d}}(P',P) & = & -\biggl \{G_{1}(Q^{2})(\xi ^{*}_{\lambda _{d}'}(P')\cdot \xi _{\lambda _{d}}(P))(P'+P)^{\mu }\nonumber \\
 & + & G_{2}(Q^{2})\Bigl [\xi ^{\mu }_{\lambda _{d}}(P)(\xi ^{*}_{\lambda _{d}'}(P')\cdot q)-\xi ^{\mu *}_{\lambda _{d}'}(P')(\xi _{\lambda _{d}}(P)\cdot q)\Bigr ]\nonumber \\
 & - & G_{3}(Q^{2}){1\over 2M_{d}^{2}}(\xi ^{*}_{\lambda _{d}'}(P')\cdot q)(\xi _{\lambda _{d}}(P)\cdot q)(P'+P)^{\mu }\biggr \}\label{invariantCurrent}
\end{eqnarray}
 where \( M_{d} \) is the deuteron mass, \( P \) and \( P' \) are the initial
and final deuteron four-momenta, \( q=P'-P \) is the virtual photon four-momentum,
\( \xi ^{\mu }_{\lambda _{d}}(P) \) and \( \xi ^{\mu *}_{\lambda _{d}'}(P') \)
are the polarization four-vectors for the inital and final deuteron states.
The \( G_{i}(Q^{2}) \) are form factors depending only upon the virtual photon
four-momentum; assuming hermiticity, they are real. Since the virtual photon
four-momentum is always spacelike for electron scattering, we use the convention
\( Q^{2}\equiv -q^{2}={\textbf {q}}^{2}-\nu ^{2} \).

The current may be expressed in terms of charge monopole, magnetic dipole and
charge quadrupole form factors. These are related to the \( G_{i}(Q^{2}) \)'s
by: \begin{eqnarray}
G_{C}(Q^{2}) & = & G_{1}(Q^{2})+{2\over 3}\eta G_{Q}(Q^{2})\nonumber \\
G_{M}(Q^{2}) & = & G_{2}(Q^{2})\nonumber \\
G_{Q}(Q^{2}) & = & G_{1}(Q^{2})-G_{2}(Q^{2})+(1+\eta )G_{3}(Q^{2})
\end{eqnarray}
 with \begin{equation}
\eta =\frac{Q^{2}}{4M_{d}^{2}}\; .
\end{equation}
 These form factors are normalized such that \begin{eqnarray}
G_{C}(0) & = & 1\, ,\nonumber \\
G_{M}(0) & = & \frac{M_{d}}{m_{p}}\mu _{d}\, ,\nonumber \\
G_{Q}(0) & = & M_{d}^{2}Q_{d}\; .\label{eq:ffnorm}
\end{eqnarray}
 The experimental values of \( G_{M}(0) \) and \( G_{Q}(0) \) are respectively
1.714 and 25.83 (see Table~\ref{tab:static}).

\subsection{Observables\label{ssec:obs}}

In the Born approximation of a one-photon exchange mechanism and neglecting
the electron mass, the cross section for elastic scattering of longitudinally
polarized electrons from a polarized deuteron target can be calculated from
the current to give in the laboratory frame~\cite{Don86}: \begin{eqnarray}
{d\sigma \over d\Omega } & = & \frac{\sigma _{M}}{1+\frac{2E}{M_{d}}\sin ^{2}\frac{\theta }{2}}\left[ v_{L}R_{L}+v_{T}R_{T}+v_{TT}R_{TT}+v_{TL}R_{TL}\right. \nonumber \\
 &  & \qquad \qquad \qquad \qquad +\left. 2hv_{T'}R_{T'}+2hv_{TL'}R_{TL'}\right] \, ,\label{eq:sigma_ed}
\end{eqnarray}
 where \begin{equation}
\sigma _{M}=\left[ \frac{\alpha \cos \frac{\theta }{2}}{2E\sin ^{2}\frac{\theta }{2}}\right] ^{2}
\end{equation}
 is the Mott cross section, \( E \) the electron beam energy, \( \theta  \)
the electron scattering angle and \( h=\pm \frac{1}{2} \) the electron helicity.
The \( R_{l} \) are response functions and the \( v_{l} \) kinematical factors
are: \begin{eqnarray}
v_{L} & = & \left( \frac{Q^{2}}{\mathbf{q}^{2}}\right) ^{2}\nonumber \\
v_{T} & = & \frac{1}{2}\frac{Q^{2}}{\mathbf{q}^{2}}+\tan ^{2}\frac{\theta }{2}\nonumber \\
v_{TT} & = & -\frac{1}{2}\frac{Q^{2}}{\mathbf{q}^{2}}\nonumber \\
v_{TL} & = & -\frac{1}{\sqrt{2}}\frac{Q^{2}}{\mathbf{q}^{2}}\left[ \frac{Q^{2}}{\mathbf{q}^{2}}+\tan ^{2}\frac{\theta }{2}\right] ^{\frac{1}{2}}\nonumber \\
v_{T'} & = & \left[ \frac{Q^{2}}{\mathbf{q}^{2}}+\tan ^{2}\frac{\theta }{2}\right] ^{\frac{1}{2}}\tan \frac{\theta }{2}\nonumber \\
v_{TL'} & = & -\frac{1}{\sqrt{2}}\frac{Q^{2}}{\mathbf{q}^{2}}\tan \frac{\theta }{2}\; .
\end{eqnarray}
 For elastic scattering, \begin{equation}
\frac{Q^{2}}{\mathbf{q}^{2}}=\frac{1}{1+\eta }\, .
\end{equation}

Each of the response functions can be written as \begin{equation}
R_{l}(Q^{2})=\sum ^{9}_{i=1}R_{l}(Q^{2},\tau _{i})\tau _{i}
\end{equation}
 where \( l=\{L,T,TT,LT,T',LT'\} \) and the \( \tau _{i} \) are members of
the set of the unique deuteron density matrix elements expressed in terms of
elements of a spherical tensor \( \rho _{kq} \). This set is represented by
\begin{eqnarray}
\tau _{i} & = & \left\{ \rho _{00},\sqrt{3\over 2}\rho _{10},{1\over \sqrt{2}}\rho _{20},\sqrt{3}Re\rho _{22},\sqrt{3}Im\rho _{22},\right. \nonumber \\
 &  & \qquad \left. \sqrt{3\over 2}Re\rho _{11},\sqrt{3\over 2}Im\rho _{11},\sqrt{3\over 2}Re\rho _{21},\sqrt{3\over 2}Im\rho _{21}\right\} \, .
\end{eqnarray}

The nonvanishing reponse functions for elastic scattering may be written in
function of the deuteron form factors~: \begin{eqnarray}
R_{L}(Q^{2},\rho _{00}) & = & \left( 1+\eta \right) ^{2}\left[ G_{C}^{2}(Q^{2})+\frac{8}{9}\eta ^{2}G_{Q}^{2}(Q^{2})\right] \nonumber \\
R_{L}(Q^{2},{1\over \sqrt{2}}\rho _{20}) & = & -\frac{8}{3}\left( 1+\eta \right) ^{2}\left[ \eta G_{C}(Q^{2})G_{Q}(Q^{2})+\frac{1}{3}\eta ^{2}G_{Q}^{2}(Q^{2})\right] \nonumber \\
R_{T}(Q^{2},\rho _{00}) & = & \frac{4}{3}\eta \left( 1+\eta \right) G_{M}^{2}(Q^{2})\nonumber \\
R_{T}(Q^{2},\frac{1}{\sqrt{2}}\rho _{20}) & = & -\frac{2}{3}\eta \left( 1+\eta \right) G_{M}^{2}(Q^{2})\nonumber \\
R_{TT}(Q^{2},\sqrt{3}Re\rho _{22}) & = & \frac{2}{3}\eta \left( 1+\eta \right) G_{M}^{2}(Q^{2})\nonumber \\
R_{TL}(Q^{2},\sqrt{\frac{3}{2}}Re\rho _{21}) & = & \frac{8}{3}\left( \eta +\eta ^{2}\right) ^{\frac{3}{2}}G_{M}(Q^{2})\, G_{Q}(Q^{2})\nonumber \\
R_{T'}(Q^{2},\sqrt{\frac{3}{2}}\rho _{10}) & = & -\frac{2}{3}\eta \left( 1+\eta \right) G_{M}^{2}(Q^{2})\nonumber \\
R_{TL'}(Q^{2},\sqrt{\frac{3}{2}}Re\rho _{11}) & = & -\frac{8}{3}\eta ^{\frac{1}{2}}\left( 1+\eta \right) ^{\frac{3}{2}}G_{M}(Q^{2})\left[ G_{C}(Q^{2})+\frac{\eta }{3}G_{Q}(Q^{2})\right] \label{eq:R}
\end{eqnarray}

It is conventional to write the cross section as\begin{eqnarray}
\frac{d\sigma }{d\Omega } & = & \frac{\sigma _{M}}{1+\frac{2E}{M_{d}}\sin ^{2}\frac{\theta }{2}}\times \left[ A(Q^{2})+B(Q^{2})\tan ^{2}\frac{\theta }{2}\right] \nonumber \\
 &  & \quad \times \left[ 1+\rho _{20}\cdot t_{20}(Q^{2},\theta )+2Re\rho _{21}\cdot t_{21}(Q^{2},\theta )+2Re\rho _{22}\cdot t_{22}(Q^{2},\theta )\right. \nonumber \\
 &  & \qquad \left. +h\rho _{10}\cdot t_{10}(Q^{2},\theta )+2hRe\rho _{11}\cdot t_{11}(Q^{2},\theta )\right] \label{eq:sigma_ed_pol}
\end{eqnarray}
 where the unpolarized elastic structure functions \( A(Q^{2}) \) and \( B(Q^{2}) \)
are defined as \begin{eqnarray}
A(Q^{2}) & \equiv  & v_{L}R_{L}(Q^{2},U)+\frac{1}{2(1+\eta )}R_{T}(Q^{2},U)\nonumber \\
 & = & G_{C}^{2}(Q^{2})+\frac{2}{3}\eta G_{M}^{2}(Q^{2})+\frac{8}{9}\eta ^{2}G_{Q}^{2}(Q^{2})\label{eq:A}
\end{eqnarray}
 and \begin{eqnarray}
B(Q^{2}) & \equiv  & R_{T}(Q^{2},U)\nonumber \\
 & = & \frac{4}{3}\eta \left( 1+\eta \right) G_{M}^{2}(Q^{2})\; .\label{eq:B}
\end{eqnarray}
 Note that the dependence of the cross section on the target polarization is
conventionally given by analyzing powers denoted \( T_{kq} \); we implicitly
use here the equivalence between analyzing powers and recoil deuteron polarizations:
\( t_{kq}=T_{kq} \) . Defining \begin{equation}
R_{0}\equiv A(Q^{2})+B(Q^{2})\tan ^{2}\frac{\theta }{2}\; ,
\end{equation}
 the tensor polarization observables are \begin{eqnarray}
t_{20}(Q^{2},\theta ) & \equiv  & \frac{1}{\sqrt{2}R_{0}}\left[ v_{L}R_{L}(Q^{2},{1\over \sqrt{2}}\rho _{20})+v_{T}R_{T}(Q^{2},{1\over \sqrt{2}}\rho _{20})\right] \nonumber \\
 & = & -\frac{1}{\sqrt{2}R_{0}}\left\{ \frac{8}{3}\eta G_{C}(Q^{2})G_{Q}(Q^{2})+\frac{8}{9}\eta ^{2}G_{Q}^{2}(Q^{2})\right. \nonumber \\
 &  & \qquad \qquad +\left. \frac{1}{3}\eta \left[ 1+2(1+\eta )\tan ^{2}\frac{\theta }{2}\right] G_{M}^{2}(Q^{2})\right\} \label{eq:t20}
\end{eqnarray}
\begin{eqnarray}
t_{22}(Q^{2},\theta ) & \equiv  & \frac{\sqrt{3}}{2R_{0}}v_{TT}R_{TT}(Q^{2},\sqrt{3}Re\rho _{22})\nonumber \\
 & = & -\frac{1}{2\sqrt{3}R_{0}}\eta \, G_{M}^{2}(Q^{2})
\end{eqnarray}
\begin{eqnarray}
t_{21}(Q^{2},\theta ) & \equiv  & \frac{1}{2R_{0}}\sqrt{\frac{3}{2}}v_{TL}R_{TL}(Q^{2},\sqrt{\frac{3}{2}}Re\rho _{21})\nonumber \\
 & = & -\frac{2}{\sqrt{3}R_{0}}\eta \sqrt{\eta +\eta (1+\eta )\tan ^{2}\frac{\theta }{2}\, }G_{M}(Q^{2})\, G_{Q}(Q^{2})
\end{eqnarray}
\begin{eqnarray}
t_{10}(Q^{2},\theta ) & \equiv  & \frac{1}{R_{0}}\sqrt{\frac{3}{2}}v_{T'}R_{T'}(Q^{2},\sqrt{\frac{3}{2}}\rho _{10})\nonumber \\
 & = & -\frac{1}{R_{0}}\sqrt{\frac{2}{3}}\eta \left( 1+\eta \right) \tan \frac{\theta }{2}\sqrt{\frac{1}{1+\eta }+\tan ^{2}\frac{\theta }{2}}\, G^{2}_{M}(Q^{2})
\end{eqnarray}
\begin{eqnarray}
t_{11}(Q^{2},\theta ) & \equiv  & \frac{1}{2R_{0}}\sqrt{\frac{3}{2}}v_{TL'}R_{TL'}(Q^{2},\sqrt{\frac{3}{2}}Re\rho _{11})\nonumber \\
 & = & \frac{2}{\sqrt{3}R_{0}}\sqrt{\eta +\eta ^{2}}\, \tan \frac{\theta }{2}\, G_{M}(Q^{2})\, \left[ G_{C}(Q^{2})+\frac{\eta }{3}G_{Q}(Q^{2})\right]
\end{eqnarray}

There are only two unpolarized elastic structure functions, but three form factors.
Since \( B(Q^{2}) \) depends only on \( G^{2}_{M}(Q^{2}) \), this form factor
can be determined by a Rosenbluth separation of \( A(Q^{2}) \) and \( B(Q^{2}) \),
or by a cross section measurement at \( \theta =180^{\circ } \). \( A(Q^{2}) \)
depends on all three form factors so that only a quadratic combination of \( G_{C}(Q^{2}) \)
and \( G_{Q}(Q^{2}) \) (the longitudinal part of \( A \)) can be determined
from the unpolarized cross section. A complete separation of the form factors
therefore requires the measurement of at least one tensor polarization observable.
The possible candidates are \( t_{20} \), \( t_{21} \) and \( t_{11} \),
since \( t_{22} \) and \( t_{10} \) depend only upon \( G_{M}^{2}(Q^{2}) \)
and the unpolarized structure functions. \( t_{21} \) and \( t_{11} \), being
both proportional to \( G_{M}(Q^{2}) \), are of smaller magnitude than \( t_{20} \)
and provide in practice a smaller {}``lever arm{}'' to determine either \( G_{C}(Q^{2}) \)
or \( G_{Q}(Q^{2}) \)~\cite{Gar90}. In addition, the measurement of \( t_{11} \)
requires intense polarized electron beams, which became available only recently,
but offers the simplification of a vector polarization measurement. In all cases
so far, this leaves \( t_{20} \) as the observable of choice to extract \( G_{C}(Q^{2}) \)
and \( G_{Q}(Q^{2}) \).

Note that all of the polarization observables depend upon the scattering angle
\( \theta  \) through kinematical factors and \( R_{0}(Q^{2},\theta ) \).
Consequently, the polarization observables measured in different experiments
under different kinematical conditions can only be compared if some convention
is assumed. Since the first \( t_{20} \) measurement~\cite{Sch84} was performed
close to \( \theta =70^{\circ } \), it has been customary to quote the observables
at that angle. For experiments not performed at \( 70^{\circ } \), the observable
is extrapolated to this angle, using the known \( A(Q^{2}) \) and \( B(Q^{2}) \).
Another convention is to use the alternate quantity: \begin{eqnarray}
\tilde{t}_{20}(Q^{2}) & \equiv  & \frac{1}{\sqrt{2}}\frac{R_{L}(Q^{2},{1\over \sqrt{2}}T_{20})}{R_{L}(Q^{2},U)}\nonumber \\
 & = & -\frac{\frac{8}{3}\eta G_{C}(Q^{2})G_{Q}(Q^{2})+\frac{8}{9}\eta ^{2}G^{2}_{Q}(Q^{2})}{\sqrt{2}\left[ G^{2}_{C}(Q^{2})+\frac{8}{9}\eta ^{2}G^{2}_{Q}(Q^{2})\right] }\; .\label{eq:t20_tilda_1}
\end{eqnarray}
 The choice of \( \tilde{t}_{20} \), which is not strictly speaking an observable,
has several advantages. \( \tilde{t}_{20} \) is independent of \( \theta  \),
and thus depends only on \( Q^{2} \). It is a purely longitudinal quantity,
and as such is independent of the magnetic form factor. It has simple properties
related to those of the charge and quadrupole form factors (see Sec.~\ref{ssec:ffphen}
and Refs.~\cite{Gar94,Abb00p}). In particular, the position of nodes in these
form factors may be determined directly from a plot of \( \tilde{t}_{20} \).
Numerically, \( \tilde{t}_{20}(Q^{2}) \) can be determined from \( A(Q^{2}) \),
\( B(Q^{2}) \) and \( t_{20}(Q^{2},\theta ) \) through \begin{equation}
\label{eq:t20_tilda_2}
\tilde{t}_{20}(Q^{2})=\frac{t_{20}(Q^{2},\theta )+\frac{\delta }{2\sqrt{2}}}{1-\delta }\quad \hbox {with}\quad \delta =v_{T}\frac{B(Q^{2})}{R_{0}(Q^{2},\theta )}\, .
\end{equation}
As illustrated in Sec.~\ref{ssec:data_review}, in all measurements to date,
the ratio \( \delta  \) is small, so that \( \tilde{t}_{20}(Q^{2}) \), \( t_{20}(Q^{2},\theta ) \)
and \( t_{20}(Q^{2},70^{\circ }) \) are not very different from each other.
For all these reasons we will use this quantity along with \( A(Q^{2}) \) and
\( B(Q^{2}) \) for comparison of theoretical predictions to data.

In closing this introduction to \( ed \) elastic scattering observables, we
refer to App.~A for a short discussion of a possible two-photon exchange contribution,
especially in view of the large \( Q^{2} \) range of available data.

\subsection{Review of elastic ed data\label{ssec:data_review}}

The first experiment to measure elastic scattering of electrons on
the deuterium was performed at the Stanford Mark III
accelerator~\cite{Mci56}. Since then, many cross section data
points have been measured at various accelerators over the world,
with ever increasing precision and at larger and larger momentum
transfers~\cite{Fri60,Dri62,Gol64,Buc65,Ben66,Gro66,Ran67,Eli69,Gal71,Gan72,Arn75,Mar77,Aki79,Sim81,Cra85,Auf85,Bos90,Pla90,Ale99,Abb99}.
Quite spectacular is the recent achievement of Jefferson Lab to
measure \( A(Q^{2}) \) up to \( Q^{2}\simeq 6 \) \gev2, making use
of a record luminosity of about \( 5\times 10^{38} \) cm\( ^{-2}
\)s\( ^{-1} \) to reach cross sections as low as \( 10^{-41} \)
cm\( ^{2} \)/sr~\cite{Ale99}. New measurements of $B(Q^2)$ for
$Q^2=$ 0.7 to 1.3 \gev2 from the same experiment will soon be
available~\cite{Sul99}. The kinematics of all these experiments
are illustrated in Fig.~\ref{fig:thq_xsect}. Forward angle
scattering yields the elastic structure function \( A \), while
backward angle scattering allows the determination of the elastic
structure function \( B \). The dashed lines in
Fig.~\ref{fig:thq_xsect} indicate what fraction of the cross
section corresponds to the contribution of \( B(Q^2)\tan
^{2}(\theta /2) \).
\begin{figure}
{\par\centering \resizebox*{!}{10cm}{\includegraphics{thq_sigma_exp.epsi}} \par}

\caption{Kinematical settings (\protect\( \theta \protect \) vs \protect\( Q\protect \))
for various cross section measurements in \protect\( ed\protect \) elastic
scattering. The small angle data from SLAC~\cite{Arn75} (open circles) and
JLab/HallA~\cite{Ale99} (filled circles) extend respectively to 10.1 and 12.4
fm\protect\( ^{-1}\protect \), beyond the scale of the horizontal axis. The
dashed lines correspond to \protect\( B\tan ^{2}(\theta /2)/R_{0}=\protect \)0.01,
0.1, 0.5, 0.9, 0.99, respectively from bottom to top.\label{fig:thq_xsect}}
\end{figure}
Other experiments measured cross section ratios \( ed/ep \)~\cite{Bum70,Ber73}.

\begin{figure}
{\par\centering \resizebox*{7cm}{!}{\includegraphics{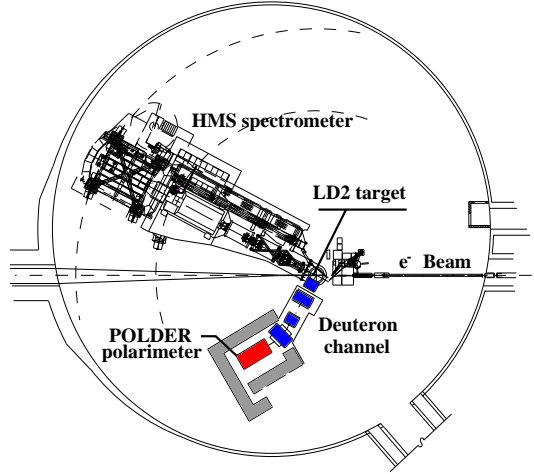}} \par}

\caption{Experimental set-up for the recent \protect\( t_{20}\protect \) double scattering
experiment~\cite{Abb00} in the Hall C (50 m diameter) of Jefferson Lab: \protect\( e+d\, \hbox {(LD}_{2})\rightarrow e'\, \hbox {(HMS)}+d\, \hbox {(Deuteron}\; \hbox {channel)}\protect \)
followed by \protect\( d+p\rightarrow (pp)n\protect \) in the POLDER polarimeter.\label{fig:hallC_setup}}
\end{figure}

\begin{figure}
{\par\centering \resizebox*{!}{8cm}{\includegraphics{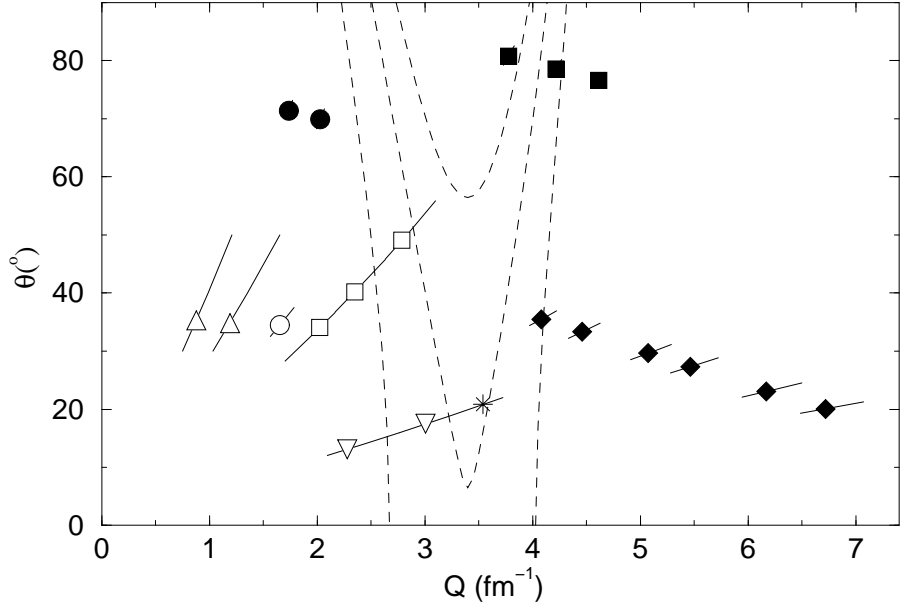}} \par}

\caption{Kinematical settings for measurements of elastic \protect\( ed\protect \)
tensor polarization observables. Filled symbols for \protect\( t_{20}\protect \)
measurements of Bates/Argonne~\cite{Sch84} (circles), Bates/AHEAD~\cite{Gar94}
(squares) and JLab/POLDER~\cite{Abb00} (diamonds), each at a fixed recoil
deuteron angle. Open symbols for \protect\( T_{20}\protect \) measurements
of Novosibirsk~\cite{Dmi85,Voi86} (triangles up), \cite{Gil90} (triangles
down), NIKHEF \cite{Fer96} (circle), \cite{Bou99} (squares) and Bonn~\cite{Bod91}
(star), each at a given electron beam energy. The tilted bars indicate the detector
acceptances. The dashed lines are a measure of the small magnetic contribution
to \protect\( t_{20}\protect \): they correspond to \protect\( \widetilde{t}_{20}-t_{20}=0.05,\protect \)
0.1, 0.15, respectively from bottom to top.\label{fig:thq_t20}}
\end{figure}

As already mentionned, the separate determination of the deuteron charge monopole
and quadrupole form factors necessitates the measurement of a polarization observable.
The observable of choice is \( t_{20} \), which is a measure of the relative
probabilities of finding the deuteron in magnetic substates \( M=-1,0,+1 \)
after the \( ed \) scattering. The Bates Linear Accelerator Center was the
first to measure \( t_{20} \)~\cite{Sch84} and to provide an experimental
evidence for the existence of a node of the charge form factor~\cite{Gar94}.
These double scattering experiments were recently brought to the (up to now)
highest possible momentum transfers at Jefferson Lab~\cite{Abb00}. For this
last experiment as well as for the previously mentioned \( A \) measurement~\cite{Ale99},
electron beams of 100 to 120 \( \mu  \)A were used in conjunction with liquid
deuterium targets up to 15 cm long, capable of dissipating 600~W of power deposited
by the beam. The combination of a record integrated luminosity, in excess of
\( 10^{9} \) pbarn\( ^{-1} \), and of a large acceptance magnetic channel
focusing recoil deuterons onto the high efficiency polarimeter POLDER (see Fig.~\ref{fig:hallC_setup}
and App.~B) allowed the measurements of \( t_{20} \) to be extended up to
\( Q^{2}\simeq 1.7 \) \gev2. The alternative measurement of the tensor analyzing
power \( T_{20} \), using a polarized target intercepting a stored electron
beam, was initiated at Novosibirsk~\cite{Dmi85,Voi86,Gil90}, and improved
at NIKHEF~\cite{Fer96,Bou99}. New preliminary results from Novosibirsk~\cite{Nik00}
are also included in this review. On the other hand, the use of a solid cryogenic
target (ND\( _{3} \)) in an external electron beam at Bonn resulted in a too
low luminosity~\cite{Bod91}. The kinematical settings of all these experiments
are illustrated in Fig.~\ref{fig:thq_t20}. In all cases, the magnetic contribution
to \( t_{20} \) is small. A further comparison between these polarization measurements
is contained in App. B.

The other tensor polarization observables \( t_{21} \) and \( t_{22} \) (or
\( T_{22} \)) were also measured~\cite{Gar94,Abb00,Fer96}.

\begin{figure}
{\par\centering \resizebox*{9cm}{!}{\includegraphics{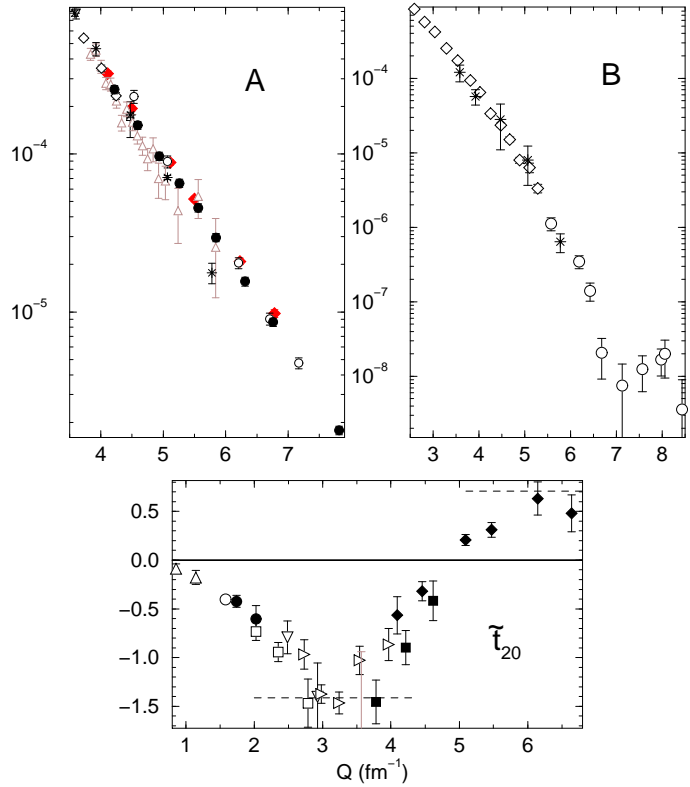}} \par}

\caption{Elastic \protect\( ed\protect \) scattering observables. The \protect\( A\protect \)
data are from Cambridge~\cite{Eli69} (triangles), SLAC~\cite{Arn75} (open
circles), Bonn~\cite{Cra85} (stars), Saclay~\cite{Pla90} (open diamonds),
JLab/HallA~\cite{Ale99} (full circles) and JLab/HallC~\cite{Abb99} (full
diamonds). The \protect\( B\protect \) data are from Saclay~\cite{Auf85}
(diamonds), SLAC~\cite{Bos90} (circles), Bonn~\cite{Cra85} (stars). For
\protect\( t_{20}\protect \) data legend, see Fig.~\ref{fig:thq_t20}.\label{fig:a_b_t20t_data}}
\end{figure}

Figure~\ref{fig:a_b_t20t_data} shows a good part of the existing
data. The \( A \) data at low and high \( Q^{2} \) will be better
illustrated in the figures of Sec.~\ref{sec:theory}, in particular
in Figs.~\ref{fig:da_db_t20t_recap} and \ref{fig:pqcd_a}. The \(
t_{21} \) data appears in Fig.~\ref{fig:pqcd_ratios}. In closing
this section, let us mention a few inconsistencies in this data
set in the light of recent measurements.\footnote{
    At the time of print of this paper, a better determination of
the beam energies at JLab/Hall C results in some corrections,
within quoted errors. The $A(Q^2)$ values~\cite{Abb99} should
decrease by 1 to 3\% with increasing $Q^2$, while the first
$t_{20}$ point~\cite{Abb00} should move down by about one third of
its error. These numbers are subject to confirmation.
    }
The \( A(Q^{2}) \) Cambridge~\cite{Eli69} and Bonn~\cite{Cra85}
data are very probably too low, since both recent measurements at
Jefferson Lab~\cite{Ale99,Abb99} agree with the {}``higher{}''
trend already given by the SLAC data~\cite{Arn75}. Still, as
apparent in Fig.~\ref{fig:da_db_t20t_recap}, these two JLab
measurements differ from each other (10-15\%) in the region \(
Q^{2}=1 \) to 2 \gev2. For a comparative discussion of these two
independent experiments, see~\cite{Pit00}. The \( t_{20} \) (or \(
T_{20} \)) data are necessarily less precise than cross section
measurements and naturally exhibit some scatter. Although all data
points are compatible with parameterizations such as discussed
below, there are some trends between different data sets. No
parameterization or model can accomodate both the
Bates~\cite{Gar94} and NIKHEF~\cite{Bou99} data sets: one or the
other is too low, or both are. The same Bates data are also
systematically lower than both the JLab~\cite{Abb00} and the
preliminary Novosibirsk data~\cite{Nik00}, and the precise low \(
Q^{2} \) NIKHEF point~\cite{Fer96} is lower than many theoretical
expectations. Though these scatters are compatible with the quoted
experimental errors, they demonstrate (together with the
theoretical models to be discussed) a need for a more accurate
measurement in the region \( Q=3 \) to 4.5 \fm. Finally, the Bates
\( t_{22} \) data point at \( Q\simeq 3.8 \) \fm~\cite{Gar94} is
obviously wrong, but the authors did not find a correlation
between \( t_{20} \) and \( t_{22} \) in their analysis .

\subsection{Empirical features of form factors\label{ssec:ffphen}}

The three deuteron electromagnetic form factors may be calculated at a fixed
value of \( Q^{2} \) from measurements of \( A \), \( B \) and \( t_{20} \).
The magnetic form factor \( G_{M} \) is readily available from the \( B \)
measurements, while the two charge form factors \( G_{C} \) and \( G_{Q} \)
are determined from \( v_{L}R_{L}(Q^{2},U)\equiv A_{L}=A-B/2(1+\eta ) \) (\ref{eq:A})
and \t20t~(\ref{eq:t20_tilda_1},\ref{eq:t20_tilda_2}). The resolution of
these equations is most simply described in Ref.~\cite{Abb00p}, together with
a discussion of possible ambiguities in the choice of different solutions in
the \( Q \)-regions where \t20t reaches its extrema. This procedure allows
a direct comparison of theoretical models with the form factors, instead of
observables, but it is limited to the domain where the three observables are
measured, which is \( Q=0-7 \) \fm (\( Q^{2}\leq 1.8 \) \gev2).

The most striking result is an experimental determination of the node of \( G_{C} \),
at \( Q=4.21 \)\( \pm 0.08 \) \fm~\cite{Abb00p}, which corresponds to \t20t\( =-1/\sqrt{2} \)
. This behaviour of \( G_{C} \), though expected from most models, could not
have been seen with cross section measurements only. The exact location of this
node is sensitive to the strength of the \( NN \) repulsive core (in the impulse
approximation, it is connected to the node of \( u(p) \) in Fig.~\ref{fig:pu})
and to the size of relativistic corrections and of the isoscalar meson exchange
contributions. A secondary maximum is also determined from the data~\cite{Abb00}.
Thus, like any other nucleus, the deuteron appears to have a charge form factor
with an oscillatory diffractive pattern. Unlike other nuclei, the sign of this
form factor is determined as well.

The quadrupole form factor \( G_{Q} \) exhibits a monotonous exponential fall-off,
and its first node, corresponding to the yet unobserved second node of \t20t,
is expected beyond \( Q=7 \) \fm.

Finally the magnetic dipole form factor \( G_{M} \) has a node at \( Q=7.2 \)\( \pm 0.3 \)
\fm, determined by \( B \) data~\cite{Bos90} and in a lesser extent by \( t_{21} \)
data~\cite{Abb00}. The position of this node is very sensitive to small non-nucleonic
components in the deuteron wave function, to nucleon-nucleon components of relativistic
origin, as well as to the description of the \( \rho \pi \gamma  \) contribution
to be discussed in the next section.

\begin{figure}
{\par\centering \resizebox*{!}{10cm}{\includegraphics{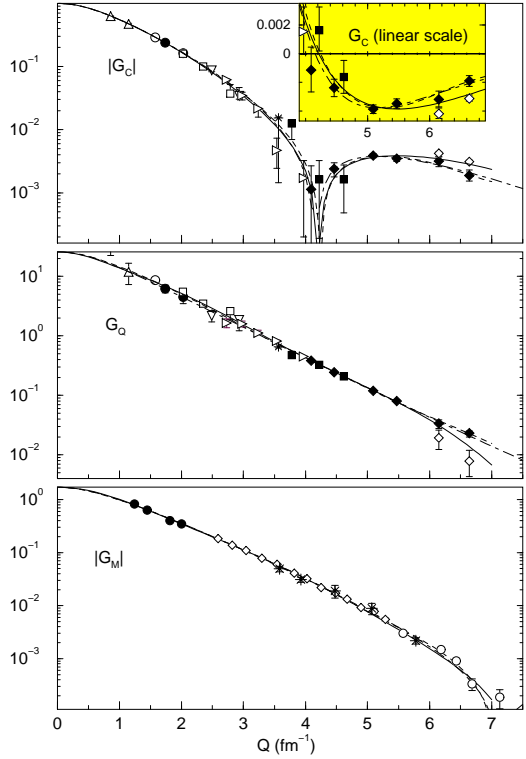}} \par}

\caption{Deuteron form factors \protect\( G_{C}\protect \), \protect\( G_{Q}\protect \)
and \protect\( G_{M}\protect \) as a function of \protect\( Q\protect \) (updated
from Fig.~1 of Ref.~\cite{Abb00p}). The data for \protect\( G_{C}\protect \)
and \protect\( G_{Q}\protect \) correspond to \protect\( t_{20}\protect \)
(or \protect\( T_{20}\protect \)) measurements (see legend of Fig.~\ref{fig:thq_t20}),
including new preliminary results from~\cite{Nik00} (triangles right). The
open diamonds correspond to a second solution of the equations \protect\( G_{C},G_{Q}=f(A,B,t_{20})\protect \)~\cite{Abb00p}.
The \protect\( G_{M}\protect \) data correspond to the \protect\( B\protect \)
measurements indicated in Fig.~\ref{fig:a_b_t20t_data}, with the addition
of \cite{Sim81} (full circles). The curves are from parameterizations I (solid
line), II (dot-dashed) and III (short dashed) discussed in the text.\label{fig:ff3}}
\end{figure}

\begin{figure}
{\par\centering \resizebox*{6cm}{!}{\includegraphics{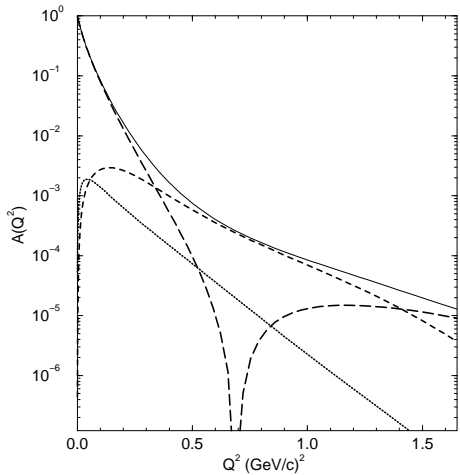}} \par}

\caption{\protect\( A(Q^{2})\protect \) (solid curve) and the contributions from \protect\( G_{C}\protect \)
(long dashed), \protect\( G_{Q}\protect \) (dashed) and \protect\( G_{M}\protect \)
(dotted), from parameterization~I. Above 1.3 \gev2, the statistical significance
of the data is such that the crossing of the \protect\( G_{C}\protect \) and
\protect\( G_{Q}\protect \) contributions, corresponding to \protect\( \tilde{t}_{20}=1/\sqrt{2}\protect \),
is not firmly established.\label{fig:a_ff}}
\end{figure}

The three form factors were parameterized in three different ways: rational
fractions (I), sum of Lorentzian functions in an helicity basis (II) and sum
of Gaussian functions (III)~\cite{Abb00p}, fitting directly the measured observables,
i.e. differential cross sections and polarizations. Figure~\ref{fig:ff3} gives
a representation of the form factors with an updated version of parameterization
I, taking into account the preliminary \( T_{20} \) data of Ref.~\cite{Nik00}.
The data base and the parameterizations are available in~\cite{Bal00}.
Another representation of the experimental knowledge of the deuteron form factors
is given in Fig.~\ref{fig:a_ff}, where the contributions from each of them
to the elastic structure function \( A \) are given, calculated using parameterization~I.

\section{Theoretical issues\label{sec:theory}}

The different classes of theoretical models of the deuteron electromagnetic
form factors are presented here, illustrated with the most recent calculations
on the subject. A summary of earlier theoretical work on the subject may be
found for instance in~\cite{Gar94,Pla90}. In all figures of this section,
the data legend is the same as in Sec.~\ref{ssec:data_review}.

\subsection{Deuteron elastic form factors in the simple potential model\label{ssec:potmod}}

\begin{figure}
{\par\centering \resizebox*{7cm}{!}{\includegraphics{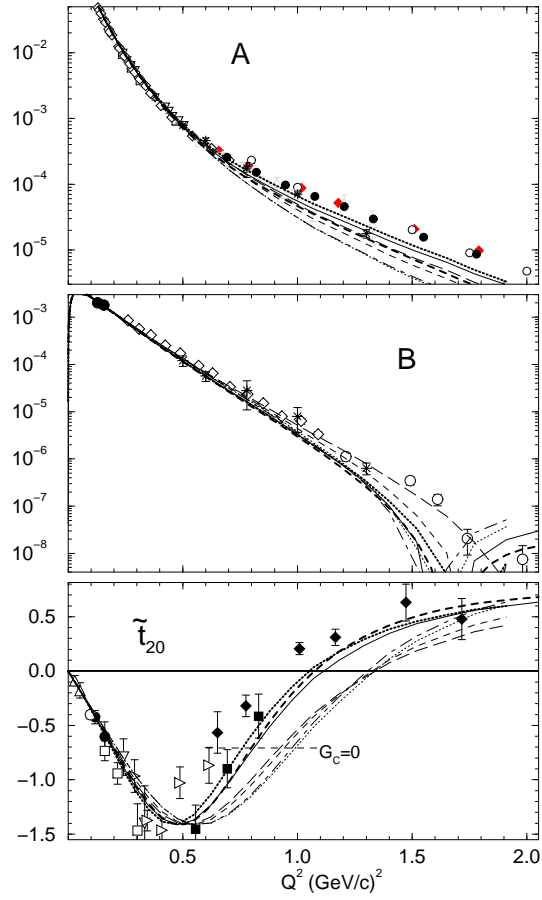}} \par}

\caption{Elastic \protect\( ed\protect \) scattering observables for a variety of phenomenological
potentials in the NRIA: Bonn-A (dotted curve), Bonn-B (dashed), Bonn-C (long-dashed),
Bonn Q (dot-dashed), Reid-SC (thick dotted), Paris (solid) and A-\protect\( v_{18}\protect \)
(thick dashed). NEMFF: MMD.\label{fig:a_b_t20t_nria}}
\end{figure}

Once the wave functions for a given potential model are obtained as discussed
in Sec.~\ref{sec:dwf}, the electromagnetic form factors may readily be calculated
in the nonrelativistic impulse approximation (NRIA) using a well established
formalism~\cite{Def66,Don75}. The virtual photon can couple to any of the
two nucleons, so that the isoscalar combinations \( G_{E,M}^{S} \) of the nucleon
form factors (NEMFF), defined in App.~C, factorize to yield~\cite{Gou66,Jan56}:\begin{eqnarray}
G_{C}(Q^{2}) & = & G_{E}^{S}(Q^{2})\times C_{E}(u,w;Q^{2})\nonumber \\
G_{Q}(Q^{2}) & = & G_{E}^{S}(Q^{2})\times C_{Q}(u,w;Q^{2})\nonumber \\
G_{M}(Q^{2}) & = & G_{E}^{S}(Q^{2})\times C_{L}(u,w;Q^{2})+G_{M}^{S}(Q^{2})\times C_{S}(u,w;Q^{2})\label{eq:ff_nria}
\end{eqnarray}
The \( C \) functions are integrals of quadratic combinations of \( u \) and
\( w \). In this NRIA, the ratio \( G_{Q}/G_{C} \), and thus \( \widetilde{t}_{20} \),
are independent of the nucleon form factors. The coupling of the virtual photon
to the moving nucleon charges and to the nucleon spins both contribute the magnetization
in \( G_{M} \), giving rise to the two terms in~(\ref{eq:ff_nria}). The elastic
structure functions \( A(Q^{2}) \), \( B(Q^{2}) \) and \( \tilde{t}_{20}(Q^{2}) \)
are illustrated in Fig.~\ref{fig:a_b_t20t_nria} for a variety of phenomenological
potentials~\cite{Hen95,AdaPr,Wir95}, using the MMD parameterization of the
nucleon form factors (see App. C).

\begin{figure}
{\par\centering \resizebox*{9cm}{!}{\includegraphics{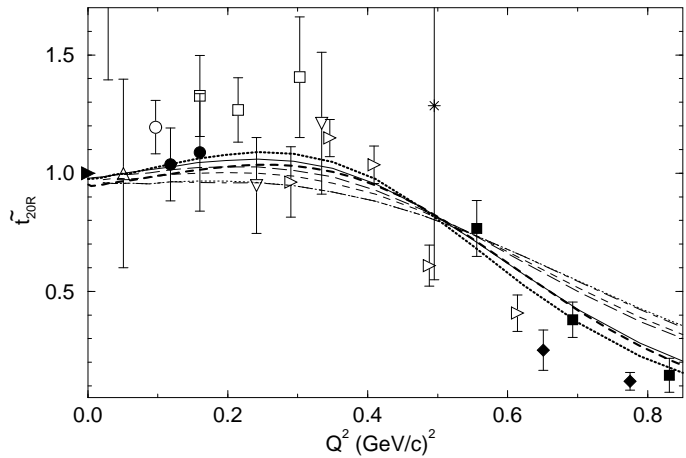}} \par}

\caption{\protect\( \tilde{t}_{20R}\protect \) from~(\ref{eq:t20tR}) for a variety
of phenomenological potentials in the NRIA. Same legend as Fig.~\ref{fig:a_b_t20t_nria}\label{fig:t20tR}.
The solid right triangle results from the definition of \protect\( \tilde{t}_{20R}\protect \).}
\end{figure}

In all cases the calculations agree with one another and with the data up to
\( Q^{2}\sim 0.5 \) \gev2 but diverge at higher \( Q^{2} \). The low \( Q^{2} \)
behaviour of the form factors is not determined with the same precision for
each of them. Since the forward cross sections at low \( Q^{2} \) depend mostly
on \( G_{C} \), the slope of this form factor at \( Q^{2}=0 \) is determined
to about 1\% (see Sec.~\ref{sssec:radius}). The slope of \( G_{M} \) at the
origin is determined by backward cross sections and is known to about 5\%. In
contradistinction, the slope of \( G_{Q} \) is known to only 15\%, in the absence
of very precise \( t_{20} \) measurements at low \( Q^{2} \). Still this slope
is model dependent. To illustrate this point, we define the reduced quantity
\begin{equation}
\label{eq:t20tR}
\tilde{t}_{20R}(Q^{2})=-\frac{3}{\sqrt{2}Q_{d}Q^{2}}\tilde{t}_{20}(Q^{2})
\end{equation}
 and show its low \( Q^{2} \) behaviour in Fig.~\ref{fig:t20tR}.
At \( Q^{2}=0 \), one gets \begin{equation}
\label{eq:t20R_slope}
\tilde{t}_{20R}(0)=\frac{G_{Q}(0)}{Q_{d}M_{d}^{2}}\quad \hbox {and}\quad \frac{\tilde{t}'_{20R}(0)}{\tilde{t}_{20R}(0)}=\frac{G'_{Q}(0)}{G_{Q}(0)}-G'_{C}(0)+\frac{G_{Q}(0)}{12M^{2}_{d}}\, .
\end{equation}
 \( G_{Q}(0) \) is the model calculation of the quadrupole moment. The slope
\( \tilde{t}'_{20R}(0) \) is rather small, due to an approximate cancellation
between the first two terms, the third one being significantly smaller. In this
connection, it is instructive to list in Table~\ref{tab:t20tR_slope} the model
dependences of these various slopes. The value of the quadrupole form factor
varies substantially for these calculations and is always smaller than the experimental
value of 25.8~(\ref{eq:ffnorm}). All of the Bonn calculations and the Reid-SC
calculation yield the same value for the derivative of the charge form factor
and the same of is true of the derivative of the quadrupole form factor with
the exception of Bonn A. Therefore the spread in \( \tilde{t}'_{20R}(0) \)
for this latter set comes only from the variation in the value of the quadrupole
form factor. However, this may not be the case in general.
\begin{table}

\caption{Dependence upon various potentials of the quantities appearing in Eq.(\ref{eq:t20R_slope}).
Slopes in (GeV/c)\protect\( ^{-2}\protect \).\label{tab:t20tR_slope}}
{\centering \begin{tabular}{|c|c|c|c|c|c|}
\hline
Potential&
\( G_{Q}(0) \)&
\( \widetilde{t}_{20R}(0) \)&
\( G'_{Q}(0) \)&
\( G'_{C}(0) \)&
\( \widetilde{t}'_{20R}(0) \)\\
\hline
\hline
Bonn A&
24.8&
.960&
-506&
-18.1&
-1.69\\
\hline
Bonn B&
25.1&
.972&
-464&
-18.1&
0.25\\
\hline
Bonn C&
25.4&
.983&
-464&
-18.1&
0.47\\
\hline
Bonn Q&
24.7&
.956&
-464&
-18.1&
-0.06\\
\hline
Reid SC&
25.2&
.976&
-464&
-18.1&
0.32\\
\hline
Paris&
25.2&
.976&
-453&
-18.5&
1.09\\
\hline
A-\( v_{18} \)&
24.1&
.932&
-444&
-18.1&
0.28\\
\hline
\end{tabular}\par}\end{table}

The increasing disagreement between these simple calculations and the data at
higher \( Q^{2} \) indicates that this approach does not contain all of the
necessary physical degrees of freedom and/or that since the momentum transfer
approaches and then surpasses the mass of the nucleon, it is necessary to consider
the impact of special relativity on the calculation of the deuteron elastic
structure functions. We will now consider the first of these possibilities.

\subsection{Limitations of the simple potential model}

For many applications in nuclear physics, the phenomenological potential approach
is completely adequate and provides a good description of various phenomena.
Its limitations were seen, however, by considering the calculation of electromagnetic
properties of the deuteron in the preceding section.

The inclusion of electromagnetic interactions with the two-nucleon system imposes
the requirement that the electromagnetic current matrix elements satisfy the
continuity equation \begin{equation}
\label{eq:cont_wf}
\nabla \cdot <\Psi _{f}|\widehat{\mathbf{J}}(\mathbf{x})|\Psi _{i}>+\frac{\partial }{\partial t}<\Psi _{f}|\hat{\rho }(\mathbf{x})|\Psi _{i}>=0
\end{equation}
 where \( \widehat{\mathbf{J}}(\mathbf{x}) \) and \( \hat{\rho }(\mathbf{x}) \)
are the current and charge density operators. The operators must then satisfy
\begin{equation}
\label{continuity}
\nabla \cdot \widehat{\mathbf{J}}(\mathbf{x})-i\left[ \hat{\rho }(\mathbf{x}),\hat{H}\right] =0\, .
\end{equation}
 Since either of the nucleons can be charged, the charge density operator is
\begin{equation}
\label{fullrho}
\hat{\rho }=\hat{\rho }_{1}+\hat{\rho }_{2}\, ,
\end{equation}
while the current density takes the general form \begin{equation}
\label{fullJ}
\widehat{\mathbf{J}}=\widehat{\mathbf{J}}_{1}+\widehat{\mathbf{J}}_{2}+\widehat{\mathbf{J}}_{ex}\, ,
\end{equation}
 where \( \hat{\rho }_{i} \) and \( \widehat{\mathbf{J}}_{i} \) are the charge
and current densities for particle \( i \) and \( \widehat{\mathbf{J}}_{ex} \)
is a possible additional contribution to the current density.

The current of a free charged particle must satisfy the continuity equation,
which implies \begin{equation}
\nabla \cdot \widehat{\mathbf{J}}_{i}(\mathbf{x})-i\left[ \hat{\rho }_{i}(\mathbf{x}),\hat{T}_{i}\right] =0\, .
\end{equation}
 Using this with (\ref{continuity}), (\ref{fullrho}) and (\ref{fullJ}) gives
\begin{equation}
\label{continuity2}
\nabla \cdot \widehat{\mathbf{J}}_{ex}(\mathbf{x})-i\left[ \hat{\rho }(\mathbf{x}),\hat{V}\right] =0\, .
\end{equation}
 Since \( \hat{\rho }_{i} \) is proportional to \( (1+\tau _{3}^{(i)}) \)
and the two-nucleon potential \( \widehat{V} \) has terms proportional to \( \tau ^{(1)}\cdot \tau ^{(2)} \),
the second term in (\ref{continuity2}), proportional to \( (\tau ^{(1)}\wedge \tau ^{(2)})_{3} \),
does not vanish and \( \widehat{\mathbf{J}}_{ex} \) must be nonzero. In addition,
since \( \widehat{V} \) involves the coordinates of both nucleons, the current
\( \widehat{\mathbf{J}}_{ex} \) must also depend upon both sets of coordinates
and is therefore a two-body operator. The physical origin of this contribution
to the current comes from the fact that the two-nucleon potential contains terms
corresponding to the exchange of charge between the nucleons, which must in
turn give rise to an associated current. These two-body currents are called
exchange or interaction currents.

Equation (\ref{continuity2}) is a symmetry constraint on the theory. It cannot,
however, be used to uniquely determine \( \widehat{\mathbf{J}}_{ex} \) since
divergenceless pieces can be added to any current satisfying (\ref{continuity2})
without modifying the constraint. A unique prediction of the current therefore
requires that the underlying dynamics of the interaction be specified. Viewing
the nuclear force in a meson-exchange model, these two-body currents are naturally
associated with contributions that involve the exchange of mesons~\cite{Vil47}.
Now that it is generally accepted that quantum chromodynamics (QCD) is the correct
description of the strong force at the scales of interest here, these currents
must ultimately be associated with the exchange of quarks.

This discussion leads to a consistent treatment in the context of a nonrelativistic
treatment of the two-nucleon problem. However, any attempt to implement this
approach in a manner that is applicable to the existing data, for instance to
the case of elastic electron-deuteron scattering, leads directly to consistency
problems. One of the most elementary indications of this problem arises from
the necessity of including electromagnetic form factors for the nucleons. At
low momentum transfers, these form factors differ from their static values by
terms of order \( Q^{2}/m_{D}^{2} \). Since the dipole mass \( m_{D} \) is
of similar magnitude as the nucleon mass, this is of leading relativistic order
\( v^{2}/c^{2} \). Whenever the presence of the nucleon electromagnetic form
factors has an appreciable effect on the calculations of the deuteron form factors,
some effects of relativistic order are already included. It then becomes necessary
to consider whether there are other relativistic effects of similar size that
can appreciably modify the calculations. In fact, there are many such effects
arising from Lorentz boosts and exchange currents that become increasingly important
in the calculations of deuteron form factors as the four-momentum transfer increases.
For this reason it is necessary to consider how relativity can be introduced
consistently in models of the deuteron. For this purpose, we will assume that
the deuteron is adequately described by nucleons and mesons.

\subsection{Construction of relativistic models}

The requirement of any truly relativistic model is that it must satisfy Poincar\'{e}
covariance: it must be covariant with respect to Lorentz boosts, spatial rotations
and space-time translations. This can be imposed by requiring that the the operators
which act as generators of these transformations satisfy the Lie algebra of
this group: \begin{equation}
\left[ J^{i},J^{j}\right] =i\epsilon ^{ijk}J^{k},\qquad \left[ K^{i},K^{j}\right] =-i\epsilon ^{ijk}J^{k}
\end{equation}
\begin{equation}
\left[ J^{i},K^{j}\right] =i\epsilon ^{ijk}K^{k}
\end{equation}
\begin{equation}
\left[ P^{\mu },P^{\nu }\right] =0
\end{equation}
\begin{equation}
\left[ K^{i},P^{0}\right] =-iP^{j},\qquad \left[ J^{i},P^{0}\right] =0
\end{equation}
\begin{equation}
\label{commute}
\left[ K^{i},P^{j}\right] =-i\delta ^{ij}P^{0},\qquad \left[ J^{i},P^{j}\right] =i\epsilon ^{ijk}P^{k}
\end{equation}
 where \( J^{i} \) are the three angular momentum operators that generate rotations,
\( K^{i} \) are the generators of the three Lorentz boosts and \( P^{\mu } \)
is a four-vector containing the energy and momentum operators that generate
space-time translations. This differs from the corresponding algebra for the
Galilean transformations in only two commutators. The differing commutators
for the Galilean transformations are: \begin{equation}
\left[ K^{i}_{G},K^{j}_{G}\right] =0
\end{equation}
 and \begin{equation}
\left[ K^{i}_{G},P^{j}\right] =-i\delta ^{ij}M
\end{equation}
 where \( K^{i}_{G} \) are the generators of the Galilean {}``boosts{}''
and \( M \) is the mass operator. The first of these implies that Galilean
boosts commute whereas Lorentz boosts do not. The second is the source of the
dynamical complexity of relativistic models and theories. While the commutators
for the Galilean boosts and the three-momentum operators are proportional to
the mass, the corresponding commutators for the Poincar\'{e} group are proportional
to the energy operator, that is the Hamiltonian. Since the Hamiltonian contains
the interactions between the constituents of the system, the first commutator
of (\ref{commute}) implies that the Lorentz boost operators or three-momentum
operators must also be dependent upon the interaction.

There are two basic approaches to constructing Poincar\'{e} invariant models.
We will start with the most familiar of these, quantum field theory.

\subsubsection{Quantum field theory}

The starting point of a quantum field theory is a Lagrangian that is constructed
to satisfy all of the required symmetries, including Poincar\'{e} invariance.
By nature field theories have an infinite number of degrees of freedom. Canonical
quantization is performed by constructing the Hamiltonian, finding the generalized
position and momentum in terms of the fields, writing the fields as an expansion
in terms of creation and annihilation operators, and then imposing canonical
equal time commutation (anticommutation) relations on the canonical variables.
This yields commutation (anticommutation) relations for the creation and annihilation
operators. An immediate consequence is that the fields commute (anticommute)
for all spacelike intervals, implying that events occuring at spacelike separations
cannot be causally connected. This is the property of microscopic causality
or microscopic locality. It should be noted here that microcausality results
from imposing the commutation (anticommutation) relations on any spacelike hypersurface.

The structure of field theories is very complex due to the
presence of an infinite degrees of freedom. Since the Hamiltonian
of an interacting system links states containing different numbers
of particles, and the time-evolution operator of the system is
given by the imaginary exponential of the interacting Hamiltonian,
the interacting system contains contributions with any number of
particles, from zero to infinity. Practical calculations in field
theory, with the exception of numerical approaches such as lattice
QCD, must then introduce some method of truncating the collection
of configurations contributing to the calculation. The most
familiar approach to this problem is Feynman perturbation theory.
Here, various configurations are carefully arranged such that all
quantities contributing to a process, such as free propagators and
vertex functions, are individually covariant with respect to
noninteracting Lorentz transformations. The truncation then
requires that there be some plausible scheme for systematically
organizing contributions in order of their relative importance.
For example, in the classic case of quantum electrodynamics where
the coupling constant is small, the terms are ordered in powers of
the coupling constant and truncated at some finite order.

The drawbacks of quantum field theory for constructing models such
as for \( NN \) interactions are related to complexities and
calculational difficulties. Most of our knowledge of field theory
is based on perturbation theory and there is no \textit{a priori}
method for determining the radius of convergence (if any) for a
given field theory. For strong coupling theories, it is no longer
plausible to construct a perturbation scheme in terms of simple
counting of coupling constants and infinite sets of contributions
must be summed. In practice, the coupling of the infinite sets of
states, or \( n \)-point functions, must be truncated if there is
to be any hope of calculation. Thus some physically reasonable
scheme for truncation must be proposed. However, the truncation of
the theory will usually violate some symmetries of the full theory
such as crossing symmetries, covariance or current conservation.
Local effective theories with the appropriate symmetries may also
be nonrenormalizable. Model calculations may also include
phenomenological elements such as form factors that are not
calculated within the context of the field theory. This also leads
to problems of consistency within the model and may also violate
symmetries.

\subsubsection{Hamiltonian dynamics}

Although Dirac was one of the founders of quantum field theory, he soon became
disillusioned with its complexity and the difficulties associated with the unavoidable
infinities. He continued for most of the rest of his life to seek an alternative
to quantum field theory. He assumed that the problems with field theory were
related to starting from an unsatisfactory relativistic classical theory. He
pointed to an alternate approach, starting with a theory with a fixed number
of degrees of freedom, as is done with the nonrelativistic Schrödinger equation.
This led to relativistic constraint dynamics~\cite{Kei91}. In this approach,
the dynamics of a model system is determined by choosing a mass operator which
contains an instantaneous interaction as in the nonrelativistic potential theory.
This basic dynamics contains a finite number of particles and has a corresponding
Hilbert space when quantized. Covariance is imposed by constructing a unitary
representation of the inhomogenous Lorentz transformations with generators that
satisfy the commutation relations of the Poincar\'{e} group. The wave functions
have the same probabilistic interpretation as in nonrelativistic quantum mechanics,
but microscopic causality is not respected: the theory must be constructed to
respect the physical requirement of cluster separability or macroscopic locality.

There are at least three different approaches for the quantization of such models.
The first is the traditional method of quantizing along constant time surfaces
(called instant form) where the evolution of the system is the usual time evolution
which is normal to the constant-time hypersurface. The second is to quantize
along spacelike surfaces with constant interval (called point form). The evolution
of the system is then along a new coordinate normal to these surfaces. The third
of these is to quantize along the light cone with evolution along the coordinate
\( x^{+}=x^{0}+x^{3} \). Since all these choices describe surfaces with spacelike
separations (or the infinite momentum limit of such a surface in the light-cone
case), they are also consistent with microscopic causality for field theories,
which may also be quantized along these surfaces.

Particularly useful in the Hamiltonian dynamics is the fact that a careful construction
of the mass operator can lead to equations of motion of the same form as the
two-body Schr\"{o}dinger equation. It is therefore possible to use, without
modification, nonrelativistic potentials that have been fitted to describe \( NN \)
scattering .

The drawbacks are related to the choice of the interaction without specification
of any underlying dynamical content. As a result, quantities such as electromagnetic
currents can be constrained by the structure of the theory, but can not be uniquely
determined from the interaction dynamics.

We will now proceed to a discussion of various approaches used in constructing
relativistic calculations of the elastic electromagnetic form factors for the
deuteron.

\subsection{\protect\( v/c\protect \) expansions\label{ssec:v/c}}

This approach is actually a hybrid of field theory and Hamiltonian dynamics.
It assumes that the basic dynamical content of the deuteron is nonrelativistic
and that the necessary relativistic effects can be described as corrections
to the nonrelativistic current matrix elements as an expansion in \( v/c \)~\cite{Fri75,Mat89,Sch91}.

\begin{figure}
{\par\centering \resizebox*{11cm}{4cm}{\includegraphics{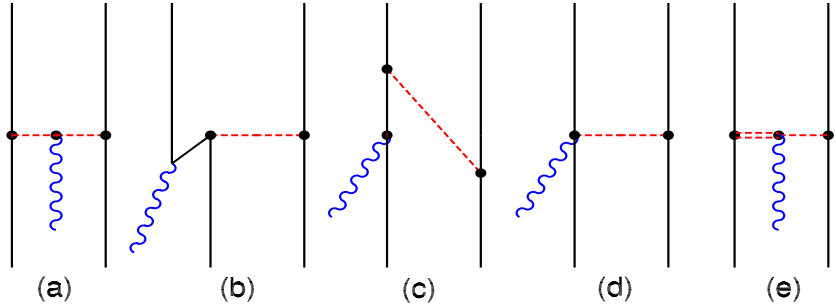}} \par}

\caption{Two-body currents arising in the \protect\( v/c\protect \) expansions (see
text for explanations).\label{fig:exchangenr}}
\end{figure}

The nucleon-nucleon interaction is taken to be a standard nonrelativistic potential
with parameters determined by fitting to the nucleon-nucleon scattering data
and to the deuteron binding energy. It is assumed that the potential is, at
least in part, represented by a one-meson-exchange model, since the meson degrees
of freedom are necessary to construct two-body exchange currents from simple
Feynman diagrams and for constructing corrections due to retardation of meson
propagators. Examples of the required two-body interaction currents are represented
in the diagrams of Fig.~\ref{fig:exchangenr}. Diagram (a) represents a contribution
due to coupling of the photon to the current of an exchanged meson, which, because
of \( G \)-parity, does not apply to isoscalar transitions such as \( ed \)
elastic scattering. Diagram (b) is a so called pair current that arises due
to the projection of the interactions onto the positive-energy space only. Contributions
that couple to the negative energy part of the nucleon propagator must then
be included in the current. Diagram (c) is a retardation correction. Since nonrelativistic
wave functions are single-time wave functions, diagrams corresponding to the
absorption of a photon while the meson is in flight are not included in the
impulse approximation and must be included in the current. Retardation corrections
can also be associated with retarded meson propagation in any of the other diagrams
in this figure. Derivative couplings of the mesons to the nucleons can also
give rise to contact interactions of the type shown in diagram (d), which is
isovector and does not apply to \( ed \) elastic scattering. It is also possible
for a photon that is absorbed by an exchanged meson to excite the meson. An
example is the \( \rho \pi \gamma  \) exchange current which is commonly included
in calculations of elastic electron-deuteron scattering. Such a current is represented
by diagram (e).

Relativity is imposed by requiring that the currents and interactions are consistent
with operator commutators of Poincar\'{e} invariance to some order in \( v/c \).
The dependence of the boost operators on the interaction also gives rise to
interaction currents in addition to those characterized above. This approach
guarantees that the interaction model can be very well constrained by data but
its application can become technically complicated. In addition, the expansion
in \( v/c \) must fail at some value of momentum transfer.

The calculations shown here are from a recent paper by Arenh\"{o}vel, Ritz
and Wilbois (ARW)~\cite{Are00}. This paper is an extension of earlier work~\cite{Hen95,Ada97}
that included relativistic corrections for the charge and quadrupole form factors,
but not the magnetic. Although there is an extensive literature on the subject,
as nicely summarized in~\cite{Are00}, only that paper and the earlier work
of Tamura~\cite{Tam92} take into account all leading order terms including
the Lorentz boost of the deuteron center of mass. The need for a realistic one-boson-exchange
potential leads ARW to use the Bonn OBEPQ models, which may however not have
the same precision in describing \( NN \) scattering data as more recent, but
more phenomenological, potentials. Finally the Galster dipole parameterization
for the nucleon electromagnetic form factors was used.
\begin{figure}
{\par\centering \resizebox*{7cm}{!}{\includegraphics{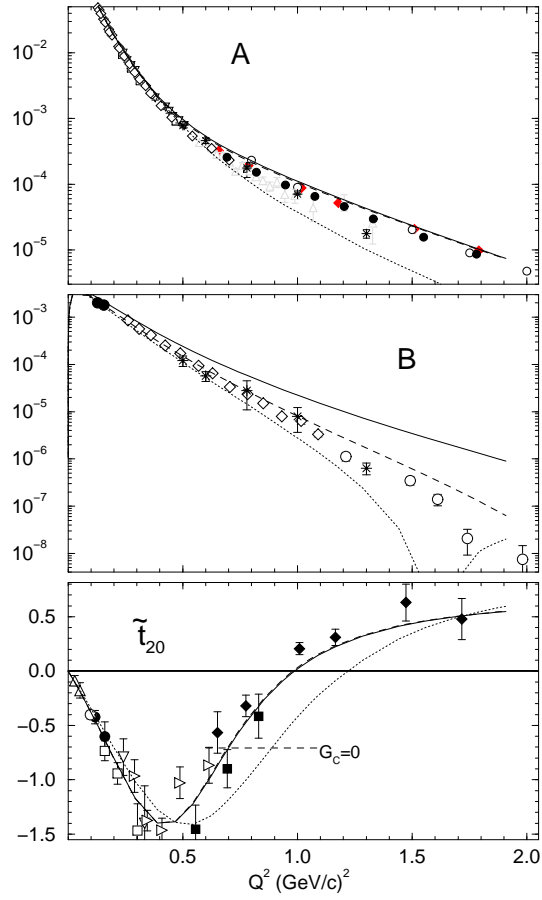}} \par}

\caption{Elastic \protect\( ed\protect \) scattering observables calculated with the
\protect\( v/c\protect \) expansion~\cite{Are00}: NRIA with the Bonn OBEPQ-B
potential (dotted curve); same plus all relativistic corrections to leading
order (dashed); all of the former plus the \protect\( \rho \pi \gamma \protect \)
exchange current (solid). NEMFF: Galster.\label{fig:a_b_t20t_ARW}}
\end{figure}
 Figure~\ref{fig:a_b_t20t_ARW} shows the observables \( A(Q^{2}) \), \( B(Q^{2}) \)
and \( \tilde{t}_{20}(Q^{2}) \) for this model. Three curves are shown for
each observable, the impulse approximation (NRIA), the impulse approximation
plus all relativistic corrections to leading order, and all of the former plus
the \( \rho \pi \gamma  \) exchange current. The \( \rho \pi \gamma  \) exchange
current is purely transverse and is not required or constrained by the interaction
model, but is the longest range isoscalar current of this type.

The elastic structure function \( A(Q^{2}) \) for the impulse approximation
is substantially below the data. The inclusion of the various relativistic corrections
brings the calculation into much better agreement with the data, while the addition
of the \( \rho \pi \gamma  \) exchange current has only a very small effect.

In the case of the magnetic structure function \( B(Q^{2}) \), the NRIA again
is below the data and has a diffraction minimum at lower \( Q^{2} \) than indicated
by the data. The addition of the relativistic corrections increases the size
of the calculated structure function and appears to move the diffraction minimum
above the value indicated by the data. The addition of the \( \rho \pi \gamma  \)
exchange current again increases the size of the structure function to a value
considerably above the data and presumably pushes the diffraction minimum to
even higher values. We will return later to the question of the effect of this
exchange current on the magnetic structure function, in the context of field-theoretical
models.

For the polarization structure function \( \tilde{t}_{20}(Q^{2}) \), the NRIA
is above the data for small \( Q^{2} \) below its minimum, and below at \( Q^{2} \)
above the minimum. The addition of the relativistic corrections brings the calculation
into much better agreement with the observed node of \( G_{C} \) and consequently
with the data. The \( \rho \pi \gamma  \) exchange current only has a very
small effect on this observable which is purely longitudinal.

The examination of the figures in Ref.~\cite{Are00} shows that the various
corrections to the NRIA are not small, they tend to be of similar magnitude,
and they contribute to the deuteron form factors with varying signs. As a result,
any calculation that includes some, but not all, of these corrections, is questionable.

These calculations clearly indicate that relativistic effects, that is meson-exchange
contributions and genuine relativistic corrections, are important even for relatively
small momentum transfers. The domain of validity of this approach is thus limited,
probably to \( Q^{2}\leq 1 \) \gev2, though it may lead in some instances to
a satisfactory description of the data up to 2 \gev2~\cite{Wir95}. The study
of the validity of the two-nucleon description of the deuteron down to very
small internucleon distances, together with new data becoming available from
Jefferson Lab, make mandatory the consideration of fully relativistic models.

\subsection{Relativistic constraint dynamics\label{ssec:HCD}}

\begin{figure}
{\par\centering \resizebox*{10cm}{!}{\includegraphics{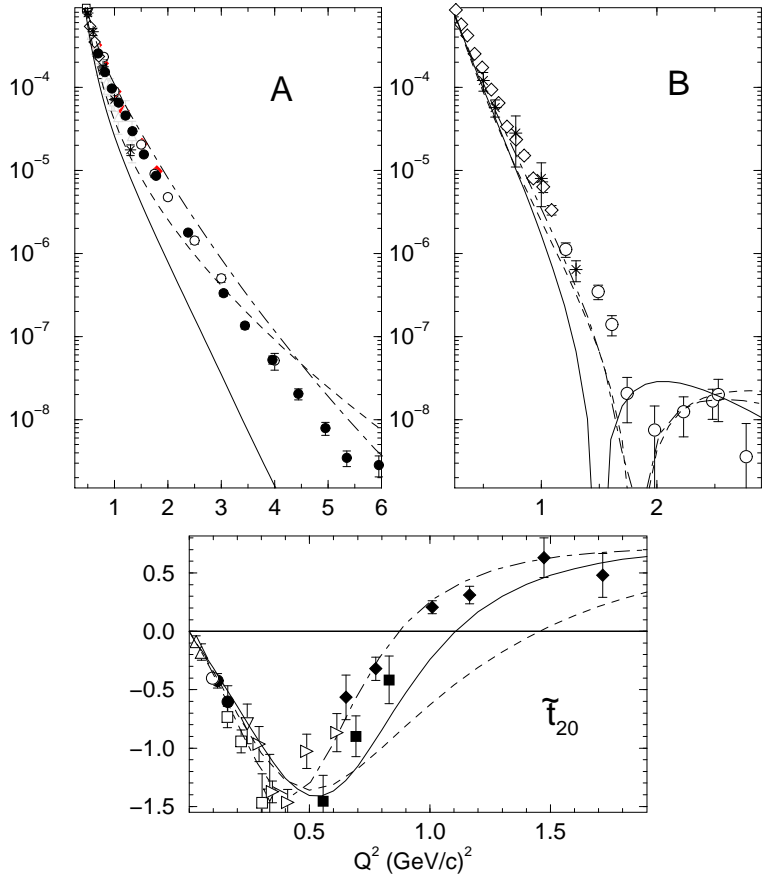}} \par}

\caption{Elastic \protect\( ed\protect \) scattering observables calculated with the
Hamiltonian constraint dynamics. AKP~\cite{All00}, NEMFF: MMD (solid curve);
LPS~\cite{Lev00}, GK85 (dashed); FS~\cite{ForPr}, H (dot-dashed).\label{fig:a_b_t20t_HCD}}
\end{figure}

Figure~\ref{fig:a_b_t20t_HCD} shows three recent calculations the deuteron
elastic structure functions. Note that in this and following figures, the different
observables are not shown in the same \( Q^{2} \) range, following the available
data. This should be kept in mind for a meaningful comparison of models to data.
The calculation of Allen, Klink and Polyzou (AKP)~\cite{All00} uses the Argonne
\( v_{18} \) potential for an interaction and is quantized in point form. The
calculation of Lev, Pace and Salmè (LPS)~\cite{Lev99,Lev00} uses the Nijmegen
II potential and is quantized in front form. The calculation of Forest and Schiavilla
(FS)~\cite{ForPr} uses the Argonne \( v_{18} \) potential and is quantized
in instant form. These three calculations involve three different approaches
to construct currents. AKP and LPS use different procedures for constructing
currents from the single-nucleon current without including any interaction dependent
two-body currents. FS assume that the potential is of one-boson-exchange origin
and construct the various currents as in the \( v/c \) expansions, but without
performing this expansion. These calculations show considerable variation in
all three observables. Although the FS calculation works quite well for all
of them, it is clear that no consensus has been reached concerning consistent
techniques for the construction of currents in this framework.

\subsection{Field theoretical models\label{ssec:FTM}}

\subsubsection{The Bethe-Salpeter equation\label{sssec:BSE}}

\begin{figure}
{\par\centering \resizebox*{12cm}{7cm}{\includegraphics{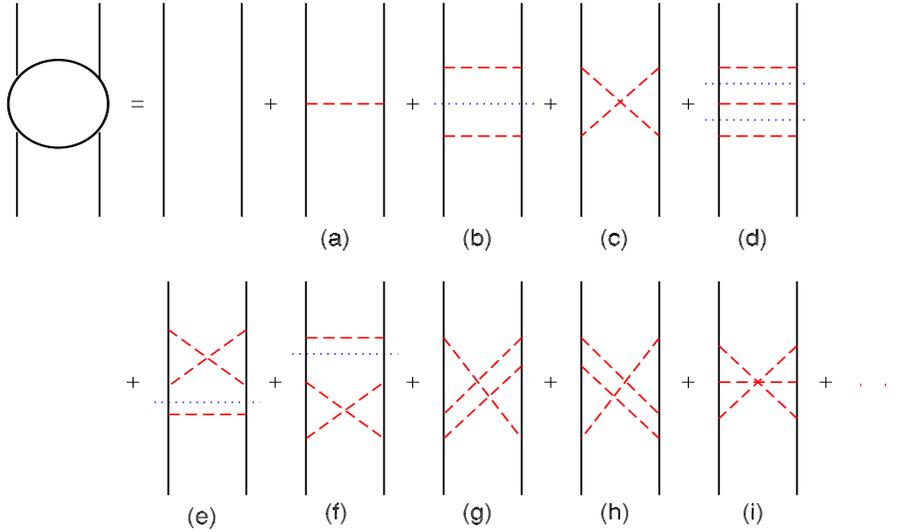}} \par}

\caption{Feynman diagrams respresenting the two-body scattering matrix \protect\( \mathcal{M}\protect \).\label{fig:scat2}}
\end{figure}

\begin{figure}
{\par\centering \resizebox*{10cm}{!}{\includegraphics{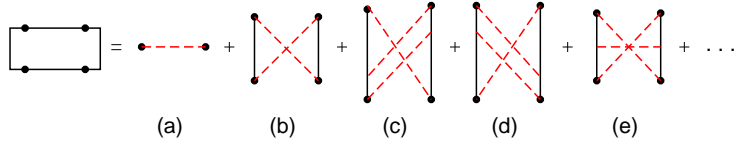}} \par}

\caption{Feynman diagrams representing the two-nucleon irreducible kernel \protect\( V\protect \).\label{fig:kernel}}
\end{figure}

\begin{figure}
{\par\centering \resizebox*{10cm}{!}{\includegraphics{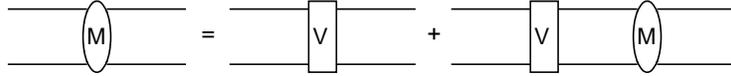}} \par}

\caption{Diagrammatic representation of the integral equation for the scattering matrix.\label{fig:four-point}}
\end{figure}

The oldest of the field-theoretical treatments of the two body-problem is due
to Bethe and Salpeter~\cite{Sal84}. The physical content of the Bethe-Salpeter
equation can be easily understood by considering the two-body scattering matrix
\( \mathcal{M} \). Figure~\ref{fig:scat2} represents the Feynman expansion
of the scattering matrix \( \mathcal{M} \)
 where the nucleons are represented by solid lines and the mesons are represented
by the dashed lines. Note that diagrams (b), (d), (e) and (f) can
be reduced to simpler diagrams by cutting across two nucleon lines
as represented by the dotted lines. These diagrams are said to be
\textsl{two-body reducible diagrams} and the remaining ones
\textsl{two-body irreducible diagrams}. The ability to classify
all of the contributing diagrams as members of one or the other of
these two classes suggests that the multiple scattering series can
be resummed by separating the irreducible diagrams into an
interaction kernel and then using an integral equation to produce
the reducible diagrams. The Feynman diagrams representing the
two-nucleon irreducible kernel are shown in Fig.~\ref{fig:kernel}.
The Bethe-Salpeter equation for the scattering matrix can then be written as
\begin{equation}
\label{eqn:BS}
\mathcal{M}(p',p;P)=V(p',p,P)-i\int \frac{d^{4}k}{(2\pi )^{4}}V(p',k;P)G_{0}(k,P)\mathcal{M}(k,p;P)
\end{equation}
 where \( P \) is the total four-momentum of the two-body system; \( p' \),
\( k \) and \( p \) are the final, intermediate and initial relative four-momenta
of the two particles; \( G_{0} \) is the free two-body propagator; and \( V \)
is a kernel consisting of a sum of all possible two-body irreducible diagrams.
The Bethe-Salpeter equation is represented by the diagrams in Fig.~\ref{fig:four-point}.

\begin{figure}
{\par\centering \resizebox*{10cm}{!}{\includegraphics{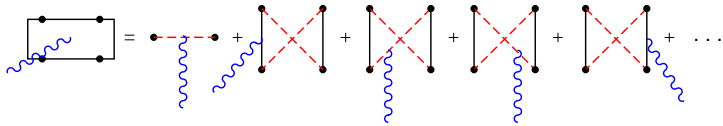}} \par}

\caption{The two-body irreducible current diagrams.\label{fig:exchange}}
\end{figure}

Since the Feynman series is organized such that all of the individual propagators
and vertices are covariant with respect to the free-particle Lorentz transformations,
the sum is manifestly covariant as well. The currents can be constructed by
combining the free one-nucleon currents with exchange currents obtained by attaching
a photon to every possible place within the irreducible interaction kernel,
as shown in Fig.~\ref{fig:exchange}. These currents will then satisfy the
Ward-Takahashi identities~\cite{War50}.

The two-body bound state is represented by the Bethe-Salpeter vertex function
\( \mathcal{O} \), which satisfies \begin{equation}
\label{eqn:BetheVertex}
\mathcal{O}(p,P)=i\int \frac{d^{4}k}{(2\pi )^{4}}V(p,k;P)G_{0}(k,P)\mathcal{O}(k,P)\, .
\end{equation}
 This vertex function for the deuteron can be written~\cite{Buc00} as\begin{equation}
\mathcal{O}_{\lambda _{d},ba}(p,P)=\left( \Gamma (p,P)\cdot \xi _{\lambda _{d}}(P){\mathcal{C}}\right) _{ba}\, .
\end{equation}
 \( P \) is the deuteron four-momentum, \( p \) the relative momentum of the
external nucleons, \( \xi _{\lambda _{d}}(P) \) the polarization four-vector
for the deuteron, \( \mathcal{C} \) the Dirac charge conjugation matrix. The
subscripts \( a \) and \( b \) are indices in the Dirac spinor space, and
\( \Gamma ^{\mu } \) can be determined by basic symmetry arguments to have
the general form: \begin{eqnarray}
\Gamma ^{\mu }(p,P) & = & g_{1}(p^{2},p\cdot P)\gamma ^{\mu }+g_{2}(p^{2},p\cdot P)\frac{p^{\mu }}{m}\nonumber \\
 &  & +\frac{\! \not p_{2}-m}{m}\left( g_{3}(p^{2},p\cdot P)\gamma ^{\mu }+g_{4}(p^{2},p\cdot P)\frac{p^{\mu }}{m}\right) \nonumber \\
 &  & +\left( g_{5}(p^{2},p\cdot P)\gamma ^{\mu }+g_{6}(p^{2},p\cdot P)\frac{p^{\mu }}{m}\right) \frac{\! \not p_{1}+m}{m}\nonumber \\
 &  & +\frac{\! \not p_{2}-m}{m}\left( g_{7}(p^{2},p\cdot P)\gamma ^{\mu }+g_{8}(p^{2},p\cdot P)\frac{p^{\mu }}{m}\right) \frac{\! \not p_{1}+m}{m}\, ,\label{eqn:vertex}
\end{eqnarray}
 where \( p_{1}=\frac{P}{2}+p \) and \( p_{2}=\frac{P}{2}-p \). A generalization
of the Pauli symmetry requires that \begin{equation}
\label{eqn:Pauli}
\Gamma ^{\mu }(p,P)=-{\mathcal{C}}\Gamma ^{\mu T}(-p,P){\mathcal{C}}^{-1}.
\end{equation}
 which places constraints on the scalar functions \( g_{i} \). Note that for
given values of \( p \) and \( P \), the vertex function depends upon eight
scalar functions whereas the Schr\"{o}dinger wave function of the deuteron
is determined by the two functions \( u \) and \( w \).

The Bethe-Salpeter equation is a complete representation of all possible contributions
to the two-body amplitudes for a given field theory. As such it retains all
of the complexities of field theories in that there is an infinite number of
contributions as represented by Feynman diagrams. The problem of particle dressing
and renormalization is present, and for strong coupling theories there is no
general scheme for organizing the renormalization program. As a practical matter,
model calculations generally assume that all propagators represent dressed particles
with physical masses. This still leaves an infinite number of contributions
to the irreducible kernel. Again, there is no \textit{a priori} means of establishing
a reasonable truncation procedure in order to obtain a tractable set of contributions.
However, the phenomenology of nuclear systems suggests that the importance of
contributions to the nuclear force depends upon the range of the contributions
due to the strong repulsion of nucleons at short distance. As a result, models
of the deuteron using field-theoretical techniques usually assume that contributions
can be ordered by range. In practice this reduces to the use of one-boson-exchange
potentials. This is referred to as the ladder approximation to the Bethe-Salpeter
equation.

The ladder approximation violates the crossing symmetry of the two-body scattering
amplitudes, which is a property of the full Bethe-Salpeter equation. This results
from the elimination of the higher-order crossed-box contributions to the kernel.
A related problem is that the two-body equation no longer reduces to a one-body
wave equation at the limit of infinite mass for one of the constituents (defined
as the static limit)~\cite{Gro93}.

A further difficulty for producing models of the deuteron using the field-theoretical
methods is that phenomenology often requires the introduction of nonrenormalizable
couplings such as the derivative coupling of the pion and tensor coupling of
vector mesons. The loop integration of the Bethe-Salpeter equation must then
be cut off to eliminate unphysically large or infinite contributions. This is
usually done by introducing form factors for strong coupling vertices, which
also attempts to include finite-size effects for the hadrons. In addition, the
hadrons have a finite electromagnetic size so that electromagnetic form factors
are also required by the phenomenology. The introduction of form factors can,
however, result in violations of gauge invariance.

The solution of the Bethe-Salpeter equation is also a calculational challenge
since it is a four-dimensional integral equation with a complicated analytical
structure. It has however been solved for the two-nucleon system in Euclidean
space~\cite{Fle75} and applied to the calculation of electron-deuteron scattering~\cite{Zui80,Umn94}.

\subsubsection{Quasipotential Equations\label{sssec:QPE}}

The Bethe-Salpeter equation is a four-dimensional integral equation. As such
it is much more difficult to solve numerically than the comparable nonrelativistic
three-dimensional Lippmann-Schwinger equation. To simplify the solution of the
relativistic equations, one resorts to the infinite class of quasipotential
equations~\cite{Bro76,Bla66,Tho70,Tod71,Erk72,Kad68,Gro69,Gro82,Wal89}. These
equations are related to the Bethe-Salpeter equation (\ref{eqn:BS}) by replacing
the free propagator \( G_{0} \) by a new propagator \( g \). The scattering
matrix equation can now be formally rewritten as \begin{equation}
\label{eqn:QuasipotentialM}
\mathcal{M}(p',p;P)=U(p',p;P)-i\int \frac{d^{4}k}{(2\pi )^{4}}U(p',k;P)g(k;P)\mathcal{M}(k,p;P)\, ,
\end{equation}
 where \( U \) is the \textit{quasipotential} defined as \begin{equation}
\label{eqn:QuasipotentialU}
U(p',p;P)\equiv V(p',p;P)-i\int \frac{d^{4}k}{(2\pi )^{4}}V(p',k;P)(G_{0}(k;P)-g(k;P))U(k,p;P)\; .
\end{equation}
 This pair of equations is equivalent to (\ref{eqn:BS}) and represents a re-summation
of the multiple scattering series. The new propagator \( g \) is usually chosen
to include a constraint in the form of a delta function involving either the
relative energy or time in such a way as to reduce (\ref{eqn:QuasipotentialM})
to three dimensions. In addition \( g \) must be chosen such that it has the
same residue as \( G_{0} \) along the righthand elastic cut. This guarantees
that the discontinuity of \( U \) produces only inelastic contributions as
in the case of (\ref{eqn:BS}).

Although it appears that the reduction of (\ref{eqn:QuasipotentialM}) to three
dimensions is of great practical advantage, it should be noted that (\ref{eqn:QuasipotentialU})
is still a four-dimensional integral equation of difficulty comparable to (\ref{eqn:BS}).
The real utility of these equations comes about when (\ref{eqn:QuasipotentialU})
is approximated by iteration and truncation. Then \( U \) can be obtained by
quadrature of (\ref{eqn:QuasipotentialU}) and used in the three-dimensional
equation (\ref{eqn:QuasipotentialM}).

It would appear that we have achieved a reduction in the difficulty of solution
of the problem at the expense of introducing considerable additional ambiguity
since there is an infinite number of possible choices for \( g \) which satisfy
the above specified requirements. However, this freedom is turned to an advantage
when noting that \( g \) can be chosen to maximize the convergence of (\ref{eqn:QuasipotentialU}).
In essence, the parts of the ladder and crossed box diagrams which tend to cancel
are being summed in the quasipotential equation (\ref{eqn:QuasipotentialU})
rather than in the equation for the scattering matrix (\ref{eqn:QuasipotentialM}).
Indeed, the lowest order calculations using a variety of quasipotential prescriptions
have been calculated for scalar particles~\cite{Nie96} and shown to be closer
to the results of the complete Bethe-Salpeter equation than is the result from
the ladder approximation to the Bethe-Salpeter equation. In addition, it has
been shown that there exists an infinite number of quasipotential equations
which have the correct static limit. This is also a reflection of the improved
convergence of these quasipotential equations~\cite{Coo89}.

The results from three different quasipotential models are
presented here. The first of these is the calculation of Hummel
and Tjon (HT)~\cite{Hum89} based on the BSLT
equation~\cite{Bla66}. In this case, the propagator \( g_{BSLT} \)
contains a term \( \delta (p_{0}) \) which forces the relative
energy of the two nucleons to zero. One can say that the two
nucleons are equally off mass-shell. The vertex functions are
calculated using the BSLT equation with a one-boson exchange
interaction. They still have eight components. The current matrix
elements are then calculated by replacing the Bethe-Salpeter
vertex functions with the BSLT vertex function, assuming that it
is energy independent. In the examples shown here the nucleon
electromagnetic form factors are taken to be the H\"{o}hler 8.2
parameterization.

The second model is that of Phillips, Wallace and Devine (PWD)~\cite{Phi98,Phi98a,Phi99}
where a one-boson-exchange interaction is used with the single-time equation
that introduces a constraint that the relative time be zero. This is very close
to the HT approach since this choice means that the propagator \( g_{ET} \)
is independent of the relative energy of the two nucleons. The single-time deuteron
vertex functions still have eight components. A consistent treatment of the
current is constructed to guarantee current conservation, but Lorentz covariance
is violated. The MMD parameterization of the nucleon electromagnetic form factor
are used.

The third model is that of Van Orden, Devine and Gross (VODG)~\cite{Van95,Van95a,Van95b},
based on the Gross equation~\cite{Gro69,Gro82}. One of the nucleons is placed
on its positive energy mass shell, which gives:\begin{equation}
g_{Gross}\propto \frac{1}{p_{2}^{2}-m^{2}+i\varepsilon }\delta ^{(+)}(p_{1}^{2}-m^{2})
\end{equation}
 Since this constraint is itself manifestly covariant, the Gross equation amplitudes
are manifestly covariant. But it is not symmetric in the nucleons and the exchange
symmetry must be recovered by symmetrizing the interaction kernel. This procedure
introduces unphysical singularities, which are treated in principal value to
avoid unitarity problems and have little numerical effect in a weakly bound
system such as the deuteron. As a result of placing one nucleon on shell in
(\ref{eqn:vertex}), the Gross equation deuteron vertex function has four-components
that can be represented in terms of an \( S \) wave, a \( D \) wave and singlet
and triplet \( P \) wave functions.

A systematic procedure for constructing the effective current operators for
the Gross equation exists~\cite{Gro87} and has been shown to be rigorous to
all orders and remarkably robust under trucation~\cite{Ada98}. The result
is that Ward identities for the Gross equation are maintained and the calculation
is gauge invariant.

A one-boson-exchange interaction is used, that has been fit to \( NN \) scattering
data up to a lab kinetic energy of about 300 MeV~\cite{Gro90,Gro92}. The fit
is reasonable, but not at the level of precision obtained with modern fully
phenomenological potentials.

Several unique features appear in this model. All of the strong vertices are
multiplied by a product of three form factors, \begin{equation}
h(p'^{2})h(p^{2})f((p'-p)^{2})\, .
\end{equation}
 Here \( p \) and \( p' \) are the initial and final four-momenta of the nucleon
and \( p'-p \) is the four-momentum of the meson. The meson form factor is
taken to be \begin{equation}
f(\ell ^{2})=\frac{(\Lambda _{\mu }^{2}-\mu ^{2})^{2}+\Lambda _{\mu }^{4}}{(\Lambda _{\mu }^{2}-\ell ^{2})^{2}+\Lambda _{\mu }^{4}}
\end{equation}
 and the nucleon form factor \begin{equation}
h(p^{2})=\frac{2(\Lambda _{n}^{2}-m^{2})^{2}}{(\Lambda _{n}^{2}-p^{2})^{2}+(\Lambda _{n}^{2}-m^{2})^{2}}\, .
\end{equation}

The prescription of Gross and Riska~\cite{Gro87} is used to construct a single-nucleon
electromagnetic current which is consistent with the strong-vertex form factors
and phenomenological single-nucleon electromagnetic form factors. This current
is constructed in such a manner that the one- and two-body Ward identities are
of the same form as for the local field theory. The simplest form that this
single-nucleon current can take is \begin{eqnarray}
J^{(i)\mu }(p',p) & = & F_{1}(Q^{2})f_{0}(p'^{2},p^{2})\, \gamma ^{\mu }+\frac{F_{2}(Q^{2})}{2m}h_{0}(p'^{2},p^{2})\, i\sigma ^{\mu \nu }q_{\nu }\nonumber \\
 &  & \quad +F_{10}(Q^{2})g_{0}(p'^{2},p^{2})\, \frac{\! \not p'-m}{2m}\, \gamma ^{\mu }\, \frac{\! \not p-m}{2m}\, ,\label{eq:gross-riska}
\end{eqnarray}
 where the nucleon form factors are \begin{equation}
F_{i}(Q^{2})=\frac{1}{2}\left( F_{i}^{S}(Q^{2})+F_{i}^{V}(Q^{2})\cdot \tau _{3}\right) \, .
\end{equation}
 The factors that depend upon the nucleon momenta are \begin{equation}
f_{0}(p'^{2},p^{2})\equiv \frac{h(p^{2})}{h(p'^{2})}\, \frac{m^{2}-p'^{2}}{p^{2}-p'^{2}}+\frac{h(p'^{2})}{h(p^{2})}\, \frac{m^{2}-p^{2}}{p'^{2}-p^{2}},
\end{equation}
 and \begin{equation}
g_{0}(p'^{2},p^{2})\equiv \left( \frac{h(p^{2})}{h(p'^{2})}-\frac{h(p'^{2})}{h(p^{2})}\right) \frac{4m^{2}}{p'^{2}-p^{2}}\, .
\end{equation}
The function \( h_{0}(p'^{2},p^{2}) \) must be equal to one when
both nucleons are on mass shell, but is otherwise unconstrained.
For simplicity, the calculations shown here use \( h_{0}=f_{0} \).
The form factor \( F_{10}(Q^{2}) \) must obey \( F_{10}(0)=1 \),
but is otherwise unconstrained and is chosen to be of a dipole
form.
\begin{figure}
{\par\centering \resizebox*{10cm}{!}{\includegraphics{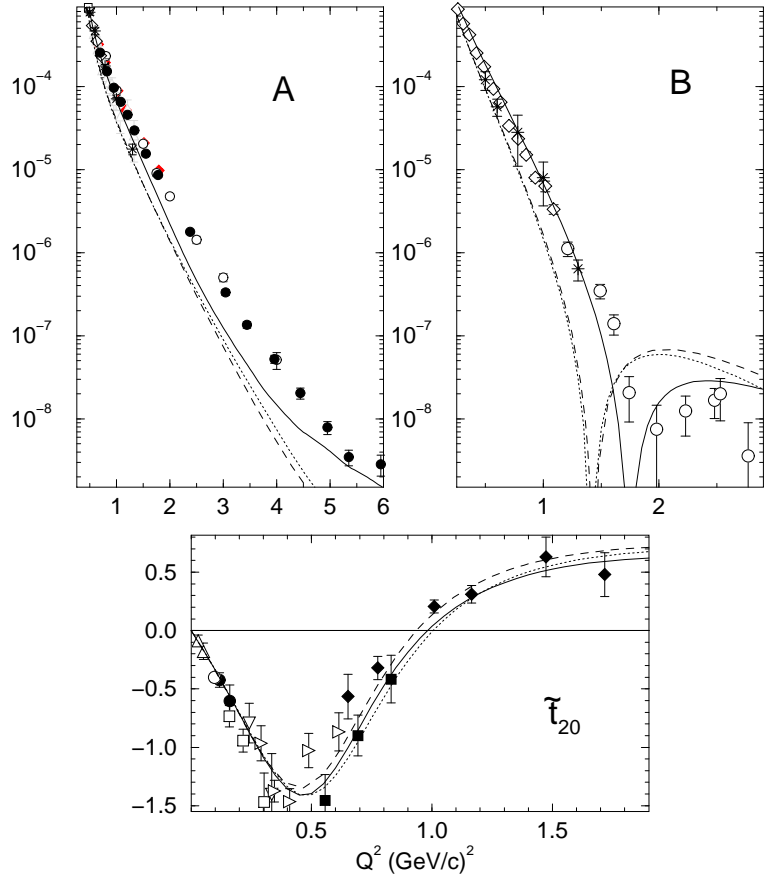}} \par}

\caption{Elastic \protect\( ed\protect \) scattering observables calculated with Quasipotential
equations. HT-RIA, NEMFF: H (dotted curve); PDW-RIA, NEMFF: MMD (dashed); VODG-CIA,
NEMFF: MMD (solid).\label{fig:a_b_t20t_RIA}}
\end{figure}

Figure~\ref{fig:a_b_t20t_RIA} shows the elastic structure functions for these
models calculated in the relativistic impulse approximation (RIA). As expected
from the nature of the quasipotential approximations, the calculations of HT
and PWD lead to very similar results for all three observables (in spite of
the fact that they use different NEMFF). The VODG calculation uses the MMD form
factors. For \( A(Q^{2}) \), it is systematically larger than the others due
to additional contributions that are necessary to conserve the current within
the context of the Gross equation. These additions yield to the so-called Complete
Impulse Approximation (CIA).

For \( B(Q^{2}) \), the HT and PWD calculations are systematically below the
data and have a minimum at a lower \( Q^{2} \) than is indicated by the data.
The VODG calculation is higher and has a minimum at higher \( Q^{2} \) than
the other two calculations. This difference has been shown to be due to a small
\( P \) wave component of relativistic origin that contributes to the normalization
of the vertex function at the level of 0.01\%~\cite{Van95b}. This clearly
indicates the sensitivity of the position of the \( G_{M} \) node to small
components.

Finally, in the case of \( \tilde{t}_{20}(Q^{2}) \), the three
models have a very similar behavior.

\begin{figure}
{\par\centering \resizebox*{10cm}{!}{\includegraphics{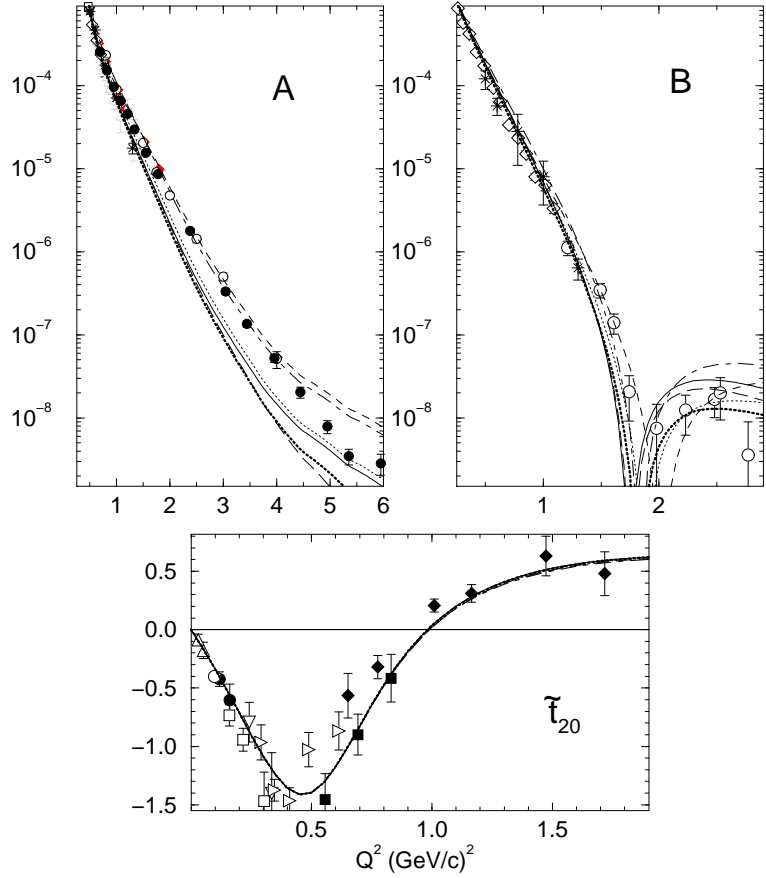}} \par}

\caption{Deuteron elastic structure functions calculated using the CIA of VODG with
various parameterizations of the nucleon form factors: Galster (dotted curve),
Höhler (long dashed), GK85 (dot-dashed), GK92 (dashed), MMD (solid), VO2 (thick
dotted).\label{fig:a_b_t20t_NFF}}
\end{figure}

The sensitivity of the observables to the choice of single-nucleon
form factors is shown in Fig.~\ref{fig:a_b_t20t_NFF}, using the
VODG calculation in the CIA approximation. All but the VO2
parameterization are standard form factors that appear in the
literature. The VO2 form factor is modeled after the MMD form
factor but adjusted to fit the new data on \( G_{E}^{p}/G_{M}^{p}
\) from Jefferson Lab~\cite{Jon00}. As expected from the
discussion in App.~C, \( A(Q^{2}) \) and \( B(Q^{2}) \) are very
sensitive to the choice of single nucleon form factor while \(
\tilde{t}_{20}(Q^{2}) \) is almost totally insensitive to it. The
various \( A(Q^{2}) \) curves in Fig.~\ref{fig:a_b_t20t_NFF} may
be directly related to the corresponding \( G_{E}^{S} \) curves in
Fig.~\ref{fig:grap_iso}. Note that for the relativistic
approaches, in contrast to the NRIA, the nucleon form factors
cannot simply be factorized, so that there is no reason that \(
\tilde{t}_{20}(Q^{2}) \) be completely independent of the NEMFF.
Much more accurate data on the single-nucleon form factors that
will be forthcoming from experimental facilities around the world
will, hopefully, provide much greater constraints on these form
factors.

\begin{figure}
{\par\centering \resizebox*{10cm}{!}{\includegraphics{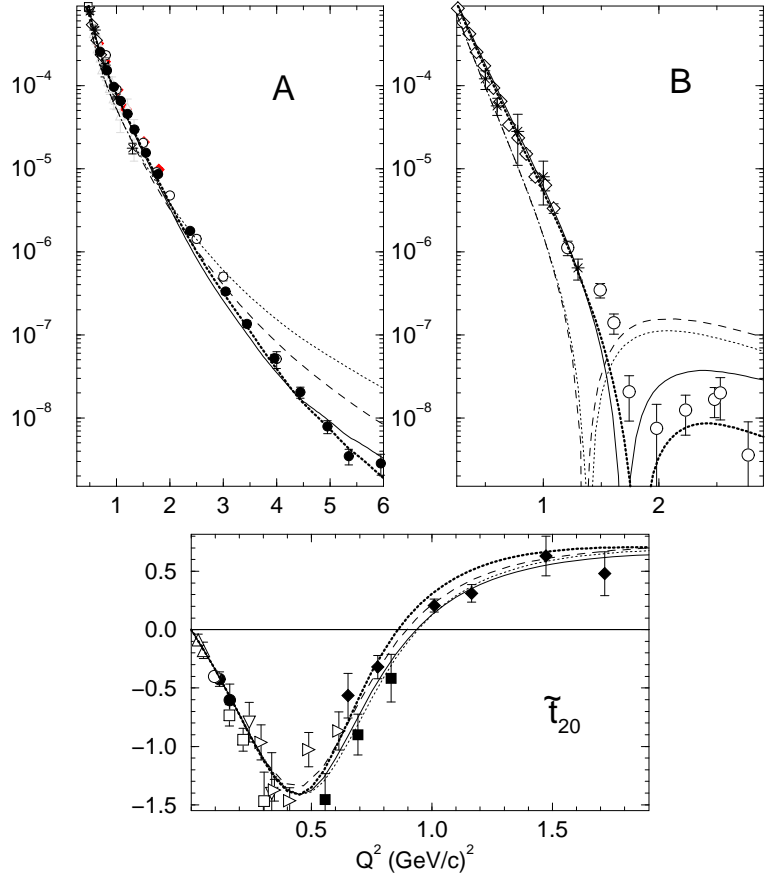}} \par}

\caption{Elastic \protect\( ed\protect \) scattering observables calculated with Quasipotential
equations, including the \protect\( \rho \pi \gamma \protect \) contribution.
HT (dotted curve), PDW (dashed), VODG-MMD (solid), VODG-SO2 with \protect\( F_{10}\protect \)
off-shell form factor adjusted (thick dotted). \label{fig:a_b_t20t_BS} }
\end{figure}

Figure \ref{fig:a_b_t20t_BS} shows the effect of the addition of \( \rho \pi \gamma  \)
exchange currents to the three calculations. HT were the first to implement
the calculation of the \( \rho \pi \gamma  \) contribution to the deuteron
form factors in a relativistic model. They also included a \( \omega \sigma \gamma  \)
exchange current, which compensates their large effect of the \( \rho \pi \gamma  \).
This last contribution is not present in the calculation presented here, because
there is considerable uncertainty about the \( \sigma  \) meson, and consequently
about its couplings and vertex form factors (see~\cite{Gok00} for a recent
discussion of the \( g_{\omega \sigma \gamma } \) coupling constant). Two sets
of calculations for the VODG structure functions are shown, one with the MMD
form factors as before, and one with the VO2 form factors. Comparing to Fig.~\ref{fig:a_b_t20t_RIA},
it is clear that this exchange current can result in considerable variation
in \( A(Q^{2}) \). The variation in \( B(Q^{2}) \) is much smaller, in part
because of the smaller \( Q^{2} \) range considered. The \( \rho \pi \gamma  \)
exchange current shifts the node of \( G_{M} \) to lower values of \( Q^{2} \)
for all calculations, in contrast with the nonrelativistic calculations using
the \( v/c \) expansion where this current has the opposite effect. This is
the result of the tensor coupling of the \( \rho  \) to the nucleon, which
appears to be a higher order contribution in the \( v/c \) expansion, but weighted
by a large coefficient. In this way, expansion schemes may lead to wrong estimates
without a careful examination of higher-order contributions. Finally, the addition
of the \( \rho \pi \gamma  \) contribution has a small effect on \( \tilde{t}_{20}(Q^{2}) \),
but brings the calculations in a somewhat better agreement with the available
data.

\begin{figure}
{\par\centering \resizebox*{7cm}{!}{\includegraphics{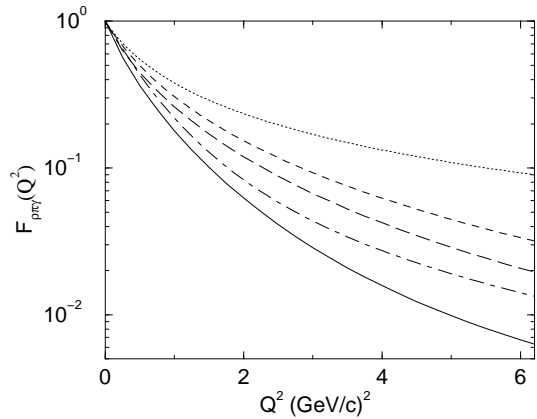}} \par}

\caption{Different prescriptions for the \protect\( \rho \pi \gamma \protect \) form
factor: VMD (dotted curve), Gross-Ito~\cite{Ito93} (dashed), Rome 1~\cite{Car95}
(long dashed), Mitchell-Tandy~\cite{MT95} (dot-dashed), Rome 2~\cite{Car95}
(solid).\label{fig:rpg}}
\end{figure}

The variation in the \( \rho \pi \gamma  \) contributions from the various
models~\cite{Hum89,Ito93,Car95,MT95} is associated with ambiguities in these
contributions. Although the coupling constant for the \( \rho \pi \gamma  \)
vertex is constrained by the decay width \( \Gamma (\rho \rightarrow \pi \gamma ) \),
little is known about the fall off of the associated form factor. Figure~\ref{fig:rpg}
shows a number of different form factors for this vertex. HT use the VMD form
factor which is the hardest of the available form factors while VODG use the
Rome 2 form factor which is the softest one. PWD use an intermediate parameterization.
The \( A(Q^{2}) \) data favor the use of the softest possible form factor.
In contradistinction, a recent theoretical reexamination of the \( \rho \pi \gamma  \)
form factor~\cite{Tru00} results in a dependence close to the VMD one. Whether
the \( \rho \pi \gamma  \) contribution to the deuteron form factors is really
small, or suppressed by other contributions is still an open question.

Another source of ambiguity, explicitly present in the the VODG calculation,
is the choice of the form factor \( F_{10}(Q^{2}) \) in (\ref{eq:gross-riska}).
In these calculations, \( F_{10}(Q^{2}) \) is adjusted to optimize the fit
of the calculation to the data for the deuteron structure functions. By using
a very hard form factor it is possible to obtain an extremely good fit to the
data. This along with ambiguities in the \( \rho \pi \gamma  \) form factor
means that no absolute predictions of the structure functions can be obtained
unless some means can be found to physically constrain the ambiguities in the
models.

\subsubsection{Light-front field theory\label{sssec:LFD}}

The above approaches based on the Bethe-Salpeter equation are generally discussed
in the context of Feynman perturbation theory with equal time quantization.
Microcausality, however, only requires that the theory be quantized on a spacelike
hypersurface. A special limiting case of such a surface is the light cone where
the spacetime interval approaches zero. Quantizing field theories on the light
cone has long been common practice in describing deep inelastic scattering where
the large four-momenta kinematically favor contributions to scattering very
near the light cone. Typically, the light-cone approach is organized as a {}``time-ordered{}''
perturbation expansion in terms of the light-front time \( x^{-}=x^{0}-x^{3} \).
The wave functions are then Fock-space wave functions with a probabilistic interpretation
as in the case of Schr\"{o}dinger wave functions.

A particular problem with light-front field theory is that the conventional
choice of a fixed light-front orientation violates manifest covariance, although
matrix elements remain covariant. One solution to this problem~\cite{Car98}
is to describe the quantization surface in terms of an arbitrary direction on
the light cone given by a light-like four-vector \( \omega ^{\mu } \) and require
that the quantization surface be described by \( \omega \cdot x=0 \). This
renders the theory manifestly covariant, at the expense of the additional complication
that the wave functions and operators become dependent on the vector \( \omega  \).
This is covariant light-front dynamics (LFD). Matrix elements and observables,
in a complete calculation, do not depend on \( \omega  \). In a practical calculation,
a definite procedure to eliminate the non physical \( \omega  \)-dependent
terms is applied.

\begin{figure}
{\par\centering \resizebox*{10cm}{!}{\includegraphics{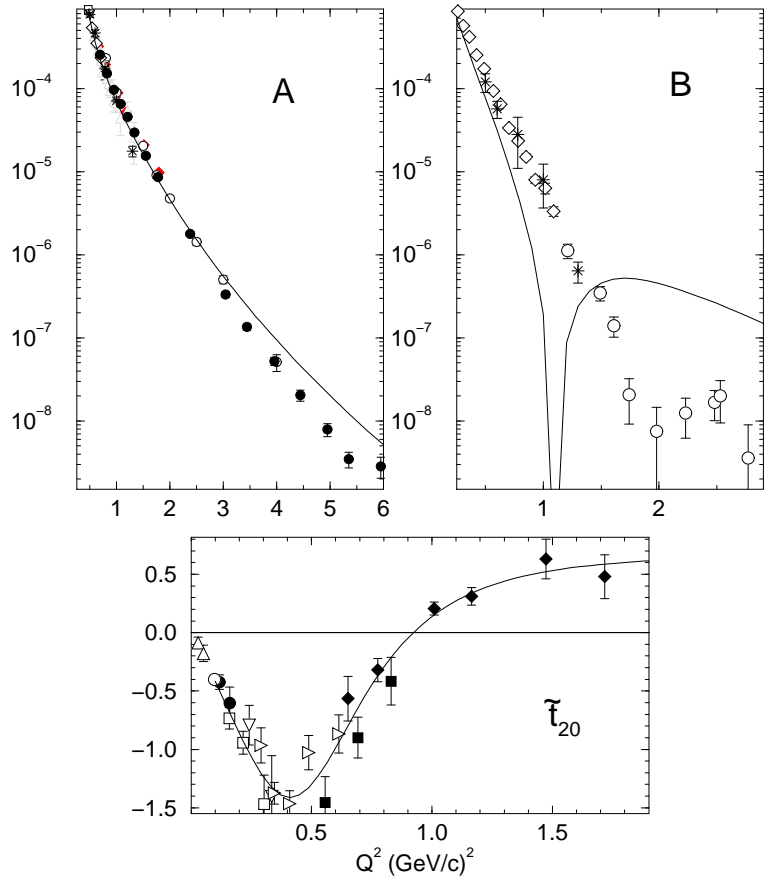}} \par}

\caption{Elastic \protect\( ed\protect \) scattering observables calculated within
light-front dynamics. \label{fig:a_b_t20t_LFD}}
\end{figure}

The deuteron elastic structure functions from this approach~\cite{Car98,Kar92,Car99a}
are shown in Fig.~\ref{fig:a_b_t20t_LFD}.

In these calculations, the deuteron wave function has six components (\( f_{i} \)),
two of which (\( f_{1} \) and \( f_{2} \)) correspond to the usual \( S \)
and \( D \) components in the nonrelativistic limit. These components are calculated
perturbatively from the usual nonrelativistic wave function. Schematically~\cite{CarK95},\begin{equation}
\psi (\mathbf{k},\mathbf{n};f_{1},f_{2},f_{3},f_{4},f_{5},f_{6})\sim \int V(\mathbf{k'},\mathbf{k},\mathbf{n},M_{d})\psi _{NR}(\mathbf{k};u,w)d^{3}\mathbf{k'}\, ,
\end{equation}
 where \( \mathbf{k} \) is the relative momentum of the two nucleons and \( \mathbf{n} \)
a unit vector along the three-vector \( \omega  \). The \( NN \) interaction
kernel \( V \) is of a one-boson-exchange type, calculated within the framework
of LFD using the parameters (such as coupling constants) of the Bonn potential.
The single-nucleon electromagnetic form factors are the MMD parameterization
(the effect of various NEMFF was considered in~\cite{Car99a}). This calculation
produces a reasonable description of the data for \( A(Q^{2}) \) and \( \tilde{t}_{20}(Q^{2}) \)
but the minimum in \( B(Q^{2}) \) occurs well below the position indicated
by the data. Given the sensitivity of the magnetic form factor to small effects,
this may simply be the result of the perturbative treatment of the small components
of the wave functions in this calculation.

Indeed, work is in progress to improve the calculation of the interaction kernel
\( V \) and of the complete wave function~\cite{CarPr}. Meson exchange contributions
are only partially included in the original calculations. The addition of recoil
terms and of the \( \rho \pi \gamma  \) contribution was also initiated: the
recoil terms move the node of \( G_{M} \) to higher \( Q^{2} \), while the
\( \rho \pi \gamma  \) contribution does not modify appreciably any observable
up to \( Q^{2}\sim 2 \) \gev2~\cite{Haf99}. A complete calculation using
LFD seems at hand and very promising.

\subsubsection{Effective field theory}

In principle, all these field theoretical techniques should yield
the same results, given the same dynamical input. This statement
applies to the calculated matrix elements which are the physical
observables of the field theory. The identification of wave
functions and operators varies from formulation to formulation and
do not have unique meanings. Furthermore, within any given
formulation, unitary transformations of the Lagrangian, such as
field redefinitions, can also move contributions between the wave
functions and operators. Such ambiguities are already present in
nonrelativistic models, which may be shown to be equivalent, at
least in part, up to a unitary transformation~\cite{Fri79,Des92}.
None of this would be of particular concern if it were not
necessary to truncate all of the approaches for reasons of
practicality. Consequently, the physical content of the various
formulations varies, resulting in a variety of ambiguities, some
of which have been discussed above. A major problem then is that
there is, in general, no organizing principle present in these
calculations which indicates the relative importance of various
physical contributions to guide the choice of truncation schemes.

A promising development that could help to resolve this problem is the application
of effective field theories to nuclear systems~\cite{Wei91,Ord92,Ord94,Kap96,Kap99,Bea98,Par98,Kol99,Epe98}.
The basic idea of effective field theory is that at low energies the observables
of a theory are largely insensitive to the details of short-range contributions
to the interactions. The long-range degrees of freedom are then treated in detail
and the short range pieces are replaced by contact interactions. The Lagrangian
is written as a sum of terms containing contact interactions with increasing
numbers of derivatives of the fields. This constitutes an expansion of the theory
in terms of momenta which are small compared to some scale chosen to separate
the short and long range physics. Contributions to observables are then ordered
according to this small momentum, providing a well defined counting scheme that
controls approximations to the theory.

For nuclear systems, the appropriate effective field theory is
Chiral Perturbation Theory (\( \chi  \)PT) since the long-range
part of the nuclear force is associated with the pion, a Goldstone
boson. A considerable amount of effort is being expended in the
application of \( \chi  \)PT to low energy nucleon-nucleon
scattering and to the deuteron. The \( NN \) system is
particularly challenging in that its scattering lengths are large,
although its bound state, the deuteron, is only weakly bound. This
implies that there is a dynamical scale in the problem in addition
to the chiral scale. The application of \( \chi  \)PT to the
deuteron has followed two approaches. The first~\cite{Wei91}
applies the chiral counting scheme at the level of the interaction
kernel. This approach appears to converge, but is cutoff dependent
at each order. The second~\cite{Kap96,Kap99} applies the counting
scheme at the level of the scattering matrix and current matrix
elements and treats the pions perturbatively while iterating the
contact interactions. It gives results that are cutoff
independent, but does not appear to be converging~\cite{Sav00}.
The problem of maintaining a consistent counting scheme in the
infinite sums of diagrams necessary to describe the bound states
has not yet been resolved. From the standpoint of the various
models used at higher momentum transfers, it is to be hoped that
effective field theory will provide some insight into the
organization of the various approaches.

\subsection{Deuteron models with nucleon isobar contributions}

In all of the approaches and models discussed to this point, the assumption
has been made that the only relevant degrees of freedom are the nucleons and
mesons. However, excitations of the nucleons to isobar states may produce contributions
to the interactions of the same range as the heavier mesons. The deuteron wave
function is in this case modified to include, in addition to the \( S \) and
\( D \) wave \( NN \) components, \( \Delta \Delta  \) and \( NN^{*} \)components.
A neutrino experiment, though subject to interpretation, indicates an upper
limit of 0.4 \% to the amount of \( \Delta \Delta  \) components in the deuteron~\cite{All86}.

\begin{figure}
{\par\centering \resizebox*{9cm}{!}{\includegraphics{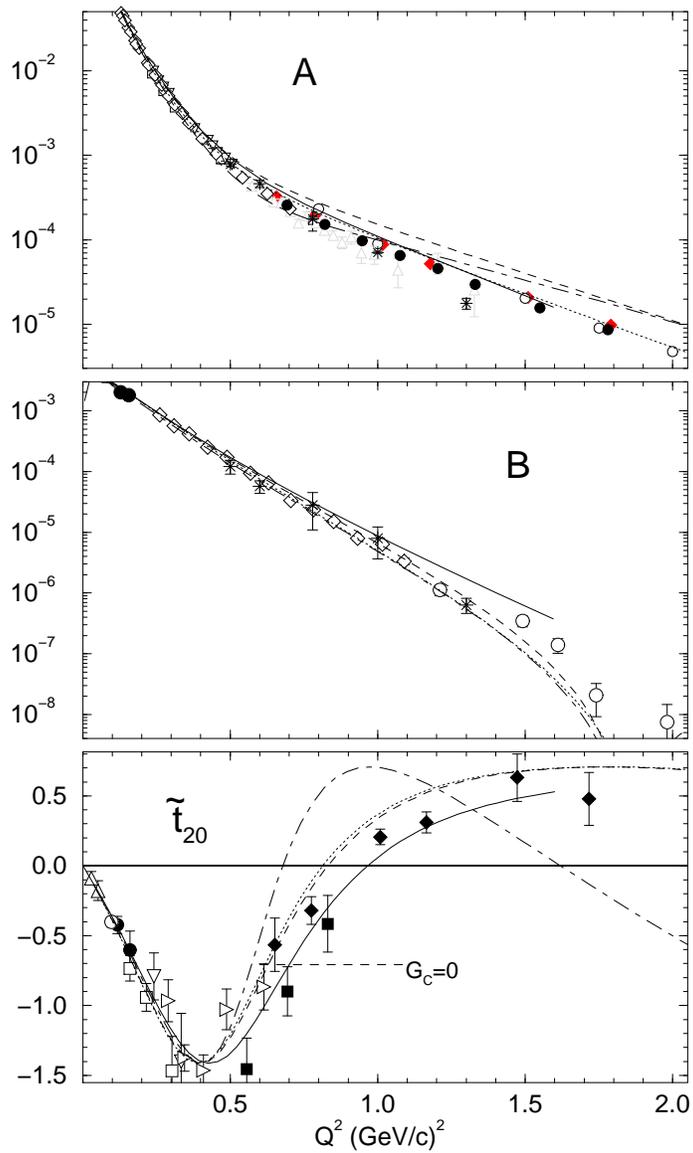}} \par}

\caption{Elastic \protect\( ed\protect \) scattering observables for models including
nucleon isobar components. DK, NEMFF: GK~\cite{Dym90} (solid curve); BGL-C',
NEMFF: H~\cite{Blu89} (dotted); BGL-C', NEMFF: GK (dashed); BGL-D', NEMFF:
GK (dot-dashed).\label{fig:a_b_t20t_isobar}}
\end{figure}

Examples of two calculations including these additional isobaric components
are shown here. In both cases, the basic model involves interactions due to
one meson exchange which can couple the nucleon-nucleon channel with channels
containing isobars. The quark model is used to relate isobar-meson couplings
(for instance \( \Delta \Delta \pi  \)) to the corresponding nucleon-meson
couplings. The fully coupled (nonrelativistic) system is then solved for the
deuteron bound state and the nucleon-nucleon scattering states. Figure \ref{fig:a_b_t20t_isobar}
shows the calculations of Dymarz and Khanna (DK)~\cite{Dym90} and Blunden
\textit{et al.} (BGL)~\cite{Blu89}. In both cases presented here, only the
\( \Delta (1232) \) isobar is included, and the \( \Delta  \) electromagnetic
form factors are assumed to be proportional to the NEMFF. The DK calculation
contains \( \Delta \Delta  \) components \( ^{3}S_{1},^{3}D_{1},^{7}D_{1},^{7}G_{1} \)
with a total probability of \( 0.36\% \). The elastic form factors include
contributions from single-nucleon currents with the GK form factors, nucleon
pair and \( \rho \pi \gamma  \) exchange currents and the isobar current contributions.
Two models are shown for BGL, both with only \( ^{3}D_{1} \) and \( ^{7}D_{1} \)
\( \Delta \Delta  \) components.. Two different quark models are assumed to
fix at a given radius \( r_{0} \) the boundary conditions used in the determination
of the wave functions. Model C' has \( r_{0}=0.74 \) fm and yields an isobar
contribution of \( 1.8\% \), while model D' uses the Cloudy Bag Model to fix
\( r_{0} \) at 1.05 fm and results in an isobar contribution of \( 7.2\% \).
Meson exchange contributions are included as well.

Within these models, \( \tilde{t}_{20} \) is very sensitive to the amount of
\( \Delta \Delta  \) components and favors the smallest probability. Although
the BGL model C' with the H\"{o}hler NEMFF seems to give an adequate description
of the data, there is considerable ambiguity in these calculations due to a
lack of knowledge of the isobar magnetic moments and form factors along with
the difficulty of completely constraining the isobar contributions from the
\( NN \) scattering data. It is clear, however, that isobar components must
ultimately be considered in any complete calculation of deuteron electromagnetic
properties in the context of hadronic models.

\subsection{Quarks and gluons}

\subsubsection{Nonrelativistic quark models\label{sssec:QCM}}

\begin{figure}
{\par\centering \resizebox*{9cm}{!}{\includegraphics{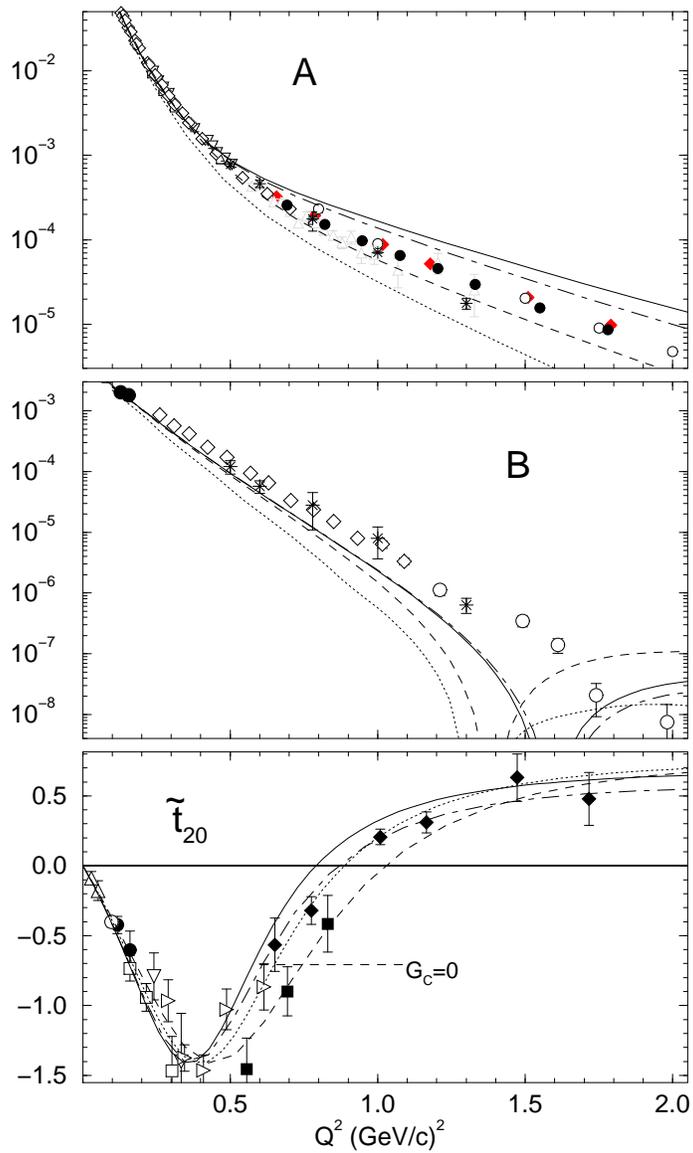}} \par}

\caption{Elastic \protect\( ed\protect \) observables calculated with quark-cluster
models. IK~\cite{Ito89} (dotted curve); BYF~\cite{Buc89}: IA (dashed); BYF:
IA + MEC (dot-dashed); BYF: former + quark exchange contributions (solid).\label{fig:a_b_t20t_QCM}}
\end{figure}

Ultimately, many of the ambiguities associated with the hadronic
model calculations can only be removed by describing the deuteron
in terms of the fundamental quark and gluon degrees of freedom. In
the absence of the capability to directly solve the QCD Lagrangian
at low energies for the deuteron, we are left to explore possible
QCD effects in the context of quark models. This greatly increases
the difficulty of treating the \( NN \) system since what is a
two-body problem in the hadronic models, becomes at least a
six-body problem for quark models. Calculations for two examples
of quark cluster models are presented here. These calculations are
very similar in concept and are based on a quark cluster model
using the resonating group method to describe the interaction of
the two three-quark clusters. The quarks interact via a quadratic
confining potential and a one-gluon-exchange potential. Long range
interactions are also provided via \( \pi  \) and \( \rho  \)
exchange between quarks. The calculation of Buchmann, Yamauchi and
Faessler (BYF)~\cite{Buc89} contains only the \( \pi  \) exchange
interaction, while that of Ito and Kisslinger (IK)~\cite{Ito89}
contains both meson-exchange contributions. This approach
naturally contains currents associated with the individual
clusters as well as exchange currents associated with quark
exchange between clusters. Since the cluster wave functions are
derived using oscillator-like confining forces, the cluster form
factors tend to have a Gaussian form. This is clearly not
consistent with the data for single-nucleon form factors and in
both cases the electromagnetic cluster form factors are replaced
by phenomenological NEMFF. The results are shown in
Fig.~\ref{fig:a_b_t20t_QCM}. Neither of these calculations
provides an adequate description of the data, but given their
necessary simplicity, they are remarkably close to the data. In
the BYF calculation, the impulse approximation is not as reliable
as in \( NN \) models because of the simple modeling of the
intermediate range interaction, but the size of the genuine quark
exchange contributions, due to the antisymmetrization of the
six-quarks wave function, is significant enough to affect a
comparison with the data.

\subsubsection{Perturbative QCD\label{sssec:PQCD}}

\begin{figure}
{\par\centering \resizebox*{!}{7cm}{\includegraphics{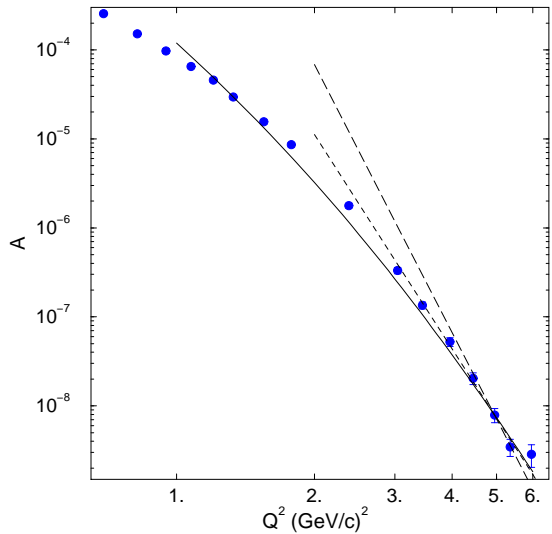}} \par}

\caption{\protect\( A(Q^{2})\protect \) in a Log-Log scale. The data is from JLab/HallA~\cite{Ale99}.
The straight lines illustrate a \protect\( Q^{-16}\protect \) (dashed) and
a \protect\( Q^{-20}\protect \) (long-dashed) dependence. The solid line represents
Eq.(\ref{eq:redff}). All curves are normalized to the point at \protect\( Q^{2}=4.95\protect \) \gev2.\label{fig:pqcd_a}}
\end{figure}

\begin{figure}
{\par\centering \resizebox*{!}{11cm}{\includegraphics{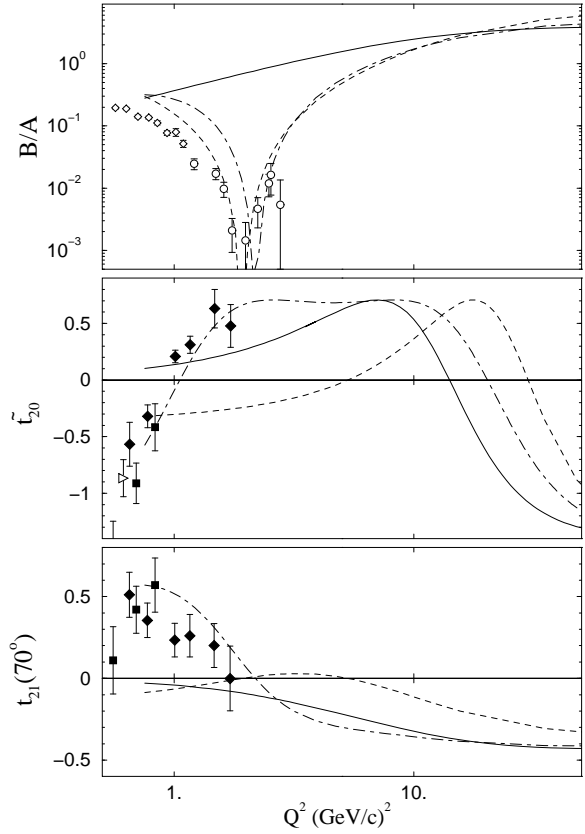}} \par}

\caption{Ratios \protect\( B/A\protect \), \protect\( \widetilde{t}_{20}\protect \)
and \protect\( t_{21}\protect \) from light-cone frame helicity amplitudes.
The curves correspond to Eq.(\ref{eq:g+}) with different values of \protect\( a\protect \)
and \protect\( b\protect \) (see text), from Refs.~\cite{Bro92} (solid line),
\cite{Kob94} (dashed) and \cite{Gar99} (dot-dashed). Note the logarithmic
scale in \protect\( Q^{2}\protect \).\label{fig:pqcd_ratios}}
\end{figure}

This subject was reviewed recently~\cite{Car97}. The success and shortcoming
of perturbative quantum-chromodynamics (PQCD) applied to the asymptotic behaviour
of form factors may be better illustrated now with recent data from Jefferson
Lab. We will only recall briefly the main predictions:

\begin{itemize}
\item Dimensional scaling~\cite{Bro75}: this property can be derived within QCD,
but is more general and was established before this theory. It is based on the
hypothesis that the momentum transfer is shared among all the constituent quarks
of the system. In the case of a hadron composed of \( n \) quarks, one would
expect the leading form factor to behave asymptotically as \( Q^{-2(n-1)} \).
Qualitatively, this can be viewed as the probability of having \( n-1 \) quarks
within a transverse distance of \( 1/Q \) of the quark struck by the virtual
photon. Since this must be true both in the initial and final states, one expects
\( F^{2}\sim Q^{-4(n-1)} \) and for the deuteron (\( n=6 \)), \( A\sim Q^{-20} \).
\item Logarithmic corrections to the leading amplitude, together with the factorization
of the nucleon form factors in the weak binding limit, yield~\cite{Bro83}:\begin{equation}
\label{eq:redff}
\sqrt{A}\sim F_{N}^{2}\left( \frac{Q^{2}}{4}\right) \cdot \frac{1}{Q^{2}}\cdot \left( \ln \frac{Q^{2}}{\Lambda ^{2}}\right) ^{-1+\epsilon }
\end{equation}
 where \( F_{N} \) is the nucleon form factor, \( \Lambda  \) an energy scale
characteristic of QCD (\( \Lambda \simeq 200 \) MeV) and \( \epsilon  \) a
calculable number much smaller than 1. An attempt to directly compute \( A \)
and its normalization in PQCD was not successfull~\cite{Car97,Far95}.
\item Helicity conservation at each photon/gluon-quark vertex implies that the dominant
contribution to elastic electron deuteron scattering should come from the configuration
where the deuteron has helicity zero in both initial and final states~\cite{Car84}.
Expressing the form factors in an helicity basis in the light-cone frame, this
is equivalent to the prediction that \( G_{00}^{+} \), the {}``\( + \){}''
component of the current matrix element between states of helicity \( 0 \),
is the leading amplitude. It was argued~\cite{Bro92} that this helicity conserving
amplitude should dominate the \( ed \) scattering for \( Q^{2}\gg 2\Lambda m_{d}\simeq 0.8 \) \gev2.
Other components \( G_{+0}^{+} \) (single helicity flip) and \( G_{+-}^{+} \)
(double helicity flip) should be suppressed by respectively one and two powers
of \( Q \):\begin{equation}
\label{eq:g+}
G_{+0}^{+}=a\left( \frac{\Lambda }{Q}\right) G_{00}^{+}\quad \hbox {and}\quad G_{+-}^{+}=b\left( \frac{\Lambda }{Q}\right) ^{2}G_{00}^{+}\quad \hbox {when}\; Q\rightarrow \infty
\end{equation}
 Since the usual form factors are linear combinations of the \( G_{hh'}^{+} \)'s,
the asymptotic behaviour of observable ratios such as \( B/A \), \( t_{20} \)
and \( t_{21} \) can easily be calculated. Note that the logarithmic corrections
such as appearing in (\ref{eq:redff}) may only be calculated for the dominant,
helicity conserving, form factor. Equation~(\ref{eq:g+}) thus assumes the
same logarithmic corrections for all helicity amplitudes.
\end{itemize}
How do these predictions compare with recent \( ed \) data ? The \( A \) data,
now extending up to \( Q^{2}=6 \) \gev2~\cite{Ale99}, is suggestive of the
expected \( Q^{-20} \) behaviour, though a \( Q^{-16} \) behaviour is still
not excluded (see Fig.~\ref{fig:pqcd_a}): a fit using the five highest \( Q^{2} \)
data points in Ref.~\cite{Ale99} to the dependence \( A\sim Q^{-2m} \) yields
\( m=8.0\pm 0.6 \). Excluding the less precise last point yields \( m=8.7\pm 0.7 \).
More interestingly, using the dipole form factor for \( F_{N} \) in (\ref{eq:redff}),
the \( Q^{2} \) behaviour of \( A \) is reproduced between 2 and 6 \gev2.
As for the helicity amplitudes, their behaviour is tested only up to \( Q^{2}\simeq 2 \)
\gev2, which is the range of the available \( B \)~\cite{Bos90} and \( t_{2j} \)~\cite{Abb00}
data. The pure dominance of the helicity \( 0\rightarrow 0 \) transition (\( a=b=0 \)
in (\ref{eq:g+})~\cite{Bro92}) does not account for this data (see Fig.~\ref{fig:pqcd_ratios}).
The prescription \( a=5 \) and \( b=0 \)~\cite{Kob94} was built to generate
a node in \( G_{M} \), but in order to qualitatively reproduce the \( t_{2j} \)
data as well, one is led to \( a=1.8 \) and \( b=38 \)~\cite{Gar99}. This
implies that the double helicity flip amplitude is as large as the non helicity
flip amplitude and contradicts the applicability of PQCD in the momentum range
considered. Fits to data using these helicity amplitudes lead to the same conclusion~\cite{Abb00p,Kob95}.
All these simple \textsl{ansatz} lead to a sharp increase of the ratio \( B/A \)
for \( Q^{2}\geq 2.5 \) \gev2, which may be an interesting feature for planned
experiments at JLab.

\subsection{Further comparison between models and data}

To conclude this section, we recapitulate some of the theoretical results, in
comparison with observables (Fig.~\ref{fig:da_db_t20t_recap}) and data on
separated form factors (Fig.~\ref{fig:gc_gq_gm_recap}). In the case of \( A \)
and \( B \), deviations with respect to an average representation of the data
(parameterization I) are presented. A summary of remaining ambiguities and foreseable
progress will be given in Sec.~\ref{sec:conclusion}.
\begin{figure}
{\par\centering \resizebox*{8cm}{!}{\includegraphics{da_db_t20t.epsi}} \par}

\caption{\protect\( \Delta A/A\protect \) and \protect\( \Delta B/B\protect \), in
\%, and \t20t as a function of \protect\( Q^{2}\protect \). The deviations
of \protect\( A\protect \) and \protect\( B\protect \) are calculated with
respect to parameterization I. \protect\( A\protect \) data legend: Saclay~\cite{Pla90}
(open diamonds), SLAC~\cite{Arn75} (open circles), JLab/HallA~\cite{Ale99}
(filled circles), JLab/HallC~\cite{Abb99} (filled diamonds). \protect\( B\protect \)
data legend: see Fig.~\ref{fig:ff3}. \t20t data legend: see Figs.~\ref{fig:thq_t20}
and \ref{fig:ff3}. Theoretical calculations: NRIA + MEC + relativistic corrections
(Sec.~\ref{ssec:v/c})~\cite{Are00} (dashed curve) and \cite{Wir95} (dotted),
CIA + MEC (Sec.~\ref{sssec:QPE}) updated from~\cite{Van95a} (solid), LFD
(Sec.~\ref{sssec:LFD})~\cite{Car99a} (dot-dashed).\label{fig:da_db_t20t_recap}}
\end{figure}

\begin{figure}
{\par\centering \resizebox*{10cm}{!}{\includegraphics{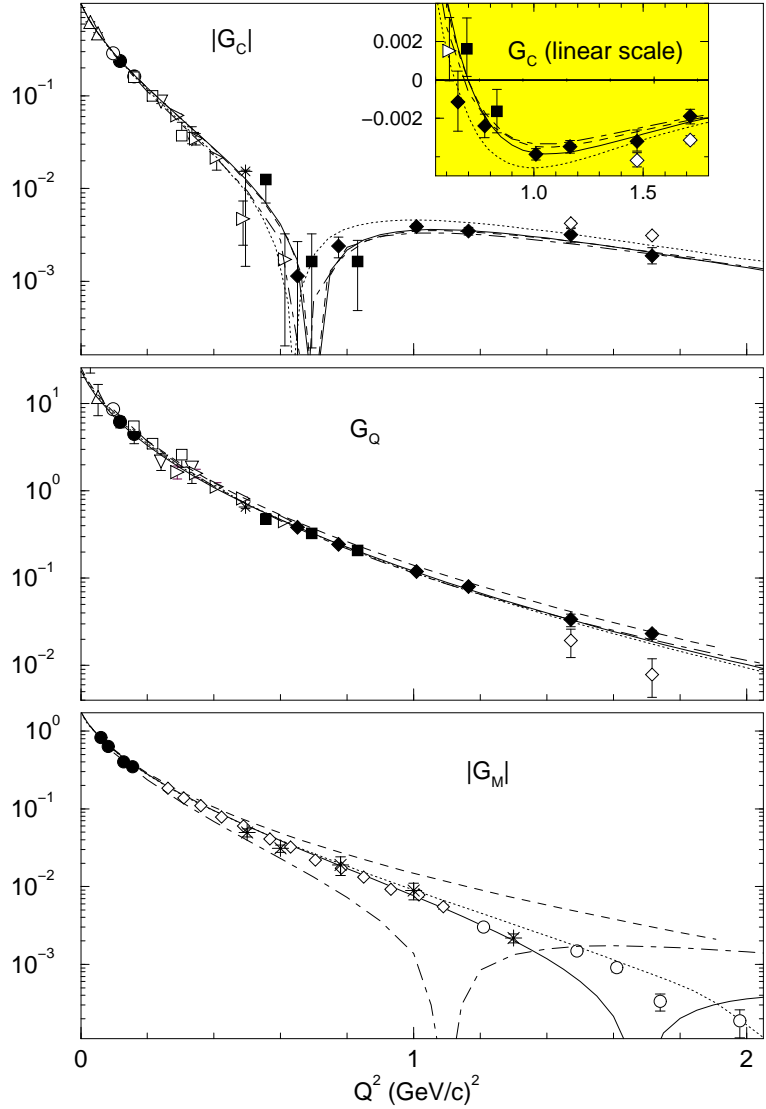}} \par}

\caption{Deuteron form factors \protect\( G_{C}\protect \), \protect\( G_{Q}\protect \)
and \protect\( G_{M}\protect \) as a function of \protect\( Q^{2}\protect \).
Same legend as Fig.~\ref{fig:da_db_t20t_recap}.\label{fig:gc_gq_gm_recap}}
\end{figure}

\section{The nucleon momentum distribution in the deuteron}

Several processes have been investigated for which the cross sections, in the
nonrelativistic impulse approximation, are proportional to the distribution
of the internal momentum \( p \) in the deuteron,\begin{equation}
\rho (p)=u^{2}(p)+w^{2}(p)\, ,
\end{equation}
 with \( u(p) \) and \( w(p) \) defined by Eq.(\ref{eq:uwp}). Note that \( \mathbf{p} \),
the conjugate variable of the relative coordinate \( \mathbf{r} \), is half
the relative momentum between the two nucleons, so that an experiment probing
a momentum \( p \) may be related to elastic electron scattering at a momentum
transfer \( q=2p \). These processes include quasi-free scattering on one of
the nucleons, \( d(p,2p)n \), \( d(e,e'p)n \) and \( d(e,e') \), or equivalently
the detection of the spectator proton in high energy deuteron hadronic break-up,
\( A(d,p) \) . Indeed, they all are proportional to \( \rho (p) \) up to \( p\simeq 200 \)
MeV/c , but deviations from the impulse approximation occur for higher values
of \( p \) . Final state interaction, dynamical excitation of the \( \Delta  \)
resonance and in inclusive reactions pion production have then to be taken account,
thus rendering the interpretation of the experiments less straightforward.

Another interesting feature of these processes is the possibility to access
the ratio \( w(p)/u(p) \) with the measurement of deuteron tensor polarization
observables. A simple relationship between the analyzing power \( T_{20} \)
and the ratio \( w(p)/u(p) \) can be derived in the (nonrelativistic) impulse
approximation, and again deviations are seen, or are to be expected, above \( p\simeq 200 \)
MeV/c .

\subsection{d(e,e'p)n measurements at high missing momenta}

\begin{figure}
{\par\centering \resizebox*{!}{5cm}{\includegraphics{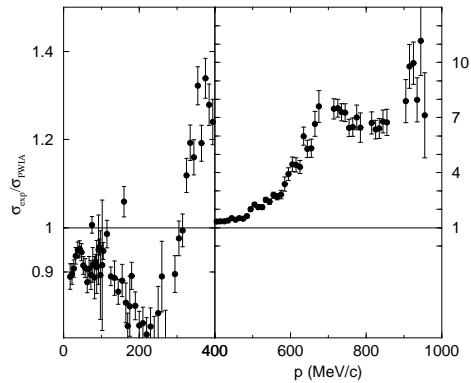}} \par}

\caption{Ratio of the experimental \protect\( d(e,e'p)n\protect \) to the calculated
PWIA (using the Paris potential) cross sections, corresponding to the data of
Ref.~\cite{Blo98}. Systematic errors are of the order of 10 \%. Note the change
of vertical scale at \protect\( p=400\protect \) MeV/c .\label{fig:ratio_rho}}
\end{figure}

The study of the single-particle properties of nuclei through \( (e,e'p) \)
reactions are the subject of an excellent review~\cite{Fru84}. Concerning
the deuteron, the (neutron) missing momentum is identified, in the plane-wave
impulse approximation (PWIA), with the internal momentum \( p \) . The experiments
reaching the highest missing momenta have been carried out at Saclay (500 MeV/c)~\cite{Tur84},
NIKHEF (700 MeV/c)~\cite{Kas98} and MAMI (950 MeV/c)~\cite{Blo98}, but in
kinematical conditions which are not always optimal to study the high momentum
components in the deuteron wave function. Deviations from the PWIA are of the
order of 50 \% at 500 MeV/c and can reach a factor 10 at 1 GeV/c (see Fig.~\ref{fig:ratio_rho}).
They can be understood, if only qualitatively at the highest missing momenta,
in terms of final state interactions, meson exchange currents and \( \Delta  \)
excitation.

Other \( d(e,e'p)n \) measurements (see the reviews~\cite{Kel96,Gil98} and,
among the most recent experiments, Ref.~\cite{Bar99}) do not address specifically
the subject of the nucleon momentum distribution in the deuteron. Their theoretical
understanding is however crucial for a comprehension of this subject.

The first \( (e,e'p) \) measurements using a tensor polarized target have been
carried out at Novosibirsk~\cite{Mos87} and NIKHEF~\cite{Zho99} . They were
limited to low values of missing momenta because of the available luminosity
and of the detector acceptance. This situation will improve with the planned
experiments using the BLAST detector in the Bates stretcher ring~\cite{Tur99}.
In the PWIA, these experiments are directly sensitive to the ratio \( w(p)/u(p) \)
(see below Eq.(\ref{eq:T20_inel.})).

\subsection{d(e,e') measurements and y-scaling}

The concept of nuclear \( y \)-scaling gives a significant insight in the determination
of momentum distribution in nuclei~\cite{Day90}. In the inclusive electron
scattering off deuterium, for sufficiently large values of momentum transfer
and in the PWIA, the cross sections are expected to be proportional to the electron-nucleon
cross sections and the proportionality factor \( f(y)=2\pi \int _{\left| y\right| }^{\infty }p\rho (p)dp \)
represents the longitudinal (along \( \mathbf{q} \)) momentum distribution.
Properly taking into account final state interactions (a correction of about
a factor 2 above 300 MeV/c), a momentum distribution was extracted in a nonrelativistic
formalism up to \( p=600 \) MeV/c~\cite{Cio87}. It is remarkably close to
the one calculated from the Paris potential, but the authors caution about the
absence of relativistic corrections.

\subsection{Hadronic deuteron break-up at high energy\label{ssec:dA}}

Several reactions using high energy (polarized) deuteron beams should allow
a study of the \( dpn \) vertex, provided the deuteron dissociates {}``cleanly{}'',
that is with only one nucleon participating in the process. In the exclusive
reactions \( p(d,p)d \), \( p(d,pp)n \) as well as in the inclusive \( A(d,p)X \),
in some kinematical conditions, the dominant process is the one where the fast
forward proton (or one of the detected protons) can be treated as a spectator.
The energies of the deuteron beams available at SATURNE and at Dubna (several
GeV), together with the relatively high hadronic cross sections, have provided
a handle on these processes involving high momentum components in the deuteron
wave function. Moreover, the tensor analyzing power of these reactions, in the
nonrelativistic impulse approximation, depends only on the \( D/S \) ratio
at a given momentum:\begin{equation}
\label{eq:T20_inel.}
T_{20}\propto \frac{w(w-2\sqrt{2}u)}{u^{2}+w^{2}}.
\end{equation}
Note the similarity between Eqs.(\ref{eq:t20_tilda_1}) and~(\ref{eq:T20_inel.}).
The former is a function of \( 2\eta G_{Q}/3G_{C} \), while the latter depends
on \( w/u \), resulting (in the impulse approximation) in similar shapes for
\( T_{20} \) in \( ed \) elastic scattering and in deuteron break-up.

A relativistic treatment using the deuteron in an infinite momentum frame allows
to define a new variable \( k \)~\cite{Str78} used as the argument of \( u \)
and \( w \) instead of \( p \) for the calculation of the observables. In
other words, the argument of the wave function is no longer equal to the momentum
of the spectator nucleon. Once this transformation is performed, all inclusive
data \( A(d,p)X \) (see the review~\cite{Per92} and Refs.~\cite{Aon95,Sit00}),
for different beam energies and target nuclei, scale approximatively as a momentum
distribution \( \rho (k) \). \( T_{20} \) however deviates significantly from
expectations within the impulse approximation as of \( k\simeq 200 \) MeV/c.
This fact has sometimes been interpretated as the signature of non conventional
components in the deuteron wave function, but this interpretation is not compatible
with the behaviour of \( t_{20} \) in elastic \( ed \) scattering. In reality,
final state interactions and pion production alter significantly the interpretation
of these experiments (see e.g.~\cite{Dol90}). The polarization transfer from
vector polarized deuterons to the fast protons has also been measured up to
\( k\simeq 600 \) MeV/c. The exclusive channels \( p(d,p)d \)~\cite{Arv87,Per92}
and \( p(d,pp)n \)~\cite{Ale94,Bel97} are likewise difficult to interpretate
unambiguously for internal momenta larger than 300 MeV/c.

\section{The deuteron as a source of {}``free{}'' neutrons}

The neutron being loosely bound in the deuteron, deuterium has often been used
as a substitute for a neutron target and deuteron beams as a source of neutron
beams. We briefly mention the list of these applications~:

\begin{enumerate}
\item \( ed \) elastic scattering to extract the neutron charge form factor \( G^{n}_{E} \)~,
\item Quasi-elastic \( \overrightarrow{d}(\overrightarrow{e},e'n)p \) or \( d(\overrightarrow{e},e'\overrightarrow{n})p \)
to measure \( G^{n}_{E}/G_{M}^{n} \)~,
\item Quasi-elastic \( d(e,e'n)p \) and \( d(e,e'p)n \) to measure \( G^{n}_{M}/G^{p}_{M} \)
,
\item Deep inelastic scattering of leptons to extract the neutron structure functions.
\item At intermediate to high energies, the break-up of a vector polarized deuteron
beam can be used to obtain a beam of polarized neutrons. A high intensity may
be achieved using the inclusive reaction \( \overrightarrow{d}+^{9}Be\rightarrow \overrightarrow{n}+X \).
For a better definition of the neutron energy, the beam may be tagged by the
detection of the spectator proton: \( p(\overrightarrow{d},p\overrightarrow{n})p \).
\end{enumerate}
In all cases but the first one~\cite{Pla90,Amg98}, corrections due to the
deuteron structure and to the reaction mechanism were shown to be either negligible
or reliably calculable (see~\cite{Umn94} for example as an application of
relativistic techniques discussed in Sec.~\ref{sssec:BSE}). When a polarized
target is used (cases 2 and 4) or when using polarized deuteron beams (case
5), the effective neutron polarization is equal to the deuteron vector polarization
multiplied by \( (1-3P_{D}/2) \) . This factor accounts for the fact that,
because of the deuteron \( D \) state, the neutron spin is not always aligned
the deuteron spin.

\section{Prospects for the future\label{sec:conclusion}}

The study of the electromagnetic properties and form factors of the deuteron,
from the birth of nuclear physics to the advent of hadronic physics, taking
into account the internal structure of the nucleons and their excitations, has
been very rich. Yet it is not completed. Ambiguities have been and still are
pointed out along this path: although the descriptions of processes involved
can be satisfactory, the calculations and observables do not in general allow
for a unique determination of the \( NN \) interaction, of the neutron charge
form factor, of the isoscalar meson exchange currents and of the precise dynamics
of the system. As already pointed out, some characteristics like the off-shell
behaviour of the \( NN \) interaction are not strictly speaking observables,
and as such cannot be determined unambiguously. As for the manifestation of
the underlying quark substructure in the nuclear properties, it is as elusive
as ever. However the progress in experiment and theory, as summarized in this
paper, is impressive and allows for promising perspectives:

\begin{itemize}
\item The nucleon electromagnetic form factors are being measured with a renewed precision.
In particular, the poorly known neutron electric form factor \( G_{E}^{n} \)
is being determined at various laboratories with polarization techniques which
make this measurement independent of the deuteron structure. In a few years,
all four NEMFF will be better known up to \( Q^{2}\simeq 2 \) \gev2. The description
of the the deuteron form factors to higher four-momentum transfers is now affected
by measurements of \( G_{E}^{p} \). Once these are completed, a small remaining
ambiguity due to the poor knowledge of \( G_{E}^{n} \) above 2 \gev2 will still
be present. Yet new parameterizations of the NEMFF, guided by theoretical models,
should become available soon and be taken into account in future calculations
of the deuteron form factors.
\item Modern \( NN \) interaction models are now fitted directly to the \( NN \)
elastic scattering data and reach a high degree of precision. But, when compared
to some models of the 1980's, these are more empirical. The lack of knowledge
(or assumption) of the underlying dynamics prevents the calculation of such
effects as the Lorentz boost of the deuteron wave function. It would be highly
desirable to have potentials of the one-boson exchange type brought to the degree
of precision of the modern phenomenological potentials. This is also true for
OBE potentials used in completely relativistic calculations. In addition, these
potentials should reproduce the elastic \( NN \) scattering data up to 600
MeV of kinetic laboratory energy in order to match the \( Q^{2} \)-range of
the existing \( ed \) elastic scattering data.
\item There is a definite procedure to construct relativistic corrections, starting
from a nonrelativistic model. Still, most calculations in the past have neglected
one correction or another.
\item From a theoretical point of view, the most satisfying success in the past twelve
years is the implementation of various fully relativistic calculations. Quasi-potentials
approximations to the Bethe-Salpeter equation have been applied with success
to the calculation of the electromagnetic deuteron form factors. The comparative
validity of each of these approximations should be studied further. Likewise,
light-front dynamics yields results which compare very favourably with data
and will still be improved. Concerning Hamiltonian constraint dynamics, calculations
using the three different forms of quantization invoked by Dirac have been developed.
The success of QPE/BSE and LFD has reaching consequences beyond the structure
of the deuteron. The relevance and applicability of relativity in nuclei can
now be explored in the A=3 systems. The same techniques are also applied to
\( q\overline{q} \) configurations.
\item The \( \rho \pi \gamma  \) isoscalar meson exchange contribution, and possibly
the \( \omega \sigma \gamma  \) or other shorter range processes, still remain
difficult to evaluate reliably, because of the lack of constraints on the associated
form factors. Within the nucleon-meson picture of the deuteron, \( ed \) elastic
scattering may provide a way to determine these form factors, provided all points
above are addressed in a systematic way.
\item The role of nucleon isobaric excitations is still very much model dependent.
Still, within the existing models, the \( ed \) elastic scattering data does
not favour very sizeable \( \Delta \Delta  \) components in the deuteron wave
function.
\item The recent \( ed \) elastic scattering data from Jefferson Lab, reaching now
the highest possible four-momentum transfers for the \( t_{20} \) and \( A \)
observables (respectively 1.7 and 6 \gev2), are still compatible with the description
of the deuteron in terms of nucleons and mesons. This is somewhat surprising
since internucleonic scales of 0.1-0.4 fm are being probed. These distances
are smaller than the size of the nucleons themselves, and presumably of the
same order of magnitude as the nucleon quark cores. Still no distinctive experimental
or theoretical signature of the manifestation of quarks in this process was
identified.
\item Quarks degrees of freedom may be explicitly taken into account at intermediate
four-momentum transfers via models, and at high \( Q^{2} \) via perturbative
QCD. If the recent \( A(Q^{2}) \) data seem to follow the \( Q^{2} \)-dependence
expected from PQCD, it is not so for \( B(Q^{2}) \) and \( t_{20}(Q^{2}) \),
albeit at lower \( Q^{2} \). An absolute determination of the leading PQCD
amplitude would clearly be desirable but this depends on a reliable description
of the soft parts of the amplitude which are not calculable in PQCD. As for
quark models of the deuteron, they seem to indicate a specific role played by
quark exchange processes between the nucleons at short distances. Unfortunately,
these models are nonrelativistic by nature and the predicted effects occur at
a scale where relativity should be taken into account. Due to the explicit appearance
of the quark-gluon degrees of freedom, this a much more difficult problem than
for the meson-nucleon models. Appreciable progress in quark models of the deuteron
will require substantial improvements in the technology of relativistic many-body
physics.
\item The planned measurements of \( t_{20} \) at Bates and Novosibirsk will locate
more precisely the position of the node of the charge monopole form factor.
while the ones of \( B(Q^{2}) \) at Jefferson Lab will be performed around
and beyond the node of the magnetic dipole form factor.
\item The recent focus on intermediate and high \( Q^{2} \) should not be detrimental
to the required precision in accounting for the low \( Q^{2} \) data and static
properties. An updated experimental status of the latter was given. The low
\( Q^{2} \) behaviour of any given calculation should be checked carefully,
for example in such representations as in Figs~\ref{fig:t20tR} and~\ref{fig:da_db_t20t_recap}.
\end{itemize}
Going beyond elastic \( ed \) scattering, the same systematic expansions in
\( v/c \), or fully relativistic models, will be applied to the calculation
of the electromagnetic form factors of the A=3 nuclei, and of the deuteron electrodisintegration.
This will provide additional contraints on the remaining ambiguities inherent
in the meson-nucleon models and will lead to a more coherent understanding of
the relativistic structure of few-body nuclei.

\subsubsection*{Acknowledgements}

One of the authors (M.G.) gratefully acknowledges the many
teachings of R.~Beurtey and A.~Boudard on the deuteron and on
polarization. He also benefited greatly from the stimulating
collaborative efforts for the Bates and Jefferson Lab {}``\(
t_{20} \){}'' experiments, in particular with E.J.~Beise,
J.~Cameron, S.~Kox and W.~Turchinetz. The Southeastern
Universities Research Association (SURA) operates the Thomas
Jefferson national Accelerator Facility under DOE contract
DE-AC05-84ER40150.

\appendix

\section{Beyond one photon exchange at high $ Q^{2} $ ?}

Since the \( ed \) elastic cross sections are now measured up to very high
momentum transfers, the question of the validity of the one-photon exchange
approximation (used throughout this review) has to be examined. As shown in
Sec.~\ref{ssec:ffphen}, the set of \( ed \) elastic scattering data, as parameterized
up to \( Q^{2}\simeq 2 \) \gev2, is compatible with Eqs.~(25-31), thus providing
evidence for the validity of the one-photon exchange approximation, at least
within experimental errors. A two-photon exchange contribution would affect
the definition of all observables presented in Sec.~\ref{ssec:obs}, introducing
additional kinematical factors and structure functions~\cite{Gou66}.

\begin{figure}
{\par\centering \resizebox*{!}{3.5cm}{\includegraphics{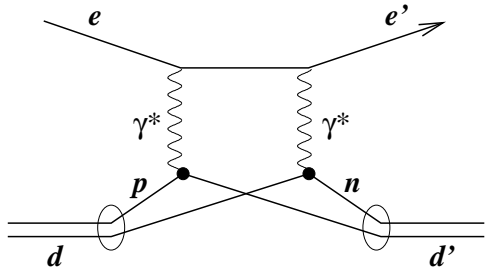}} \par}

\caption{Two-photon exchange diagram.\label{fig:two_photon}}
\end{figure}

Several papers have addressed this question in the early 1970's~\cite{Gun73,Fra73,Boi73,Lev75}:
the two-photon exchange process may be viewed in Glauber theory as the double
scattering illustrated in Fig.~\ref{fig:two_photon}. Estimates of this amplitude
were made with the assumption that each of the exchanged photons carries half
of the momentum transfer. With such approximations, and depending on the relative
phase of the two amplitudes, the double scattering amplitudes could change the
\( ed \) forward cross sections by as much as 10\% for \( Q^{2}\simeq 1-2 \)
\gev2. A complete calculation would involve an integration over all possible
four-momenta of the exchanged photons and over all possible intermediate \( pn \)
states. This has not been attempted yet, except for small \( Q^{2} \) and for
a spinless deuteron~\cite{Her98}. In this particular case, definite corrections
to the deuteron charge radius could be made. Finally, the high \( Q^{2} \)
experimental data on the structure function \( A \)~\cite{Arn75,Ale99,Abb99}
were recently compared, but no conclusive evidence for a two-photon exchange
contribution could be reached~\cite{Rek99}. New dedicated experiments and
calculations would certainly clarify this important question.

\section{Polarized deuteron targets - Polarimeters}

The parallel development of polarized atomic beam sources and solid polarized
targets originated in the late 1950's. The use of adiabatic transitions between
hyperfine energy levels of atomic hydrogen or deuterium for the former and the
dynamic polarization of nuclei for the latter were decisive improvements due
in good part to the work of A. Abragam~\cite{Abr??}. Their use in electron
scattering experiments is however much more recent.

\begin{figure}
{\par\centering \resizebox*{!}{6cm}{\includegraphics{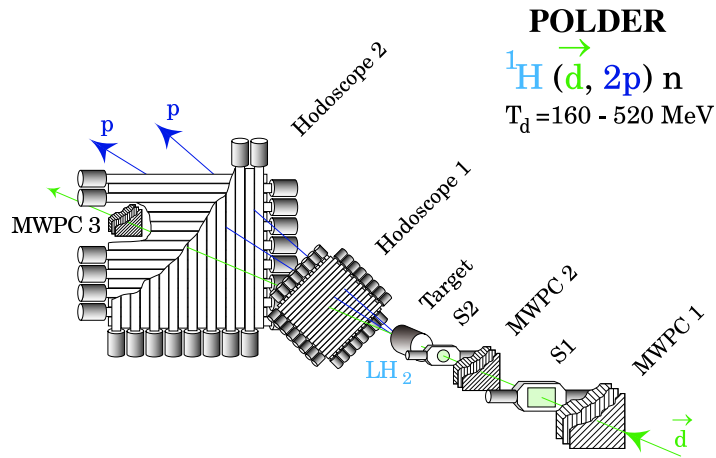}} \par}

\caption{The polarimeter POLDER~\cite{Kox94} as used in the Jefferson Lab \protect\( t_{20}\protect \)
experiment~\cite{Abb00}.\label{fig:polder}}
\end{figure}

The polarized sources led to the acceleration of polarized deuteron beams, which
in turn were used to conceive and calibrate deuteron polarimeters. There are
however very few exemples of efficient deuteron tensor polarimeters at intermediate
energies (in \( ed \) elastic scattering, the recoil deuteron energy is related
to the four-momentum transfer by \( T_{d}=Q^{2}/2M_{d} \)). The most recent
ones were developed at the synchrotron SATURNE-2~\cite{Gar00}: the AHEAD polarimeter~\cite{Cam91}
was based on \( dp \) elastic and inelastic scattering in the 100-200 MeV range,
while POLDER (Fig.~\ref{fig:polder} and \cite{Kox94}), using the charge exchange
reaction \( p(d,pp)n \), was operated between 160 and 520 MeV. There is at
present no good concept to develop a tensor polarimeter for deuteron energies
above 500 MeV. In contradistinction, vector polarimeters are available~\cite{Gar00},
but they are less useful for the determination of the deuteron form factors
(see Sec.~\ref{ssec:obs} and~\cite{Gar90}). A peculiar feature of the double
scattering experiments using a tensor polarimeter is related to the nature of
the \( t_{20} \) moment~: contrarily to other moments and to the more familiar
vector polarization, it does not induce any azimuthal dependence in the distribution
of counts in the polarimeter, which goes like\[
N_{0}(\theta ^{pol})\times [1+t_{20}T^{pol}_{20}(\theta ^{pol})+2t_{21}T^{pol}_{21}(\theta ^{pol})\cos \varphi ^{pol}+2t_{22}T_{22}^{pol}(\theta ^{pol})\cos 2\varphi ^{pol}]\, .\]
 One then needs to know the absolute response of the polarimeter to both polarized
and unpolarized deuterons in order to extract \( t_{20} \). This requires the
separate polarimeter calibration with a deuteron beam of known polarization.
In some instances, the angular dependence of \( T_{20}^{pol} \), if large enough,
may be used to avoid or check the determination of an absolute normalization~\cite{Gar90,Gar94,Abb00}.

Atomic beams have also been used as internal targets in storage rings. In this
case the thinness of the target is compensated by the multiple passage of the
circulating electron beam. The target thickness was later increased by a factor
of about 100 by accumulating the polarized atomic beam in a cell~\cite{Pri94}.
Alternative plans to use an internal target based on the spin-exchange optical
pumping technique~\cite{Cou92} were developed at Argonne and Novosibirsk,
but not implemented~\cite{Pop95}.

Cryogenic solid polarized targets (ND\( _{3} \) or deuterated buthanol) cannot
withstand much electron beam intensity, but this has significantly improved
in the past few years~\cite{Cra97}. The target tensor polarization or alignment
(\( P_{zz} \) or \( A \)) is generally rather small: it is related to the
vector polarization by \( P_{zz}=2-\sqrt{4-3P_{z}^{2}} \)~\cite{Bod91} and
the obtained values of \( P_{z} \) are at best of the order of 0.5.

All these techniques were applied to \( ed \) elastic scattering for the measurement
of \( t_{20} \) or \( T_{20} \). They are compared quantitatively in Table~\ref{tab:fom},
using a figure of merit defined for polarimeter experiments as \( F=E^{2}L\Omega _{e}l\varepsilon (T_{20}^{pol})^{2} \)
and for polarized target experiments as \( F=E^{2}L\Omega _{e}A^{2}/2 \). All
quantities are defined in Table~\ref{tab:fom}. The factor \( E^{2} \) accounts
for the approximate energy dependence of the \( ed \) cross section for a fixed
value of \( Q \). We have also indicated in that table our estimation of the
presently achievable (but not planned) highest figure of merit with polarized
targets, either internal in the upgraded HERMES~\cite{Rit99} configuration
at HERA, or external at Jefferson Laboratory. With present day technology and
ideas, it seems very difficult to extend the tensor polarization measurements
beyond the four-momentum transfer values reached in the JLab/POLDER experiment~\cite{Abb00}.
But upcoming experiments~\cite{Nik00,Tur99} will cover an intermediate \( Q \)-range
with different systematic uncertainties than in the past polarimeter experiments~\cite{Gar94,Abb00},
and in some cases better statistical precision.

\begin{table}

\caption{Figure of merit and maximum four-momentum transfer for different polarization
experiments in \protect\( ed\protect \) elastic scattering.\label{tab:fom}}
{\centering \begin{sideways}
\begin{tabular}{|c||c|c|c|c|c|c|c|c|c|}
\hline
 &
\multicolumn{3}{|c|}{ \textbf{Recoil polarimeters} }&
\multicolumn{6}{|c|}{\textbf{Polarized targets} \( e\overrightarrow{d}\rightarrow ed \)}\\
\cline{5-10}
&
\multicolumn{3}{|c|}{\( ed\rightarrow e\overrightarrow{d} \)}&
\multicolumn{4}{|c|}{\textbf{Internal} }&
\multicolumn{2}{|c|}{\textbf{External} }\\
\cline{2-4} \cline{5-8} \cline{9-10}
&
Bates/&
Bates/&
JLab/&
VEPP-3&
NIKHEF&
Bates/&
\textsl{Achievable}&
Bonn&
\textsl{Achievable}\\
&
Argonne&
AHEAD&
POLDER&
&
&
BLAST&
&
&
\\
&
\cite{Sch84}&
\cite{Gar94}&
\cite{Abb00}&
\cite{Voi86}/\cite{Nik00}&
\cite{Bou99}&
\cite{Tur99}&
&
\cite{Bod91}&
\\
\hline
\hline
Beam energy \( E \) (GeV)&
.371&
.850&
4.05&
2.0&
.704&
\textsl{1.0}&
\textsl{27.5}&
2.0&
\textsl{6.0}\\
\hline
Beam intensity&
30 \( \mu  \)A&
30 \( \mu  \)A&
100 \( \mu  \)A&
200/100 mA &
150 mA&
\textsl{100 mA}&
\textsl{30 mA}&
0.4 nA&
\textsl{80 nA}\\
\hline
Target thickness (d/cm\( ^{2} \))&
\( 4\times 10^{21} \)&
\( 2.4\times 10^{23} \)&
\( 6\times 10^{23} \)&
\( 6\times 10^{11} \)\textsl{/\( 10^{14} \)}&
\( 2\times 10^{13} \)&
\textsl{10\( ^{14} \)}&
\textsl{10\( ^{14} \)}&
\( 9\times 10^{22} \)&
\textsl{2\( \times  \)10\( ^{23} \)}\\
\hline
Luminosity \( L \) (cm\( ^{-2} \)s\( ^{-1} \))&
\( 8\times 10^{35} \)&
\( 4.5\times 10^{37} \)&
\( 4\times 10^{38} \)&
\( 7\times 10^{29} \)\textsl{/\( 5\times 10^{31} \)}&
\( 2\times 10^{31} \)&
\textsl{5\( \times  \)10\( ^{31} \)}&
\textsl{2\( \times  \)10\( ^{31} \)}&
\( 2\times 10^{32} \)&
\textsl{10\( ^{35} \)}\\
\hline
Solid angle \( \Omega _{e} \) (msr)&
20&
18&
6&
120/160&
150&
\textsl{100}&
\textsl{150}&
5&
\textsl{6}\\
\hline
Loss factor \( l \)&
.5&
.25&
.5&
-&
-&
-&
-&
-&
-\\
\hline
Polarimeter efficiency \( \varepsilon  \)&
\( 10^{-4} \)&
\( 2\times 10^{-3} \)&
\( 2\times 10^{-3} \)&
-&
-&
-&
-&
-&
-\\
\hline
Analyzing power \( T^{pol}_{20} \)&
\( -0.8 \)&
\( -0.4 \)&
\( -0.2 \)&
-&
-&
-&
-&
-&
-\\
\hline
Target polarization \( A/\sqrt{2} \)&
-&
-&
-&
.46/.85&
.83&
\textsl{.85}&
\textsl{.85}&
.12&
\textsl{.21}\\
\hline
\hline
Figure of merit \( F \) &
\( 3\times 10^{28} \)&
\( 5\times 10^{31} \)&
\( 1.4\times 10^{33} \)&
\( 8\times 10^{28} \)\textsl{/\( 2\times 10^{31} \)}&
\( 9\times 10^{29} \)&
 \textsl{5\( \times  \)10\( ^{30} \)}&
\textsl{1.5\( \times  \)10\( ^{33} \)}&
\( 6\times 10^{28} \)&
\textsl{10\( ^{33} \)}\\
\hline
\hline
\( Q_{max} \) (fm\( ^{-1} \))&
2.0&
4.6&
6.7&
2.9/4.0&
3.2&
\textsl{(4.5)}&
\textit{(7)}&
(3.6)&
\textsl{(6.5)}\\
\hline
\end{tabular}
\end{sideways}\par}\end{table}

\section{Nucleon electromagnetic form factors}

We present here the most commonly used parameterizations of the nucleon form
factors (NEMFF) in various deuteron form factors calculations. In this case
the quantities of interest are the isoscalar combinations \( G_{E,M}^{S}=G_{E,M}^{p}+G_{E,M}^{n} \).
We recall that in the NRIA, the deuteron charge form factors are proportional
to \( G_{E}^{S} \) ; in this case, the \( A_{L} \) elastic structure function
(Sec.~\ref{ssec:ffphen}) is proportional to \( \left( G_{E}^{S}\right) ^{2} \)
while \t20t is independent of any nucleon form factor.

\subsection*{The dipole parameterization}

The proton form factors and the neutron magnetic form factor can be written
approximatively as\begin{equation}
\label{eq:gpgmngd}
G_{E}^{p}=\frac{G_{M}^{p}}{\mu _{p}}=\frac{G_{M}^{n}}{\mu _{n}}=G_{D}=\left( 1+\frac{Q^{2}}{0.71}\right) ^{-2}
\end{equation}
while the neutron electric form factor is taken to be zero or following the
Galster parameterization~\cite{Gal71}:\begin{equation}
\label{eq:gen}
G_{E}^{n}=-\frac{a\eta }{1+b\eta }\mu _{n}G_{D}
\end{equation}
 where \( a \) and \( b \) are two free parameters adjusted on data. Frequently
used values of these parameters are given in Table~\ref{tab:gen}.

\begin{table}

\caption{Neutron electric form factor parameters for Eq.(\ref{eq:gen}).\label{tab:gen}}
{\centering \begin{tabular}{|c|c|c|}
\hline
a&
b&
Ref.\\
\hline
\hline
1.&
5.6&
\cite{Gal71}\\
\hline
1.25&
18.3&
\cite{Pla90}\\
\hline
1.&
3.4&
\cite{Her99}\\
\hline
\end{tabular}\par}\end{table}

An alternative parameterization of \( G_{D} \) is a product of two different
monopole form factors~\cite{Bil72}. \( G_{E}^{p} \) and \( G_{M}^{p} \)
can also be described as a sum of four monopole form factors~\cite{Sim80}.

\subsection*{The IJL-G parameterization}

The IJL parameterization~\cite{Iac73} rests on the vector meson dominance
(VMD) model: the interaction of the photon with the nucleon is mediated by the
\( \rho  \) (isovector) and by the \( \omega  \) and \( \phi  \) (isoscalar)
mesons. The parameters are adjusted on data available in 1972 and the form factors
do not have the asymptotic behaviour anticipated from PQCD, so that this parameterization
should not be used beyond 1 \gev2. In several calculations of the deuteron form
factors, a parameterization using IJL for \( G_{E}^{p} \) ,\( G_{M}^{p} \)
and \( G_{M}^{n} \), and the Galster parameterization (\ref{eq:gen}) for \( G_{E}^{n} \)
was often used. It is denoted by IJL-G.

\subsection*{The Höhler parameterization}

This parameterization~\cite{Hoh76} also rests on VMD, but uses a larger and
corrected data set, as well as additional heavier mesons.

\subsection*{The GK parameterization}

In addition to the VMD model (with the \( \rho  \) and \( \omega  \) mesons),
this parameterization (GK85)~\cite{Gar85} incorporates scaling laws at high
momentum transfer compatible with PQCD. A later version (GK92)~\cite{Gar92}
includes the \( \phi  \) meson but is in worse agreement with recent precise
data on \( G_{E}^{p} \)/\( G_{M}^{p} \)~\cite{Jon00}.

\subsection*{The MMD parameterization}

This more recent parameterization~\cite{Mer96} includes four isovector and
three isoscalar mesons and offers in addition a more precise description of
the low \( Q \) behaviour of the form factors.

\begin{figure}
{\par\centering \resizebox*{!}{9cm}{\includegraphics{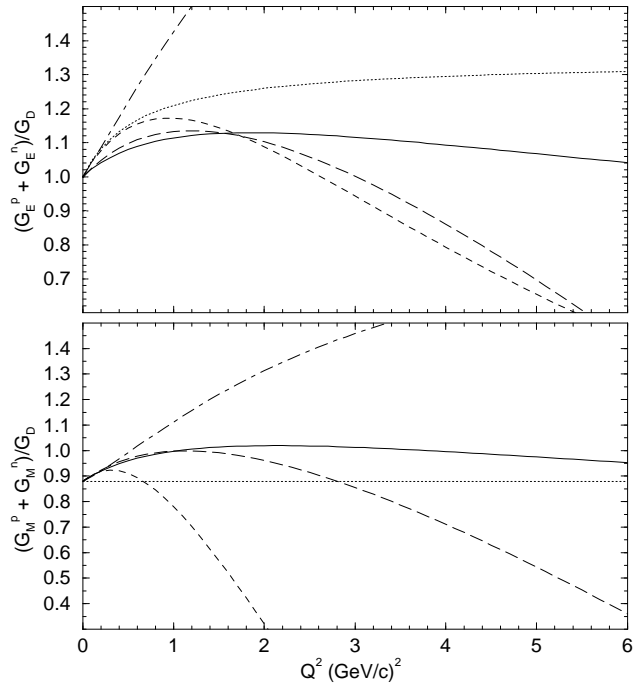}} \par}

\caption{Ratios of the isoscalar nucleon form factors (electric and magnetic) to the
dipole form factor. The parameterizations are the dipole-Galster of Eqs.(\ref{eq:gpgmngd},\ref{eq:gen})
(dotted line), IJL-G (dashed), Höhler (long dashed), GK85 (dot-dashed) and MMD
(solid). See text for notations and references.\label{fig:grap_iso}}
\end{figure}

Very recent data (see for example the reviews in~\cite{Bates25}) on the proton
and neutron form factors have not yet been incorporated into any of these parameterizations,
which differ appreciably from each other. At the present time, the MMD parameterization
seems the most satisfactory, but more precise data and parameterizations of
the NEMFF are certainly needed. Finally, in Fig.~\ref{fig:grap_iso} are plotted
the isoscalar nucleon form factors, which are the relevant combinations for
the deuteron form factors calculations.

\bibliographystyle{unsrt}
\bibliography{bibfile}

\end{document}